\documentclass[12pt]{article}
\newlength\mytemplength

\newcommand{\argmax}{\mathop{\rm arg\,max}}

\NewDocumentCommand{\cmt}{om}{%
    \IfNoValueTF{#1}%
    { %
        \textcolor{DefaultColorComment}{[ #2 ]}%
    }%
    { %
        \textcolor{DefaultColorComment}{[ #1 | #2 ]}%
    }%
}

\usepackage{xcolor}
\usepackage{xparse} %
\definecolor{DefaultColorComment}{RGB}{120,120,120}
\definecolor{ThanqingColorComment}{RGB}{0,120,180} %

\NewDocumentCommand{\cmthanqing}{om}{%
    \IfNoValueTF{#1}%
    {%
        \textcolor{ThanqingColorComment}{[ #2 ]}%
    }%
    {%
        \textcolor{ThanqingColorComment}{[ #1 | #2 ]}%
    }%
}

\usepackage{xparse}
\usepackage{multirow}
\usepackage{makecell}
\usepackage[todonotes={textsize=tiny}, defaultcolor=red]{changes}
\usepackage{setspace,graphicx,epstopdf,amsmath,amsfonts,amssymb,amsthm}
\makeatletter

\@ifundefined{verbatim@@@}{\input{verbatim.sty}}{}

\def\includeversion#1{%
  \expandafter\def\csname #1\endcsname{}%
  \expandafter\def\csname end#1\endcsname{}%
  \expandafter\let\csname not-#1\endcsname\comment
  \expandafter\let\csname endnot-#1\endcsname\endcomment
  \expandafter\def\csname if#1\endcsname##1##2{##1}
}
\def\excludeversion#1{%
  \expandafter\let\csname #1\endcsname\comment
  \expandafter\let\csname end#1\endcsname\endcomment
  \expandafter\def\csname not-#1\endcsname{}%
  \expandafter\def\csname endnot-#1\endcsname{}%
  \expandafter\def\csname if#1\endcsname##1##2{##2}
}
\makeatother
\usepackage{marginnote,datetime,enumitem,subfigure,rotating,fancyvrb}
\usepackage{float}
\usepackage[
  colorlinks=true,
  linkcolor=blue,
  citecolor=blue]{hyperref}
\usepackage{booktabs}
\usepackage{natbib}
\usepackage{mathtools}
\usepackage{lscape}
\usepackage[toc,page]{appendix}

\usepackage{tikz}
\usetikzlibrary{arrows.meta}
\usetikzlibrary{positioning}
\usepackage{pgfplots}
\pgfplotsset{compat=newest}
\pgfplotsset{plot coordinates/math parser=false}
\newlength\figureheight
\newlength\figurewidth

\usepackage{forest}

\usetikzlibrary{trees}
\usdate
\excludeversion{notes}        %
\includeversion{links}          %
\iflinks{}{\hypersetup{draft=true}}
\usepackage[margin=1in]{geometry}%
\makeatletter\let\chapter\@undefined\makeatother %

\usepackage{bbm}
\usepackage{indentfirst} %
\makeatletter

\newtheoremstyle{jf}%
{6pt}%
{6pt}%
{\itshape}%
{}%
{}%
{:}%
{.5em}%
{}%

\theoremstyle{jf}

\renewcommand{\@seccntformat}[1]{{\csname the#1\endcsname}.\hspace{1em}}

\def\thesection       {\Roman{section}}
\def\thesubsection    {\Alph{subsection}}
\def\thetable         {\Roman{table}}

\renewcommand{\section}{\@startsection
  {section}{1}{0mm}{-3.5ex \@plus -1ex \@minus -.2ex}{2.3ex \@plus.2ex}{\centering\normalfont\Large\bfseries}}

\renewcommand{\subsection}{\@startsection
  {subsection}{2}{0mm}{-3.25ex \@plus -1ex \@minus -.2ex}{1.5ex \@plus.2ex}{\normalfont\large\itshape}}

\renewcommand{\p@subsection}{\thesection .}
\renewcommand{\p@subsubsection}{\thesection .}

\expandafter\def\csname jf@appprefix@section\endcsname{Appendix }
\expandafter\def\csname jf@appprefix@subsection\endcsname{}
\expandafter\def\csname jf@appprefix@subsubsection\endcsname{}
\expandafter\def\csname jf@appprefix@paragraph\endcsname{}

\def\appendix{\par
  \setcounter{section}{0}%
  \setcounter{subsection}{0}%
  \gdef\thesection{\@Alph\c@section}%
  \renewcommand{\theequation}{\thesection\arabic{equation}}%
  \renewcommand{\@seccntformat}[1]{%
    \csname jf@appprefix@##1\endcsname
    {\csname the##1\endcsname}.\hspace{1em}}
  \renewcommand{\section}{\setcounter{equation}{0}\@startsection
    {section}{1}{0mm}{-3.5ex \@plus -1ex \@minus -.2ex}{2.3ex \@plus.2ex}{\centering\normalfont\Large\bfseries}}
}

\makeatother
\usepackage[labelfont=bf,labelsep=period]{caption}   %

\newdateformat{monthyeardate}{%
  \monthname[\THEMONTH], \THEYEAR}
\usepackage{float} %

\definecolor{DefaultColorComment}{rgb}{0.0, 0.0, 0.85} 

\usepackage{listings} %

\usepackage{tcolorbox} %

\tcbset{
    colframe=black,     %
    colback=white,      %
    boxrule=0.5pt,      %
    arc=0mm,            %
    left=5pt,
    right=5pt,
    top=5pt,
    bottom=5pt
}

\floatstyle{plain}
\newfloat{Prompt}{htbp}{lop} %
\floatname{Prompt}{Prompt}

\DeclareCaptionLabelFormat{promptlabel}{Prompt~#2}
\usepackage{placeins}
\usepackage{booktabs}
\usepackage{threeparttable}
\usepackage[normalem]{ulem} %

\begin{document}
\setlist{noitemsep}  %
\onehalfspacing      %
\parskip 3pt

\setlist{noitemsep}  %
\onehalfspacing      %
\parskip 3pt

\title{A Financial Brain Scan of the LLM}
\author{Hui Chen, Antoine Didisheim, \\ Mo Pourmohammadi, Luciano Somoza, Hanqing Tian\footnote{Hui Chen is at MIT Sloan and NBER (\href{mailto:huichen@mit.edu}{huichen@mit.edu}). Antoine Didisheim is at the University of Melbourne (\href{mailto:antoine.didisheim@unimelb.edu.au}{antoine.didisheim@unimelb.edu.au}). Mo Pourmohammadi is at Yale School of Management (\href{mailto:mo.pourmohammadi@yale.edu}{mo.pourmohammadi@yale.edu}). Luciano Somoza is at ESSEC Business School (\href{mailto:somoza@essec.edu}{somoza@essec.edu}). Hanqing Tian is at the University of Melbourne (\href{mailto:hanqing.tian1@unimelb.edu.au}{hanqing.tian1@unimelb.edu.au}). We thank participants at the Toronto Asset Pricing and Investments Workshop 2026 and the Stanford Institute for Theoretical Economics (SITE) 2026, as well as seminar participants at Oxford University and the University of Technology Sydney for their valuable comments. Any remaining errors are our own. The authors have nothing to disclose.}}

\date{\today}

\maketitle
\begin{center}
\href{https://ssrn.com/abstract=5412277}{Latest version}
\end{center}
\bigskip

\begin{abstract}
We use sparse autoencoders (SAEs) to ``brain scan'' large language models, mapping internal activations to interpretable concepts and enabling targeted ``steering'' of specific features. We show that prompting models to adopt specific preferences activates many unintended concepts, complicating causal inference. In contrast, SAEs permit smooth, more isolated steering of target concepts. Furthermore, our scans reveal a stark disconnect between the model's internal feature contributions and its self-generated explanations, showing that these explanations are unreliable guides to its reasoning. Finally, SAEs provide a lightweight intervention for existing models, producing interpretable embeddings and allowing researchers to attenuate specific economic biases.

\end{abstract}

\medskip

\thispagestyle{empty}

\newpage
\clearpage

\setcounter{page}{1}

\section{Introduction}
\label{sec:intro}

Large language models (LLMs) are increasingly used in financial analysis, research workflows, and investment decisions \citep{cheng2024does,lu2026markets}. Their appeal is clear: they can process and summarize large volumes of unstructured text. Among the challenges these models present, two are central to our analysis. First, their internal representations are dense and difficult to interpret, making it hard to identify which concepts drive their outputs \citep{ludwig2025large}. Second, their outputs can reflect latent biases, including systematic differences associated with demographic characteristics \citep{fedyk2024chatgpt}. We address these challenges by making the model's internal representations interpretable and controllable.

We apply sparse autoencoders (SAEs), an emerging interpretability technique from computer science \citep[see, e.g.,][]{cunningham2023sparse,gao2024scaling,shi2025route}, to economic applications. An SAE is a small auxiliary network attached to an already trained LLM. It transforms the model's dense internal state into a larger, sparse representation in which only a small subset of features activates for any given input. Each feature is associated with an individual concept. Researchers can therefore observe which concepts activate as the model processes an input and intervene by increasing or suppressing selected activations without retraining the LLM or modifying its underlying weights.

Neuroscientists can study the human brain by asking people to perform different cognitive tasks during a brain scan and comparing which brain regions become active as the task changes. Our SAE enables an analogous exercise for the LLM. To create economic counterparts to brain regions, we need to identify and organize the features associated with economic concepts. Off-the-shelf SAEs are general-purpose, so many of their features capture concepts unrelated to economics and finance. We therefore develop a finance-specific labeling and selection procedure that maps 7,278 SAE features to 46 economic topics organized into five high-level themes. These topics and themes serve as our economic ``brain regions,'' allowing us to examine how the model processes the same information when answering different economic questions.

Unlike neuroscientists studying the human brain, we can directly manipulate individual features within these economic ``regions,'' increasing or suppressing their activation with precise numerical control. This allows us to compare two methods of constructing economic agents: prompt steering and SAE steering. Prompt steering induces a trait, preference, or bias by describing it in the input, for example, ``You are a highly risk-averse investor.'' We use risk aversion as a concrete application and show that this prompting approach fails to isolate the targeted concept. Asking the model to become more risk-averse simultaneously activates many unintended concepts, making it difficult to attribute behavioral changes to the intended preference. Prompt steering also produces abrupt changes rather than smooth, calibrated adjustments in behavior. By contrast, direct steering of the SAE's \textit{Risk Aversion} feature changes the targeted concept continuously while producing substantially smaller changes in other measured features. These properties provide a more credible basis for causal analysis in economic agent simulations.

\paragraph{Constructing and Validating the Brain Scan.} The standard approach to labeling SAE features is bottom-up. Researchers pass a general-purpose text corpus through the model, identify the passages that most strongly activate each feature, and ask another LLM to describe their shared theme. Because both the SAE and the corpus are general-purpose, the resulting labels span subjects ranging from cooking recipes to sports, and most are of little use for economic research. Our first contribution is a finance-specific labeling and selection pipeline. We define 46 granular economic topics in advance, including \textit{Leverage}, \textit{Monetary Policy}, and \textit{Risk Aversion}. We organize these topics into five broad themes: \textit{Corporate Finance}, \textit{Investment \& Portfolio Management}, \textit{Risk Management \& Derivatives}, \textit{Financial Markets \& Instruments}, and \textit{Macro-Finance \& Economics}. We use an LLM to generate representative texts for each topic and assign each SAE feature to the topic for which it activates most strongly and selectively. This procedure both labels the features and isolates the subset relevant to our analysis, reducing the feature space from 131,072 features to 7,278 economically relevant ones.

To assess both our labeling procedure and the broader usefulness of SAEs for financial analysis, we conduct two initial validation exercises. First, we give the model the same neutral prompt, ``Give me 3 finance/economic concepts without explanation,'' while steering its internal state toward each of our 46 topics in turn. With the prompt held fixed, the model consistently generates vocabulary associated with the targeted topic. For example, steering toward \textit{Derivatives} produces terms such as ``option,'' ``call,'' and ``straddle,'' whereas steering toward \textit{Governance} produces ``board,'' ``succession,'' and ``activist.'' The alignment between the steered topic and the generated vocabulary provides direct validation of our topic assignments.

Second, we validate the model's ability to form economically meaningful expectations. Following \citet{bybee2023surveying}, we prompt the SAE-augmented LLM to read news articles and form expectations about fourteen macroeconomic variables. For thirteen of the fourteen variables, the correlations between the LLM and human survey expectations are broadly in line with the baseline results reported by \citet{bybee2023surveying}. These results alleviate concerns that the LLM we use lacks the capacity to form economically meaningful expectations or that the SAE procedure impairs this capacity.\footnote{The portfolio-construction exercise in Section~\ref{sec:data_reuters} provides further validation: portfolios based on SAE features match or exceed the performance of the dense-embedding benchmark following \citet{chen2022expected}.}

Our method allows direct comparisons at the article level, revealing stark contrasts in how the LLM's ``brain'' activates when reading the same news for different forecasting exercises. Aggregating the gradient-based contributions into our five high-level themes shows how these differences extend across the news sample. AAA bond rate forecasts draw most heavily on \textit{Financial Markets \& Instruments}, particularly \textit{Fixed Income}. Real consumption forecasts emphasize \textit{Macro-Finance \& Economics}, especially \textit{GDP} and \textit{Quantitative Easing}, while real nonresidential investment forecasts draw primarily on \textit{Corporate Finance}, particularly \textit{Investments} and \textit{CAPEX}.

\paragraph{Steering \& Agent Simulation.}
Researchers increasingly use LLMs to simulate economic agents \citep[e.g.,][]{horton2023homo}. A growing literature shows that these models can reproduce decision-making patterns observed across different human populations and economic settings \citep[e.g.,][]{fedyk2024chatgpt,aher2023simulate,argyle2023outofone}. Researchers can therefore use a single LLM to construct heterogeneous simulated agents by assigning different traits through the prompt. For example, a prompt may assign the agent a demographic profile (``You are a female investor...''), a psychological preference (``You are highly risk-averse...''), or a cognitive bias. The underlying model remains fixed, and differences in behavior across simulated agents are attributed to the traits assigned through these prompts.

This approach relies on the assumption that changing the prompt alters only the assigned trait. We show that this assumption fails. Prompting the model to adopt a particular trait changes not only the targeted concept but also many unrelated concepts. Behavioral differences across prompts therefore cannot be attributed solely to the intended preference, compromising causal interpretation in prompt-steered agent simulations.

We demonstrate this limitation in a controlled simulation in which the LLM chooses between two investment opportunities. We increase the LLM's risk aversion using two methods. Prompt steering adds explicit instructions that tell the model to become more risk-averse, whereas SAE steering directly increases the activation of a selected \textit{Risk Aversion} feature. Our brain scan reveals that both methods increase the targeted concept, but their broader effects differ substantially. SAE steering directly modifies the selected feature, while prompt steering also changes many unrelated concepts, some by magnitudes comparable to the change in risk aversion. Prompt-induced changes in behavior therefore combine the effect of risk aversion with changes in other dimensions of the model's reasoning.

This contamination has direct consequences for the model's economic reasoning. We demonstrate this with a second experiment. We present the LLM with a choice between two projects with identical risk but different net present values (NPVs). Because the higher-NPV project strictly dominates, the correct choice should be invariant to the LLM's level of risk aversion. Under SAE steering, the model's decision remains unchanged across all steering strengths we test, and it consistently selects the higher-NPV project. Under prompt steering, however, the model eventually reverses its decision and selects the dominated, lower-NPV project as the instruction to become more risk-averse intensifies. Our brain scan shows that this reversal coincides with changes in unrelated concepts, including a reduction in the activation associated with \textit{Capital Budgeting}. We obtain the same general pattern across alternative prompts and decision environments. Prompt steering changes concepts unrelated to the intended preference and can overturn otherwise correct financial decisions, whereas SAE steering preserves the correct response throughout the range of interventions we examine.

A second limitation of prompt-based steering is its inability to produce smooth, continuous changes in LLM behavior. To demonstrate this, we present the LLM with a binary choice between a safe and a risky investment. Following \citet{chen2024out}, we examine the next-token probability that the model assigns to each decision. Rather than observing only the model's final choice, these probabilities reveal how strongly the model favors one option over the other. The probability assigned to the risky asset can range from nearly 0\%, when the model strongly favors the safe asset, to nearly 100\%, when it strongly favors the risky asset, with intermediate values capturing less decisive preferences. As we increase the activation of the SAE's \textit{Risk Aversion} feature, this probability changes continuously, allowing us to move the model gradually from the risky asset toward the safe asset. Prompt steering does not provide the same granularity. Even when we vary the instruction incrementally by asking the model to become ``1\%,'' ``2\%,'' or ``3\% more risk-averse,'' its decision probability changes through abrupt, discrete jumps rather than gradual adjustments.

Controlled experiments require interventions that can be calibrated and applied across decision settings. SAE steering provides a numerical coefficient that researchers can calibrate on one prompt and transfer to related prompts, yielding similar behavioral responses. Prompt steering lacks this portability: the same instruction can produce small, abrupt, or negligible changes depending on the setting.

\paragraph{Steering and Bias Correction.}
A central problem in applying LLMs to economics is their systematic biases \citep{bini2026behavioral} and tendency to produce spurious economic reasoning. We show how SAEs provide an interpretable way to identify the internal concepts underlying these errors and intervene on them directly.

A compelling example emerges from our earlier analysis of macroeconomic expectations: the three-month Treasury bill rate is the only one of the fourteen forecasts that fails to match the baseline results reported by \citet{bybee2023surveying}. During the 2008 financial crisis, the LLM predicts a rate increase while professional forecasters predict a decline. The model justifies its prediction through a ``flight-to-safety'' narrative, overlooking the central bank's interest-rate cuts. We correct this omission by increasing the activation of the SAE feature most strongly associated with \textit{Quantitative Easing}. The model then predicts a rate decline and cites the central bank's response in its justification.

To show bias correction, we turn to the optimism bias associated with post-training alignment \citep{LiuKirshner2024}. Following \citet{lopez2023can} and \citet{chen2024out}, we ask the LLM to classify firm-level news as positive or negative. Increasing the activation of an SAE feature associated with positive sentiment monotonically raises the share of positive classifications; decreasing it reduces this share and attenuates the model's baseline optimism. Downward steering makes the model more selective: fewer news items receive positive classifications, but those that do are followed by larger positive returns. The resulting long-short portfolio's annualized Sharpe ratio rises from 4.24 to 4.51. By contrast, prompting the model to be ``less optimistic'' produces abrupt classification changes and Sharpe ratios well below the unsteered baseline. SAE steering therefore attenuates optimism while preserving the economic information in the model's classifications.

\paragraph{Interpretable Embeddings.}
An emerging literature uses LLM embeddings as inputs to predictive models \citep{chen2022expected,didisheim2026inefficient,zhang2026large}. This approach retains information from the model's internal representations and is generally more tractable and stable than prompt-based classification. Its main limitation is interpretability: dense embeddings distribute economic concepts across many dimensions, obscuring which concepts drive predictions. SAEs transform these representations into sparse activations mapped to economically meaningful concepts, preserving the tractability of embedding-based methods while revealing the economic content underlying their predictions.

We next examine whether interpretability comes at the cost of predictive performance. Following \citet{chen2022expected}, we use embeddings extracted from firm-level news to predict subsequent stock returns. We compare our 7,278 economically labeled SAE features with standard dense embeddings from the same LLM. Interpretability does not reduce predictive content: the long-short portfolio using all labeled SAE features achieves an annualized Sharpe ratio of 5.22, compared with 3.67 for the dense benchmark. Even the 50 most predictive labeled features yield a Sharpe ratio of 3.84, slightly exceeding the benchmark.

The SAE representation reveals the economic sources of return predictability. Among the ten most predictive features, \textit{Macro-Finance \& Economics} and \textit{Financial Markets \& Instruments} each account for approximately 40\%. As we include more features, the composition shifts toward \textit{Corporate Finance}, which becomes the largest theme at approximately 1,000 features. Thus, macroeconomic and market-wide information provides a few strong individual signals, while firm-specific predictive content is distributed across a broader set of features.

\paragraph{Related Literature.}
Our paper contributes to four strands of literature. The first uses LLMs as empirical tools in finance and economics, including applications to news-based return prediction \citep{lopez2023can,chen2022expected}, financial-statement analysis \citep{kim2024financial}, macroeconomic expectations \citep{bybee2023surveying}, investor adoption \citep{cheng2024does}, and economic research methodology \citep{korinek2023generative,ludwig2025large}. Most of this work takes the model's generated output or embedding as the empirical object. We instead examine the model's internal representations to identify which economic concepts contribute to a given answer.

A second strand uses LLMs to simulate economic agents whose characteristics and roles are assigned through prompts. Applications range from synthetic survey respondents and generative agents \citep{horton2023homo,aher2023simulate,argyle2023outofone,park2023generative} to decision makers assigned particular financial advisory roles, risk preferences, or other economic preferences \citep{fieberg2025financial,liu2025risk,kim2024choice,cook2026want}. Our results identify an important limitation of this approach. A prompt intended to change one trait also activates many unrelated concepts, so prompt-assigned personas do not isolate the intended treatment. Observed behavioral changes may therefore reflect both the assigned trait and collateral changes in other dimensions of the model's reasoning.

A third strand documents biases and fragility in LLM behavior, including demographic and perception biases, forecast optimism, and look-ahead or memorization artifacts \citep{fedyk2024chatgpt,gallegos2024bias,LiuKirshner2024,sarkar2024lookahead,lopez2025memorization}. We complement this literature by identifying internal concepts associated with these failures and testing whether intervening on those concepts directly can correct the resulting predictions and reasoning.

A fourth strand uses textual data to predict returns and price assets. Early work develops measures of media sentiment and textual content \citep{tetlock2007giving,loughran2011liability,engelberg2011causal}, while recent studies use dense LLM embeddings as predictive inputs \citep{chen2022expected,didisheim2026inefficient,zhang2026large}. Related work uses SAEs to construct interpretable news-based firm characteristics and asset-pricing factors \citep{didisheim2026factorzoo}. We show that economically labeled SAE features preserve and improve upon the predictive content of dense embeddings while revealing the economic concepts underlying return predictability.

The remainder of the paper is organized as follows. Section~\ref{sec:methodology} introduces the SAE methodology, presents the finance-specific labeling procedure, and validates the resulting labels through a steering exercise. Section~\ref{sec:methodology_belief} applies the brain scan to the formation of macroeconomic expectations and decomposes the model's forecasts into interpretable economic concepts. Section~\ref{sec:steer_and_agent_simulation} compares SAE and prompt steering in economic agent simulations using a risk-aversion case study. Section~\ref{sec:steering_bias_correction} applies SAE steering to two settings: a Treasury bill forecasting error during the 2008 financial crisis and optimism bias in news-based return prediction. Section~\ref{sec:data_reuters} evaluates whether interpretable SAE embeddings retain predictive content in a firm-level news return-prediction exercise. Section~\ref{sec:conclusion} concludes.

\section{Methodology}\label{sec:methodology}
\subsection{Sparse Autoencoders: An Intuitive Overview}

\begin{figure}[!htbp]
  \centering
  \includegraphics[width=\linewidth]{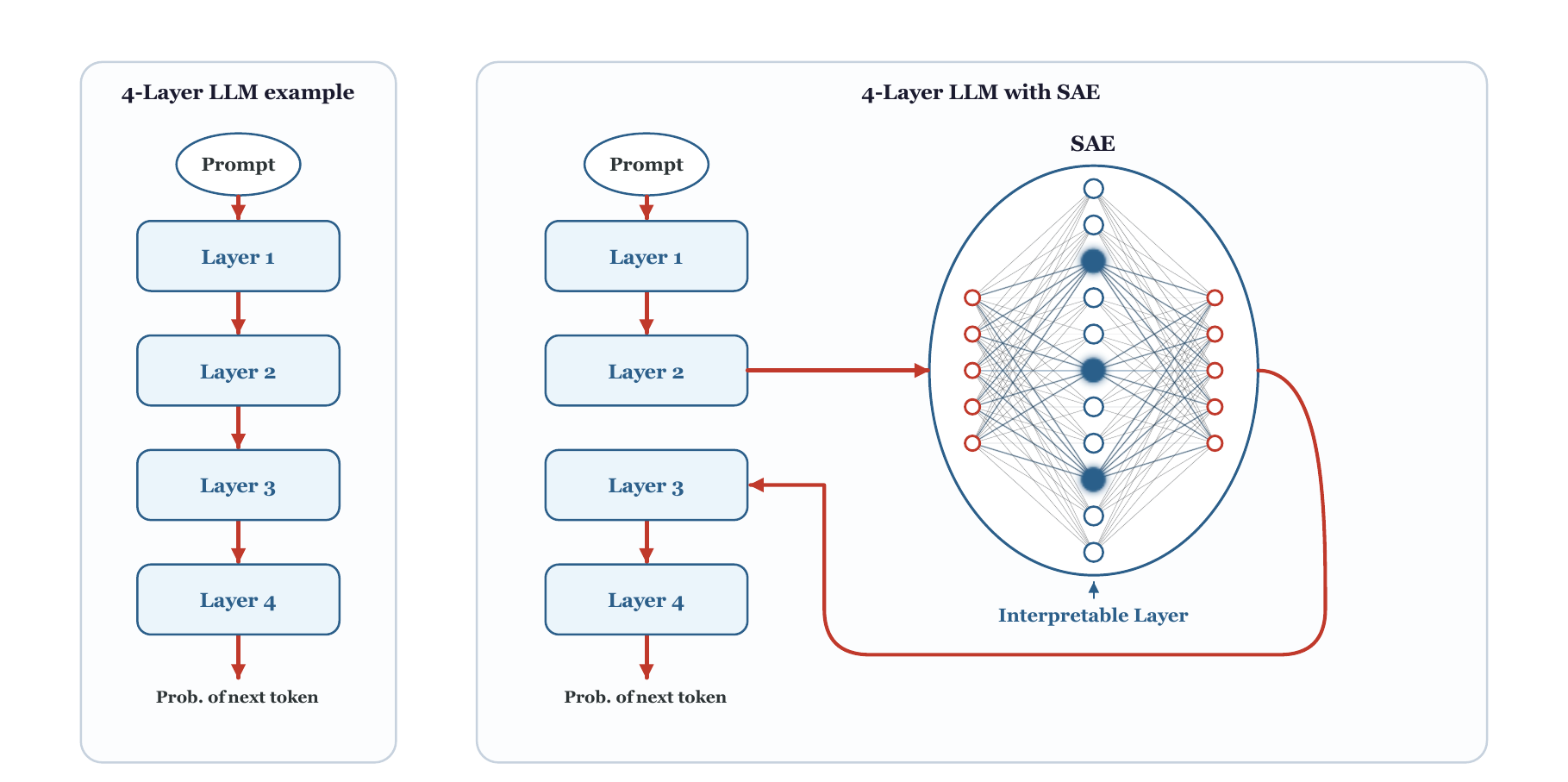}
  \caption{\textbf{Sparse Autoencoder Representation of the Residual Stream.}\\
  \footnotesize{The left panel shows a stylized four-layer language model. A prompt passes through successive transformer layers to produce a probability distribution over the next token. The red arrows trace the \textit{residual stream}, the running hidden representation updated and passed from one layer to the next. The right panel shows the same model with an SAE inserted between Layers~2 and~3. The encoder maps the dense residual-stream vector into a substantially wider feature layer. Only a small number of these features are active for a given input, as shown in dark blue. The decoder then reconstructs the vector in its original dimensions and passes it to Layer~3.}}
  \label{fig:sae_illus}
\end{figure}

LLMs process text through successive transformer layers. At each token position, a vector called the \textit{residual stream} carries information forward and is updated by each layer (left panel of Figure~\ref{fig:sae_illus}). Although the residual stream may appear large, its few thousand coordinates are relatively few compared with the number of concepts it must represent. Concepts therefore span many coordinates in overlapping patterns, with each coordinate contributing to several concepts. This overlap is central to the LLM's ability to compress rich textual information into a compact vector. Such compression allows LLMs to process rich information efficiently and perform well across a wide range of tasks. The downside is that the compressed representation is difficult to interpret: individual coordinates have no clear meaning.

A more interpretable representation would associate each concept with one or a few distinct coordinates. We can move toward this by decompressing the residual stream into a larger, sparse vector. This expansion separates concepts that overlap in the original representation, giving them distinct coordinates that activate when relevant. The intuition is similar to a prism separating a mixed beam of light into its constituent colors. We can then examine which concepts activate each coordinate and assign it an interpretable label.

We construct this interpretable representation using a sparse autoencoder, or SAE \citep{cunningham2024sparse,bricken2023towards}. An SAE is a separate neural network that takes the residual stream at a selected transformer layer and maps it into a much larger, sparse vector. It is trained to preserve as much information as possible from the original representation while penalizing feature activation, encouraging most coordinates in the expanded vector to remain zero for a given input.

Figure~\ref{fig:sae_illus} illustrates this architecture. The left panel shows the original dense residual stream, with concepts distributed across overlapping coordinates. The right panel introduces an SAE whose larger intermediate layer allows individual coordinates to specialize in distinct concepts. The first half of this network, called the encoder, maps the dense vector into this sparse intermediate layer. The second half, called the decoder, reconstructs the vector in its original dimensions before passing it to the next transformer layer. Training the SAE to reproduce its input encourages it to preserve information, while the sparsity constraint makes the intermediate representation easier to interpret. A growing machine-learning literature demonstrates the usefulness of SAEs for interpreting LLMs \citep[see][and references therein]{shu-etal-2025-survey}.\footnote{Appendix~\ref{app:sae_details} provides formal definitions of the SAE architecture and our concept-steering methodology.}

A key practical strength of SAEs is that they provide a lightweight addition to an existing LLM. The underlying language model remains unchanged, making an SAE substantially less expensive to train than the model itself. During SAE training, the LLM's parameters remain fixed. A text corpus is passed through the frozen model, residual-stream activations are collected at a selected layer, and the SAE is trained on those activations.

Once trained, an SAE makes the model's internal state both interpretable and controllable. First, it acts as a ``brain scan'' of the model's reasoning. By monitoring active features and their gradients, we can identify the concepts that contribute to the model's next-token predictions and quantify their relative importance. Second, the SAE enables \textit{concept steering}. By amplifying or suppressing selected features during text generation, we can directly shift the model's internal focus toward or away from a targeted concept. These two capabilities allow us both to observe the economic concepts underlying the model's output and to intervene on them directly.

Throughout the paper, we use the pretrained Gemma Scope SAEs released by Google DeepMind for \textit{Gemma-2-9B-IT} \citep{lieberum2024gemmascopeopenspars}. This setup combines a powerful instruction-tuned model with the computational efficiency required for extensive experimentation. Our subsequent empirical exercises show that it reproduces or exceeds established performance benchmarks in financial applications. The central limitation of the off-the-shelf SAE is therefore not performance, but domain relevance. Its 131{,}072 features form a large, general-purpose dictionary spanning concepts from finance to cooking, sports, and everyday conversation. Only a small fraction directly captures the economic mechanisms relevant to our analysis. Our first contribution transforms this general-purpose representation into an economic one through a finance-specific procedure that labels and selects the features associated with economically meaningful concepts.

\FloatBarrier

\subsection{Finance-specific Feature Labeling and Selection}\label{sec:methodology_implementation}

To label each feature, the standard procedure is bottom-up: researchers collect the text passages that most strongly activate the feature and ask a separate LLM to summarize their common theme. A passage activates a feature when processing that text produces a high value at the corresponding coordinate of the SAE's sparse representation. Because the passages come from generic text corpora, many labels describe everyday subjects such as recipes, sports, or casual conversation rather than the concepts needed for financial and macroeconomic analysis.

We reverse the standard labeling logic. Rather than starting with existing text, grouping passages by the SAE features they activate, and asking an LLM to label those groups, we begin with the labels we want to study. We then use an LLM to generate example texts for each label and identify the SAE features those texts activate.\footnote{Appendix~\ref{app:sae_details_implementation} provides the formal assignment and selection procedure.} This approach is highly customizable: researchers can choose labels tailored to their specific questions. In this paper, we define 46 economic topics grouped into five broad themes: \textit{Corporate Finance}, \textit{Investment \& Portfolio Management}, \textit{Risk Management \& Derivatives}, \textit{Financial Markets \& Instruments}, and \textit{Macro-Finance \& Economics}. Table~\ref{tab:economic_themes} presents the full taxonomy.

\begin{table}[!htbp]\centering\caption{\textbf{Taxonomy of Economic and Financial Topics}}\label{tab:economic_themes}
\renewcommand{\arraystretch}{1.5}\begin{tabular}{@{}p{0.25\textwidth}p{0.69\textwidth}@{}}\toprule\textbf{Theme} & \textbf{Topics}\\\midrule
\textbf{Corporate Finance} & Leverage; Financial Distress; CAPEX; Investments; Revenues growth; Mergers \& Acquisitions (M\&A); Dividend Policy; Cost of Capital; Working Capital; Inventory; Governance; Capital Budgeting; Human resources\\[5pt]
\textbf{Investment \& Portfolio Management} & Risk Aversion; Asset Allocation; Alpha; Beta; Sharpe Ratio; Passive investing; Arbitrage\\[5pt]
\textbf{Risk Management \& Derivatives} & Value at Risk (VaR); Hedging; Credit Default Swaps (CDS); Counterparty Risk; Systemic Risk; Stress Testing; Operational Risk; Regulation\\[5pt]
\textbf{Financial Markets \& Instruments} & Equity Markets; Fixed Income; Derivatives; Liquidity; Initial Public Offering (IPO); Short Selling; High-Frequency Trading (HFT); Foreign Exchange (FX); Commodities\\[5pt]
\textbf{Macro-Finance \& Economics} & Inflation; Unemployment; Monetary Policy; Fiscal Deficit; Gross Domestic Product (GDP); Quantitative Easing (QE); Yield Curve; Balance of Payments; Tariffs\\[5pt]
\bottomrule\end{tabular}\par\vspace{8pt}{\footnotesize\textit{Note.} The table presents the 46 economic topics used in our finance-specific feature-labeling procedure, organized into five broad themes. We generate synthetic text for each topic and use the resulting SAE activations to assign features to topics and select those relevant to economic analysis.\par}\end{table}
\FloatBarrier

The procedure performs labeling and dimensionality reduction in a single step. Across the 46 topics, this procedure identifies 7{,}278 of the SAE's 131{,}072 features (5.6\%) as economically relevant. Another benefit of the procedure is that it uses only synthetic topic-specific articles, avoiding data leakage from the empirical news sample or realized returns into feature labeling and selection.

\subsection{Validating Labels}
\label{sec:validating_labels}

A natural way to test whether these features capture their assigned economic topics is through \textit{concept steering}. The idea is to artificially increase the activations of SAE features associated with a specific concept and pass the modified residual stream back into the LLM. This effectively ``steers'' the model's thought process toward a selected set of concepts without changing the prompt.

We use this approach in a simple controlled experiment. We prompt the LLM to generate three financial or economic concepts, keeping the prompt identical in every trial. The only change is which features we activate: for each economic topic, we increase the activations of the features identified by our brain scan. Differences in the generated text can therefore be directly attributed to the features being steered. We then examine whether the model's responses reflect the assigned topic.

\begin{Prompt}[!htbp]
    \begin{tcolorbox}
        \begin{lstlisting}
Give me 3 finance/economic concepts without explanation.
        \end{lstlisting}
    \end{tcolorbox}
    \caption{\textbf{Word-Cloud Probe} \\
    {This neutral prompt elicits a short completion from the LLM while its residual stream is steered along the features assigned to a given topic.}}
    \label{prompt:wordcloud}
\end{Prompt}
\FloatBarrier

\begin{figure}[!htbp]
\begin{center}
\singlespacing
\setlength{\tabcolsep}{3pt}
\begin{tabular}{@{}ccc@{}}

\footnotesize Panel A: Capital Budgeting & \footnotesize Panel B: Derivatives & \footnotesize Panel C: Dividend Policy \\
\includegraphics[width=0.32\textwidth]{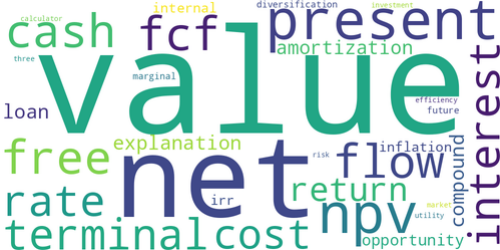} &
\includegraphics[width=0.32\textwidth]{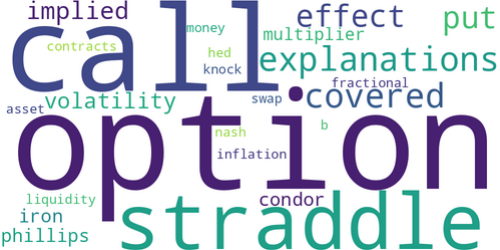} &
\includegraphics[width=0.32\textwidth]{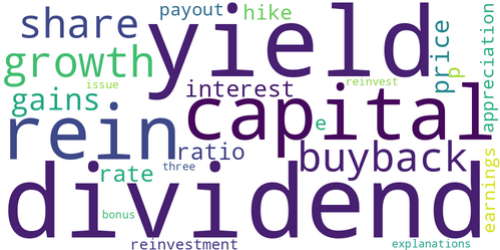} \\[4pt]

\end{tabular}
\end{center}
\caption{\textbf{Word Clouds for Example Economic Topics}}
\label{fig:example_wordclouds}
\vspace{4pt}
\noindent\begin{minipage}[c]{1\textwidth}%
\singlespacing
\textbf{\footnotesize Note.} \footnotesize
The figure shows the most frequent terms generated in response to Prompt~\ref{prompt:wordcloud} when the LLM is steered toward \textit{Capital Budgeting}, \textit{Derivatives}, and \textit{Dividend Policy}. The prompt remains fixed across the three exercises, and only the SAE features being steered change. Font size is proportional to the frequency with which each term appears.
\medskip
\end{minipage}
\end{figure}

Figure~\ref{fig:example_wordclouds} displays the word clouds for three topics: \textit{Capital Budgeting}, \textit{Derivatives}, and \textit{Dividend Policy}. The generated vocabulary differs sharply across the three steering interventions and aligns closely with the assigned topic labels. For example, steering the features assigned to \textit{Capital Budgeting} produces terms such as ``NPV'' and ``net value,'' while steering those assigned to \textit{Derivatives} produces terms such as ``option'' and ``call.'' Figure~\ref{fig:cloud_1} in Appendix~\ref{appendix:wordclouds} presents the complete set of 46 word clouds. The same pattern holds across the full taxonomy: steering produces vocabulary closely aligned with each assigned topic.

\FloatBarrier

\section{Macroeconomic Brain Scan}
\label{sec:methodology_belief}
\label{sec:macro_correlation}

We first apply our brain scan to examine how an LLM forms macroeconomic forecasts. We adopt the empirical setting of \citet{bybee2023surveying} and compare the LLM's forecasts of fourteen macroeconomic quantities based on \textit{Wall Street Journal} news with survey expectations from professional forecasters and individual investors.\footnote{Our sample covers 2000--2021. For thirteen variables, we use the quarterly Survey of Professional Forecasters: employment, federal government consumption growth, housing starts, industrial production, real consumption growth, real nonresidential investment growth, real residential investment growth, state and local government consumption growth, the 3-month Treasury bill rate, the AAA bond rate, the GDP price deflator, the consumer price index, and real GDP growth. For the S\&P~500, we use the monthly American Association of Individual Investors survey of individual-investor sentiment.}

For each macroeconomic variable, we prompt the LLM to classify a news article's effect as an increase, a decrease, or uncertain. Averaging these forecasts across news articles yields a point estimate that we compare with the corresponding survey expectation. The model sees exactly the same news across all fourteen variables. Only the name of the quantity to forecast changes in the prompt. Differences in internal activations therefore reveal how the LLM shifts its focus when analyzing the same news to forecast different economic outcomes.

Because we use a relatively small LLM, \textit{Gemma-2-9B-IT}, it is unclear ex ante whether its expectations are suitable for our analysis. Before looking at the brain scan, we therefore check whether the correlations between our model's forecasts and survey expectations match those reported by \citet{bybee2023surveying}. Our results are broadly in line with his for thirteen of the fourteen variables. Table~\ref{tab:macro_corr} reports the correlations for the thirteen SPF variables, while the S\&P~500 comparison appears in Appendix~\ref{appendix:survey_comparison}. The exception is the 3-month Treasury bill rate, which we diagnose and correct through targeted SAE steering in Section~\ref{sec:steering_qe}.\footnote{Our objective is to decompose the model's expectations rather than establish out-of-sample forecasting performance. Training-data contamination may affect the measured agreement between LLM and survey expectations \citep{sarkar2024lookahead,didisheim2025ai}. We use these correlations only to establish that the model produces economically meaningful expectations for which an internal decomposition is informative.}

\begin{table}[!htbp]
\centering
\caption{\textbf{Expectation Correlation of Macro Variables} \\
    \footnotesize{This table reports the correlation between LLM (Gemma-2-9B-IT) expectations and Survey of Professional Forecasters (SPF) expectations. Following \cite{bybee2023surveying}, we report the correlation separately for level and revision forecasts. The frequency is quarterly and the sample period is 2000--2021.}}
    \label{tab:macro_corr}
\resizebox{0.9\textwidth}{!}{%
  \begin{tabular}{p{7cm} >{\centering\arraybackslash}p{3.5cm} >{\centering\arraybackslash}p{3.5cm}}
\toprule
\textbf{Variable} & \textbf{Corr. (Level)} & \textbf{Corr. (Revision)} \\
\midrule
Employment & 0.277** & 0.256** \\[4pt]
\addlinespace[2pt]
Federal Gov. Consumption Growth & 0.197* & 0.064 \\[4pt]
\addlinespace[2pt]
Housing Starts & 0.246** & 0.396*** \\[4pt]
\addlinespace[2pt]
Industrial Production & -0.155 & 0.356*** \\[4pt]
\addlinespace[2pt]
Real Consumption Growth & 0.418*** & 0.326*** \\[4pt]
\addlinespace[2pt]
Real Nonresidential Inv. Growth & 0.590*** & 0.465*** \\[4pt]
\addlinespace[2pt]
Real Residential Inv. Growth & 0.375*** & 0.476*** \\[4pt]
\addlinespace[2pt]
Local Gov. Consumption Growth & 0.136 & 0.293*** \\[4pt]
\addlinespace[2pt]
3-Month T-Bill Rate & -0.399*** & -0.198* \\[4pt]
\addlinespace[2pt]
AAA Bond Rate & 0.088 & -0.024 \\[4pt]
\addlinespace[2pt]
GDP Price Deflator & 0.158 & 0.219** \\[4pt]
\addlinespace[2pt]
CPI & 0.207* & 0.282*** \\[4pt]
\addlinespace[2pt]
Real GDP Growth & 0.471*** & 0.404*** \\[4pt]
\bottomrule
\end{tabular}\hspace{\fontdimen2\font}
  }
\end{table}
\FloatBarrier

\subsection{Brain Scan}
\label{sec:wsj_gradient}
Neuroscientists study the human brain by asking people to perform different tasks and comparing the resulting patterns of activity. Our SAE enables an analogous exercise for the LLM. We present each news article to the model fourteen times, changing only the macroeconomic variable it is asked to forecast. We then examine which economic features are active and how they influence its directional assessment. Individual SAE features serve as the ``neurons'' in this analogy, while the topics and themes defined above form the economic ``regions'' into which we group them.

To give intuition, we start with a single news article. In the exercise described above, the model reads each article fourteen times, once for each forecast target. Here we focus on two: the S\&P~500 index and employment. We use the same news in both cases and change only the target variable in Prompt~\ref{prompt:macro_direction}. The aim is to visualize how the SAE brain scan changes when the model considers two different economic implications of the same information.

\begin{tcolorbox}[title={BNY Mellon Loses Key ETF Client}]
\footnotesize\singlespacing
\textit{Wall Street Journal, 23 August 2019.}\par
Bankof New York Mellon Corp. lost VanEck Associates as a client of its servicing business for exchange-traded funds to rival State Street Corp., people familiar with the matter said. VanEck is consolidating its custody work, the people said. [\ldots] The loss puts a dent in BNY Mellon's push to become a bigger player in the market for servicing ETFs. The rise of ETFs has transformed large parts of Wall Street. These products trade on exchanges like stocks but are cheaper, more transparent and often more tax-advantageous than mutual funds. [\ldots] As money shifted away from traditional mutual-fund managers and other investment firms, ETF providers emerged as important customers to custody firms like BNY Mellon. During a conference call last month, BNY Mellon Chief Executive Charles Scharf said that ETF assets under custody at the firm had risen by about 50\% in the past year. [\ldots] ``Not a huge revenue impact in the short term, but important strategically for us." [\ldots] BNY Mellon and other custody banks have successfully streamlined their operations since the financial crisis by slashing costs. Separately, BNY Mellon began another wave of job cuts this week, people familiar with the matter said. About 200 employees were affected by the cuts, part of the continuing push to shed expenses.
\end{tcolorbox}

\begin{figure}[!htbp]
\centering
{\footnotesize Panel A: Topic Regions}\par
\includegraphics[width=\textwidth]{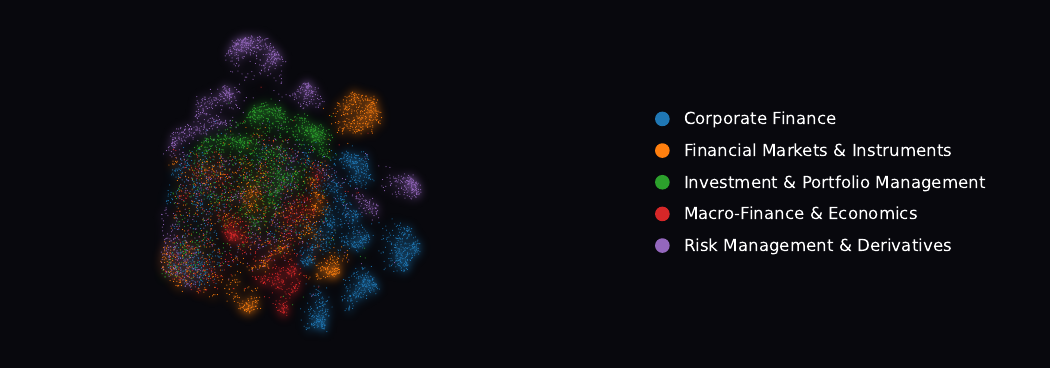}\par
\begin{tabular*}{\textwidth}{@{}c@{\extracolsep{\fill}}c@{}}
\parbox[b]{0.485\textwidth}{\centering\footnotesize Panel B: Asked About the S\&P 500 Index} &
\parbox[b]{0.485\textwidth}{\centering\footnotesize Panel C: Asked About Employment}\\[2pt]
\includegraphics[width=0.485\textwidth]{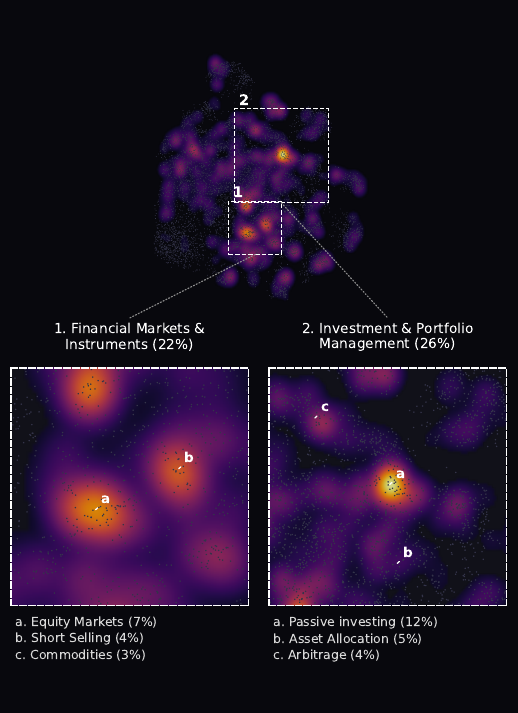} &
\includegraphics[width=0.485\textwidth]{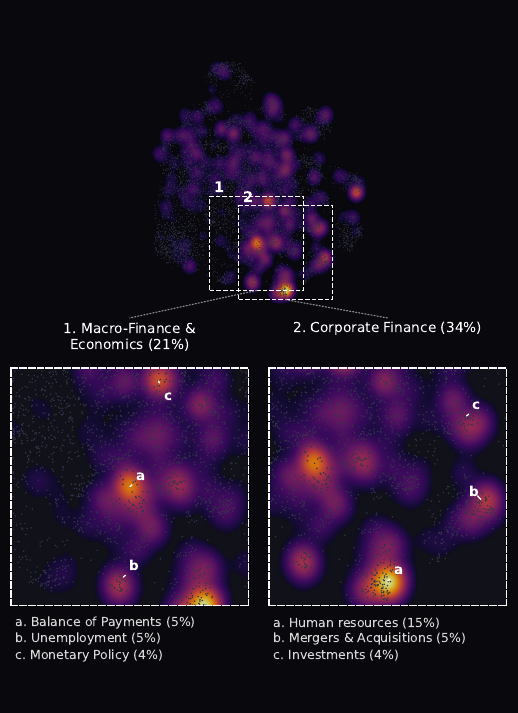}
\end{tabular*}
\caption{Brainscan: Example of a Single News Article.}
\label{fig:brain_visual_map}
\label{fig:brain_visual_comparison}
\begin{minipage}{\textwidth}
\footnotesize\singlespacing
\textbf{Note.} Panel A shows the reference feature map, colored by economic theme. Panels B and C show activations for the same BNY Mellon article under S\&P~500 and employment forecasting prompts. The activation maps contain 7,278 features assigned to 46 economic topics. We subtract each target's mean activation across 2019 articles, then subtract the article's average adjusted activation across its fourteen prompts and clip at zero. The glow displays this excess activation after smoothing and square-root scaling. Dashed boxes identify the two leading themes, enlarged below using the same glow and brightness scale as their parent map. Letters mark their three leading topics; percentages are shares of total excess activation. Brightness is scaled separately across panels. No forecast gradients enter the plots. Appendix~\ref{appendix:brain_visualization} provides the construction details.
\end{minipage}
\end{figure}

The brain scan records activity across 7,278 economic features. To make these measurements readable, we first arrange the features on a two-dimensional map. Each feature has a decoder direction describing how it enters the LLM's internal representation. We project these directions into two dimensions using a standard dimensionality-reduction technique that groups features by economic theme and proximity in decoder space.\footnote{We use Uniform Manifold Approximation and Projection (UMAP), which places features with similar decoder directions near one another, using topic labels to help organize the map into economic regions. Appendix~\ref{appendix:brain_visualization} provides the construction details.} Panel A of Figure~\ref{fig:brain_visual_comparison} shows the reference representation of the model's ``brain'' before article-specific activity is overlaid. Each dot is a feature, and its color identifies one of our five economic themes.

Keeping feature positions fixed across the two forecasting prompts lets us compare activation patterns directly. Panels B and C of Figure~\ref{fig:brain_visual_comparison} show the same article under the S\&P~500 and employment prompts. We remove both each forecast target's usual activation and the activity shared across the article's fourteen prompts. The enlarged views identify the leading themes and their most active topics.

\FloatBarrier

The contrast is striking. The LLM reads exactly the same news, yet different regions of its economic ``brain'' light up depending on the question. When asked about the S\&P~500, the model activates features associated with \textit{Passive Investing}, \textit{Equity Markets}, and \textit{Asset Allocation}, consistent with the article's discussion of exchange-traded funds. When asked about employment, its focus shifts toward \textit{Human Resources}, consistent with the reported job cuts. Appendix~\ref{appendix:brain_examples} provides two further single-article examples, illustrating workforce-related concepts in a GE restructuring article and shifts between equity-market and interest-rate concepts in a silver-market article, where the scans also differ visibly in the concentration of illuminated regions. Next, we propose a simple methodology to link SAE activations directly to the probability of making a specific forecast.

\subsection{From Feature Activations to Forecast Attribution}
\label{sec:forecast_attribution}

Our starting point is the LLM's probabilities of generating each possible next token, which we call its ``inner probabilities.'' Prompt~\ref{prompt:macro_direction} restricts the model's answer to a single token ($P$, $N$, or $U$), so its inner probabilities map directly to an increase, a decrease, or uncertainty. Following \citet{chen2024out}, we extract the probabilities assigned to these three responses directly from the LLM.

\begin{Prompt}[!htbp]
    \begin{tcolorbox}
        \begin{lstlisting}
Here is a piece of news: `{news}'. Do you think this news will
increase or decrease {variable}? Output exactly one character:
P, U, or N, and nothing else.

P  (increase)
U  (uncertain)
N  (decrease)

Your answer:
        \end{lstlisting}
    \end{tcolorbox}
    \caption{\textbf{Macro-Direction Forecast} \\
    {The prompt asks the LLM to forecast the directional impact of a given news item on a target macroeconomic variable, with the response constrained to a single token from the set $\{P,U,N\}$. The brackets $\{\cdots\}$ indicate dynamic inserts where we place the \textit{news} (text of the news article) and the \textit{variable} (e.g., the S\&P~500, CPI, real GDP).}}
    \label{prompt:macro_direction}
\end{Prompt}
\FloatBarrier

For article $a$, we summarize the LLM's preference for an increase relative to a decrease by
\begin{equation}
\ell_a = \log \frac{p_a(P)}{p_a(N)}.
\label{eq:logodds_objective}
\end{equation}
Positive values indicate a preference for an increase, and negative values indicate a preference for a decrease. To simplify notation, we do not include an index for the macroeconomic variable, although we conduct the analysis separately for each variable.

Let $z_{jas}$ denote the activation of feature $j$ in the SAE representation at token position $s$ in article $a$. Our goal is to measure how sensitive the LLM's preference for an increase relative to a decrease is to this activation:
\begin{equation}
 g_{jas} = \frac{\partial \ell_a}{\partial z_{jas}}.
 \label{eq:logodds_grad}
\end{equation}
A positive gradient means that strengthening the feature at that position raises the probability of an increase relative to a decrease. A negative gradient shifts the assessment toward a decrease.

We calculate these derivatives by backpropagation through the remaining LLM layers and the SAE decoder, evaluating them at the model’s original internal state and holding all learned weights fixed. This yields the sensitivities across features and token positions without separately perturbing each activation and rerunning the LLM.\footnote{The SAE's decoder connects feature activations back to the LLM's internal vector. Each feature has a learned decoder direction that determines how changing its activation changes this vector. The remaining LLM layers carry that change forward to the answer probabilities, providing a differentiable path from feature activations to the directional assessment.}

For each article, we therefore observe two quantities for every feature at every token position: its activation and the sensitivity of the assessment to it. We multiply the two at each position and sum across positions to obtain the feature's contribution to the article's assessment:
\begin{equation}
 c_{ja} = \sum_{s=1}^{L_a} z_{jas}\,g_{jas},
 \label{eq:article_contrib}
\end{equation}
where $L_a$ is the number of article tokens. To a first-order approximation, $z_{jas}\,g_{jas}$ is the change in the assessment $\ell_a$ produced by feature $j$ at position $s$, so a feature contributes only where it is both active and influential. Summing over positions gives $c_{ja}$, the feature's total contribution to the assessment of article $a$.

We then average these article-level contributions across the articles in month $t$ to define the monthly feature attribution score:
\begin{equation}
 c_{jt} = \frac{1}{|\mathcal A_t|}\sum_{a\in\mathcal A_t} c_{ja},
 \label{eq:monthly_contrib_method1}
\end{equation}
where $\mathcal A_t$ is the set of articles in month $t$.
This score combines how strongly the feature is activated with how sensitive the model's directional assessment is to that activation. Positive scores indicate influence toward an increase, while negative scores indicate influence toward a decrease.

\label{sec:wsj_ts}
\label{sec:macro_decomposition}
We group the feature scores into the 46 economic topics identified by our labeling procedure. The monthly score for topic $k$ is
\begin{equation}
 C_{kt} = \sum_{j \in \mathcal F_k} w_{kj} c_{jt},
 \label{eq:topic_score}
\end{equation}
where $\mathcal F_k$ contains the features assigned to the topic. The weights are proportional to the labeling scores and sum to one within each topic, giving greater weight to features that respond strongly and specifically to that topic.

\begin{figure}[!htbp]
\centering
\begin{tabular}{@{}c@{}}
Panel A: AAA Bond Rate \\
\includegraphics[width=0.90\textwidth]{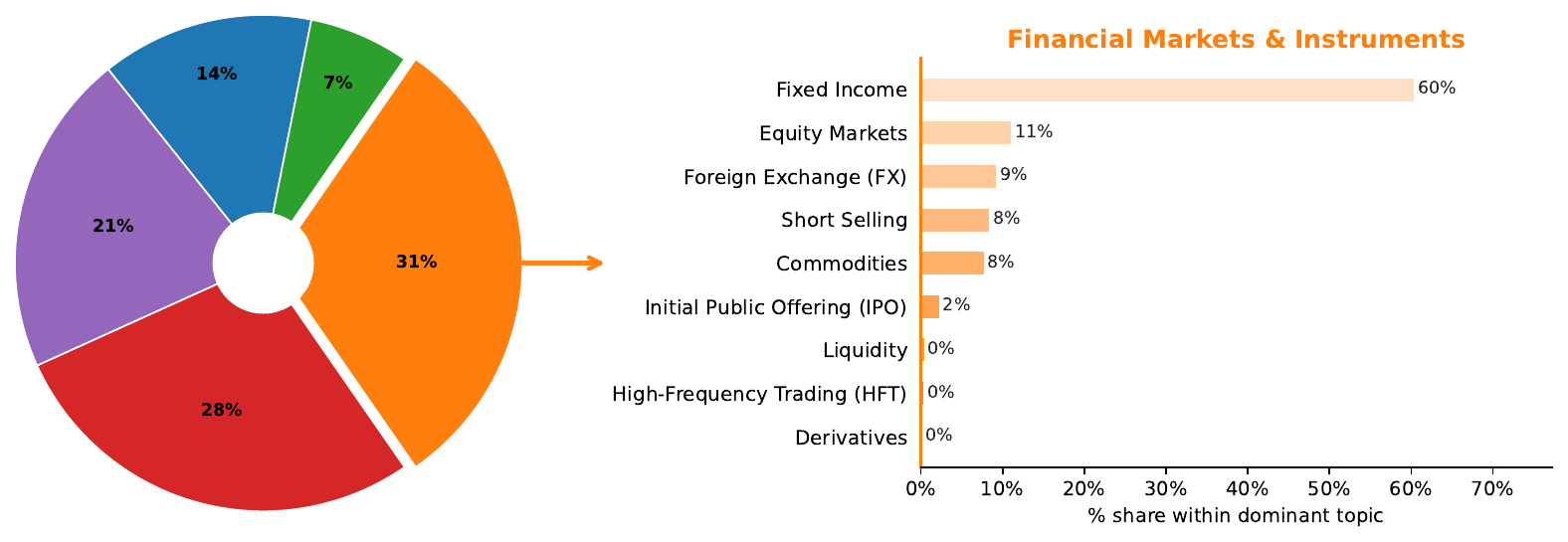} \\[4pt]
Panel B: Real Consumption \\
\includegraphics[width=0.90\textwidth]{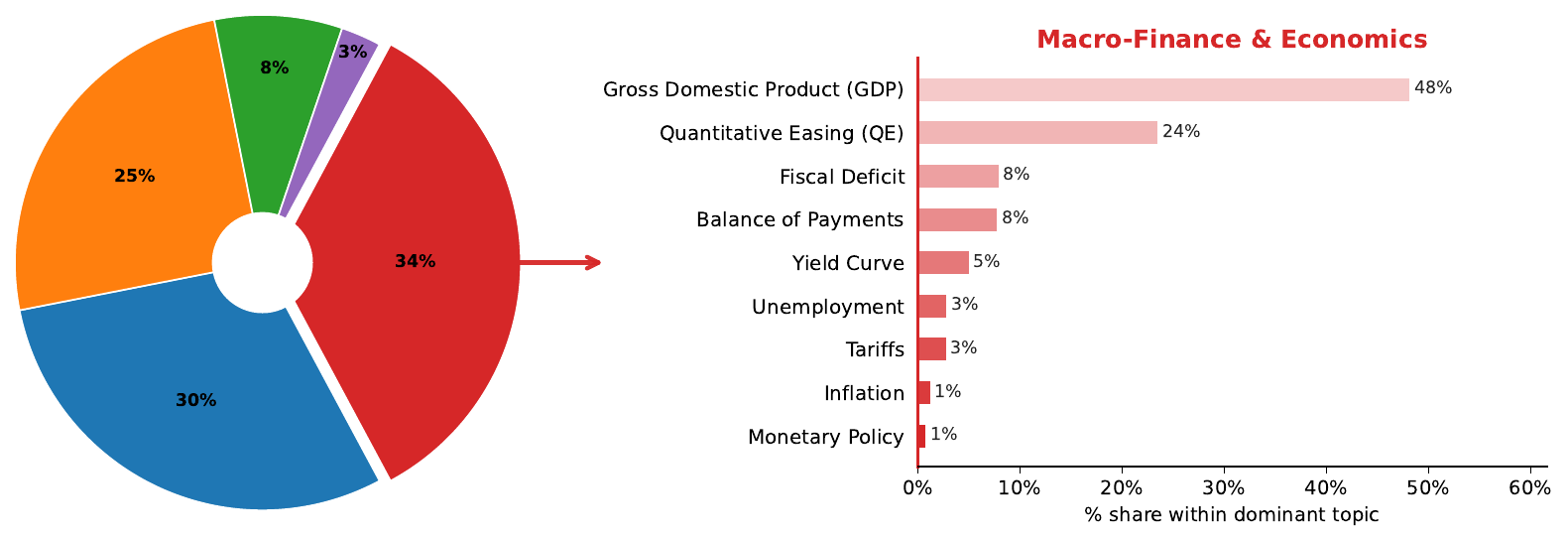} \\[4pt]
Panel C: Real Nonresidential Investment \\
\includegraphics[width=0.90\textwidth]{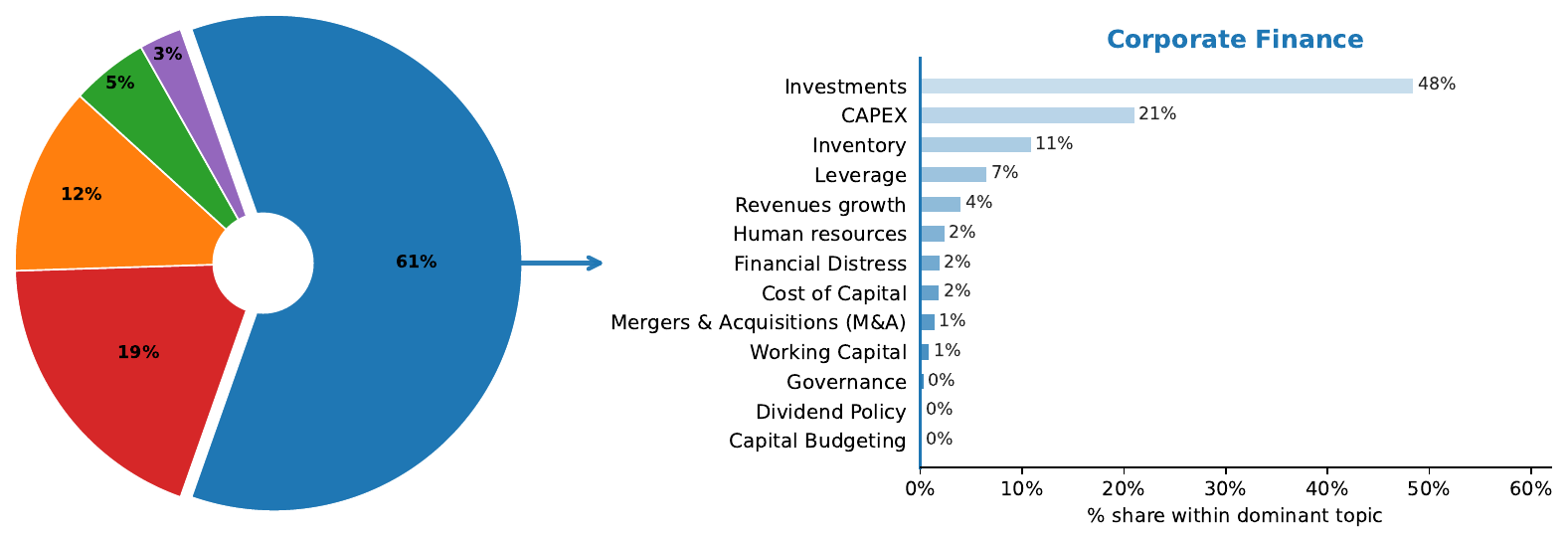} \\[2pt]
\includegraphics[width=0.85\textwidth]{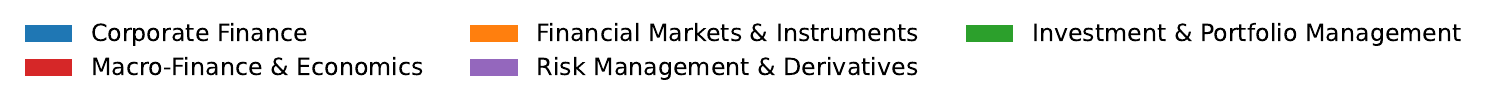}
\end{tabular}
\caption{\textbf{Brainscan: Aggregation at Economic Concepts}\\
\footnotesize{Panels A, B, and C report the decompositions of AAA bond rate, real consumption, and real nonresidential investment expectations, respectively. For each variable, the pie chart on the left shows each of the five economic themes' share of total contribution magnitude, averaged over the sample. The bar chart on the right decomposes the dominant theme into its constituent topics, expressed as shares of that theme's contribution. Contributions combine SAE feature activations with the gradients of the LLM's directional assessment, as described in the text.}}
\label{fig:macro_cpi_employment_decomp}
\end{figure}

Figure~\ref{fig:macro_cpi_employment_decomp} shows the results for AAA bond rates, real consumption, and real nonresidential investment, providing an aggregate counterpart to the single-article brain scan in Figure~\ref{fig:brain_visual_comparison}. For each forecast, we take the absolute value of each monthly topic contribution, sum these magnitudes within each economic theme, and average over the sample. The pie chart reports each theme's share of this total contribution magnitude. The accompanying bar chart decomposes the dominant theme into its constituent topics. The leading theme differs across the three forecasts: \textit{Financial Markets \& Instruments} for AAA bond rates, \textit{Macro-Finance \& Economics} for real consumption, and \textit{Corporate Finance} for real nonresidential investment.

\FloatBarrier
The topic decomposition sharpens this interpretation. For AAA bond rates, \textit{Financial Markets \& Instruments} accounts for 31\% of total contribution magnitude, followed closely by \textit{Macro-Finance \& Economics} at 28\%. Within the leading theme, \textit{Fixed Income} accounts for 60\% of the contribution, consistent with the model focusing on bond-market concepts when assessing news about bond yields. Smaller contributions from \textit{Equity Markets}, \textit{Foreign Exchange}, \textit{Short Selling}, and \textit{Commodities} connect this assessment to broader financial-market conditions.

For real consumption, \textit{Macro-Finance \& Economics} is the leading theme at 34\%, alongside substantial contributions from \textit{Corporate Finance} (30\%) and \textit{Financial Markets \& Instruments} (25\%). Within the leading theme, \textit{GDP} accounts for 48\% and \textit{Quantitative Easing} for 24\%, followed by \textit{Fiscal Deficit} and \textit{Balance of Payments} at 8\% each. These concepts connect consumption expectations to aggregate activity, policy, and external economic conditions.

For real nonresidential investment, \textit{Corporate Finance} dominates, accounting for 61\% of total contribution magnitude. Within this theme, \textit{Investments} and \textit{CAPEX} account for 48\% and 21\%, respectively, followed by \textit{Inventory} (11\%) and \textit{Leverage} (7\%). The model focuses on individual firms' investment strategies and their future capacity to invest.

Appendix~\ref{appendix:macro_decomp} reports the corresponding decompositions for the remaining forecast targets. Together, these examples show how the brain scan distinguishes the economic concepts underlying different forecasts, from bond-market conditions to aggregate consumption and firms' investment decisions.

\subsection{Self-Reported Reasoning versus the Brain Scan}
\label{sec:macro_selfreport}

Continuing the analogy with neuroscience, an alternative to scanning the brain is to ask people what they feel. Rather than measuring activity in regions associated with anger, for example, we can ask someone how angry they are. A common approach to understanding LLM reasoning in financial applications follows this logic: researchers prompt the model to explain or justify its assessments \citep{lopez2023can,hansen2023fedspeak,bybee2023surveying,jha2024corporate,fedyk2024chatgpt}. Yet, as with humans, the ability to produce a plausible explanation does not guarantee an accurate account of the internal processes behind an answer.

Our setup provides a natural way to assess these self-reports by comparing them directly with the brain scan. Using Prompt~\ref{prompt:topic_weighted_macro_direction}, we ask the LLM to assign importance scores to the economic topics it considers relevant when assessing each article. The prompt requires positive scores for selected topics that sum to 100.\footnote{We also renormalize the returned scores to sum to 100 in case the LLM does not satisfy this constraint exactly.} Let $\widetilde c_{ka}$ denote the LLM's self-reported importance score for topic $k$ in article $a$.

\begin{Prompt}[!htbp]
    \begin{tcolorbox}
        \begin{lstlisting}
Here is a piece of news: `{news}'.

Task: predict whether this news will increase or decrease {variable}.
From the numbered list below, select only the topics you consider
genuinely relevant for making this prediction, and assign each topic
a weight reflecting its importance.

Use the topic number as the key. Weights must be positive and sum to 100.

{topic_menu}

Respond with a JSON object and nothing else, using this exact format:

{"topics": {"<topic_id>": <weight>, ...},
 "prediction": <1 increase | 0 uncertain | -1 decrease>}
        \end{lstlisting}
    \end{tcolorbox}
    \caption{\textbf{Topic-Weighted Macro-Direction Forecast} \\
    {The prompt asks the LLM to forecast the directional impact of a given
    news item on a target macroeconomic variable while also identifying the
    topics that are genuinely relevant to the prediction.}}
    \label{prompt:topic_weighted_macro_direction}
\end{Prompt}
\FloatBarrier

To compare the two accounts, we aggregate the self-reported scores to the same monthly
frequency as the brain scan. Let $\widetilde C_{kt}=\frac{1}{|\mathcal A_t|}\sum_{a\in\mathcal A_t}\widetilde c_{ka}$
denote the average self-reported score for topic $k$ in month $t$, where $\mathcal A_t$ is the
set of articles in that month. For each macroeconomic variable, we then compute the Pearson
correlation across topics between the sample-average magnitude of the brain-scan topic
contribution and the sample-average self-reported score:
\begin{equation}
    \rho = \operatorname{Corr}_{k}\!\left(
    \frac{1}{T}\sum_{t=1}^{T}\lvert C_{kt}\rvert,\;
    \frac{1}{T}\sum_{t=1}^{T}\widetilde C_{kt}\right),
    \label{eq:selfreport_correlation}
\end{equation}
where $T$ is the number of months in the sample and the correlation is computed across the
46 topics in the brain-scan map.

\begin{figure}[!htbp]
  \centering
  \includegraphics[width=\textwidth,height=0.9\textheight,keepaspectratio]{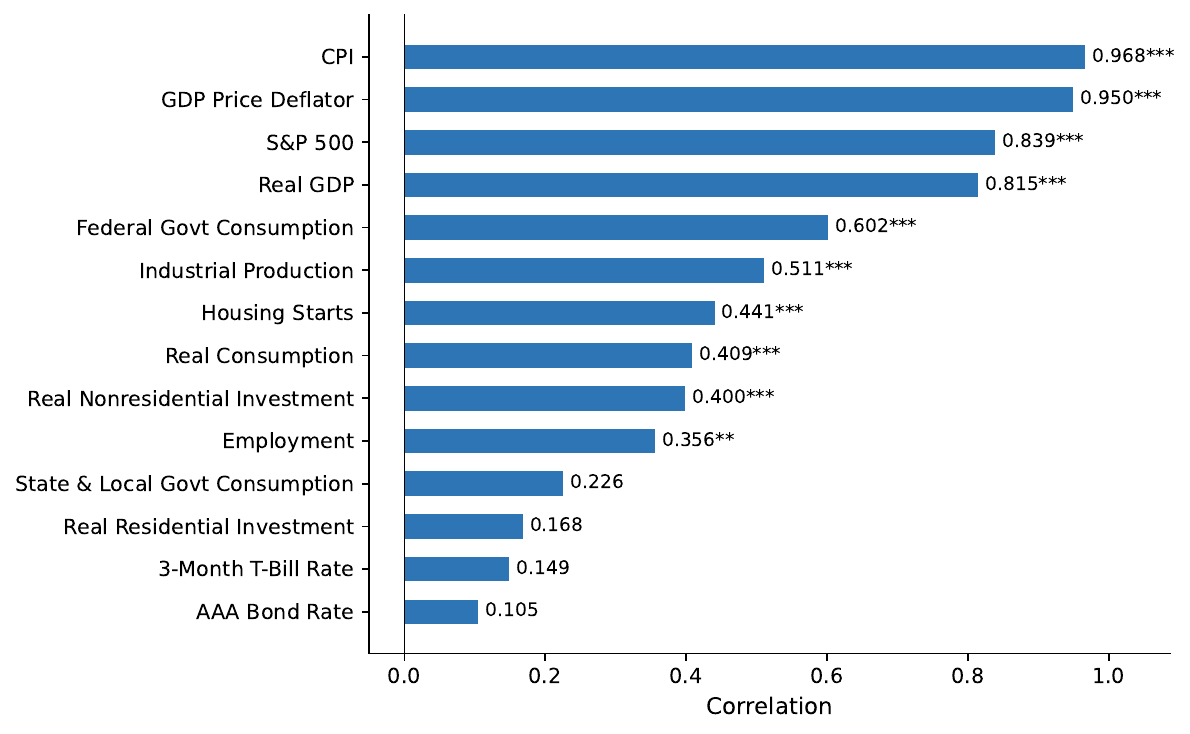}
  \caption{\textbf{Mechanistic versus self-reported topic decomposition.} \\
  \footnotesize{\footnotesize{For each macroeconomic variable, we compute the Pearson correlation across the 46 economic topics between the sample-average absolute brain-scan topic contribution, $\frac{1}{T}\sum_t \lvert C_{kt}\rvert$, and the sample-average self-reported importance score, $\frac{1}{T}\sum_t \widetilde C_{kt}$, as defined in Equation~\eqref{eq:selfreport_correlation}. Each bar reports this correlation for the corresponding forecast target. A higher value indicates greater agreement between the brain scan and the LLM's self-reports about which topics matter for that target. ***, **, and * denote significance at the 1\%, 5\%, and 10\% level, respectively, based on the Pearson correlation test across the 46 topics.}}}
  \label{fig:contrib_corr}
\end{figure}

Figure~\ref{fig:contrib_corr} reports these correlations for each macroeconomic variable. The two accounts agree only partially: some forecasts show substantial alignment, but the correlations are not statistically significant for state and local government consumption, real residential investment, the three-month Treasury bill rate, and the AAA bond rate.

Even partial agreement is a remarkable capability. Yet the weak and statistically insignificant correlations for important forecast targets raise doubts about whether self-reported reasoning provides a reliable foundation for rigorous econometric analysis.

\FloatBarrier

\section{Steering \& Agent Simulation}
\label{sec:steer_and_agent_simulation}
Large language models are increasingly used to simulate economic agents whose decisions resemble human behavior \citep{horton2023homo}. Researchers typically construct these agents through prompt steering, specifying preferences such as heightened risk aversion in the input text. As we show in this section, this approach has two limitations: it activates unrelated concepts that can distort economic reasoning, and it struggles to produce smooth, continuous changes in behavior. We examine SAE steering as an alternative that can alleviate these problems.

\paragraph{Steering Risk Aversion.} We start with a controlled experiment in which both prompt steering and SAE steering induce the desired increase in risk aversion, but SAE steering allows a smoother, more gradual adjustment. Prompt~\ref{prompt:risk_aversion_sp500_treasuries} asks the LLM to choose between an S\&P500 ETF and US Treasuries. We repeat this choice while gradually increasing the strength of the risk-aversion intervention.

\begin{Prompt}[!htbp]
    \begin{tcolorbox}
        \begin{lstlisting}
This is a simulation for a university assignment.

You must choose one of two options:

1. An ETF tracking the S&P500 - higher risk, potentially higher return.
2. US Treasuries - low risk, low return.

This is just a simulation. You are not giving any financial advice, and nobody will invest anything based on your answer.

Write your answer as one of the following integers with no explanation and nothing else:

1 (S&P500 ETF)
2 (US Treasuries)
        \end{lstlisting}
    \end{tcolorbox}
    \caption{\textbf{Risk Aversion and Investment Choice: S\&P500 vs Treasuries} \\
    {The prompt asks the LLM to choose between a higher-risk option, an ETF tracking the S\&P500, and a lower-risk option, US Treasuries. We use it to assess whether the model's choice aligns with expected risk preferences under a forced-choice investment setting.}}
    \label{prompt:risk_aversion_sp500_treasuries}
\end{Prompt}
\FloatBarrier

For prompt steering, we add an instruction asking the model to be ``10\%'', ``20\%'', ``30\%'', and so on, more risk-averse than its baseline. The model then translates this instruction into a decision.

SAE steering allows a more direct intervention. We select the feature most closely associated with risk aversion and gradually increase its activation.\footnote{We use Neuronpedia's label as an additional criterion to select among features associated with risk aversion. The selected feature is labeled ``avoidance and caution in various contexts.'' Appendix~\ref{app:sae_details_steering} provides the technical details.} The SAE decoder maps this increase into a change in the LLM's residual stream:
\begin{equation}
    r_{\mathrm{steer}} = r + \alpha d_j,
    \label{eq:steering_main}
\end{equation}
where $r$ is the original residual stream, $d_j$ is the selected feature's decoder direction, and $\alpha$ is the increase in its activation. This mapping is linear: doubling $\alpha$ doubles the added perturbation. The LLM then continues from this modified state with the prompt unchanged.

\begin{figure}[!htbp]
\centering
\begin{tabular}{@{}c@{}}
\textbf{Panel A: No Steering} \\[2pt]
\includegraphics[width=\textwidth]{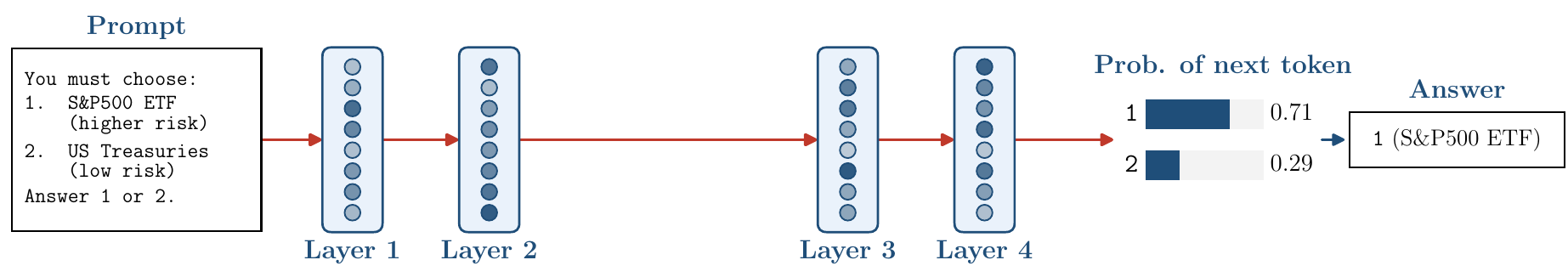} \\[10pt]
\textbf{Panel B: SAE Steering of the Risk Aversion Feature} \\[2pt]
\includegraphics[width=\textwidth]{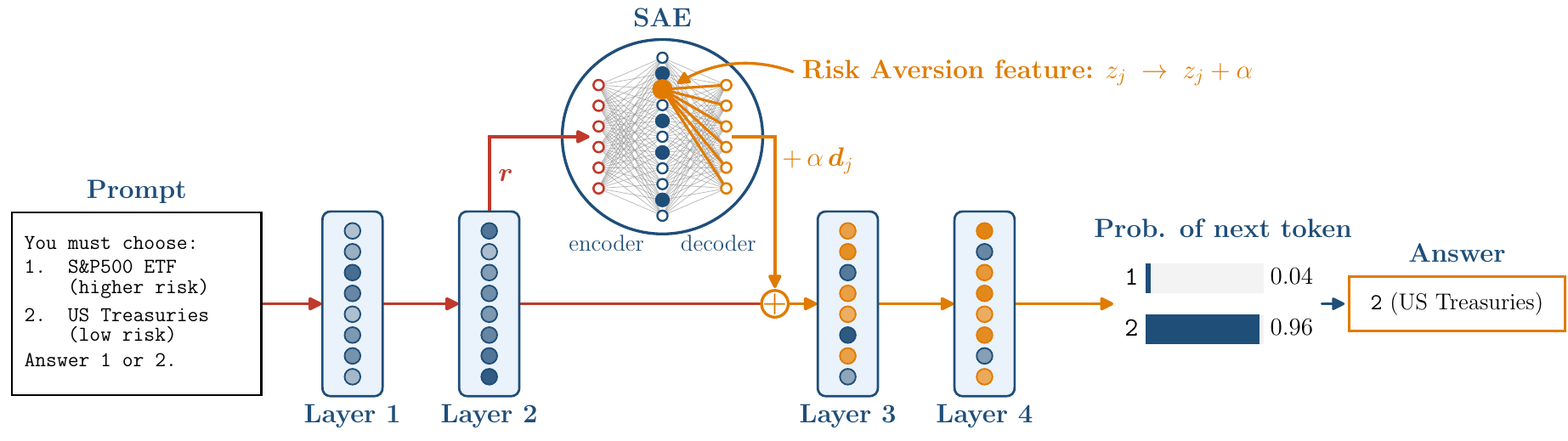} \\[10pt]
\textbf{Panel C: Prompt Steering} \\[2pt]
\includegraphics[width=\textwidth]{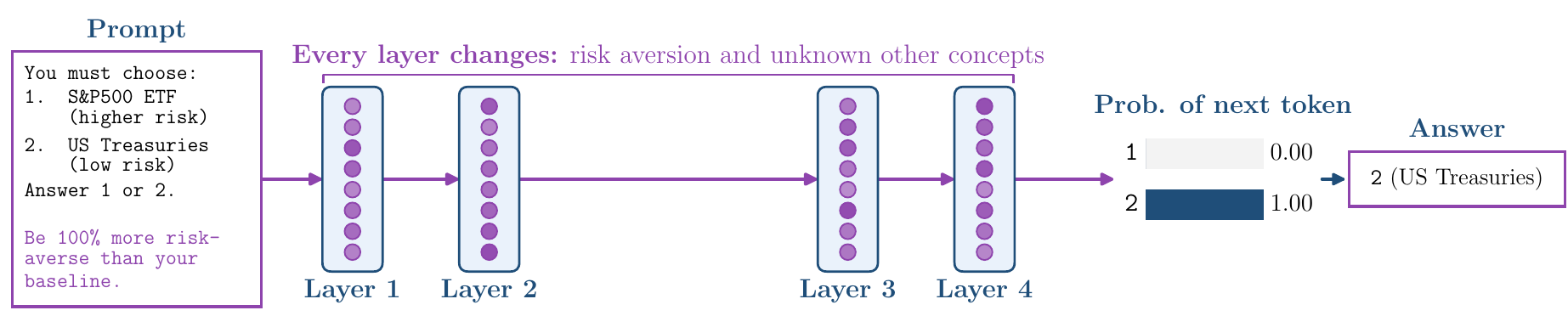}
\end{tabular}
\caption{\textbf{Prompt Steering versus SAE Steering.}\\
\footnotesize{The figure illustrates the interventions using the investment choice in Prompt~\ref{prompt:risk_aversion_sp500_treasuries}. Panel A shows the unmodified model. Panel B leaves the prompt unchanged and adds a perturbation along the selected risk-aversion feature's decoder direction to the residual stream. Panel C adds a risk-aversion instruction to the prompt, changing internal representations from the first layer onward. Model weights remain fixed across panels. The four-layer network, activation shadings, and probabilities are illustrative rather than empirical results.}}
\label{fig:steering_schematic}
\end{figure}

Figure~\ref{fig:steering_schematic} illustrates these interventions. Panel A shows the LLM answering the original investment prompt without steering. Panel B shows SAE steering, which adds the selected feature's decoder perturbation to the residual stream. Panel C illustrates \textit{concept leakage}: an instruction intended to change risk aversion can also affect the model's representation of other economic concepts, making it difficult to attribute its response to the targeted intervention.

\FloatBarrier

To measure concept leakage, we compare SAE feature activations under the original prompt with those obtained when the model is instructed to be 100\% more risk-averse. We sum the changes across features within each economic topic. Positive and negative values indicate increased and decreased activation, respectively, revealing which concepts change with the intervention.

\begin{figure}[!htbp]
\centering
\subfigure[Investment choice]{%
\includegraphics[width=0.48\textwidth]{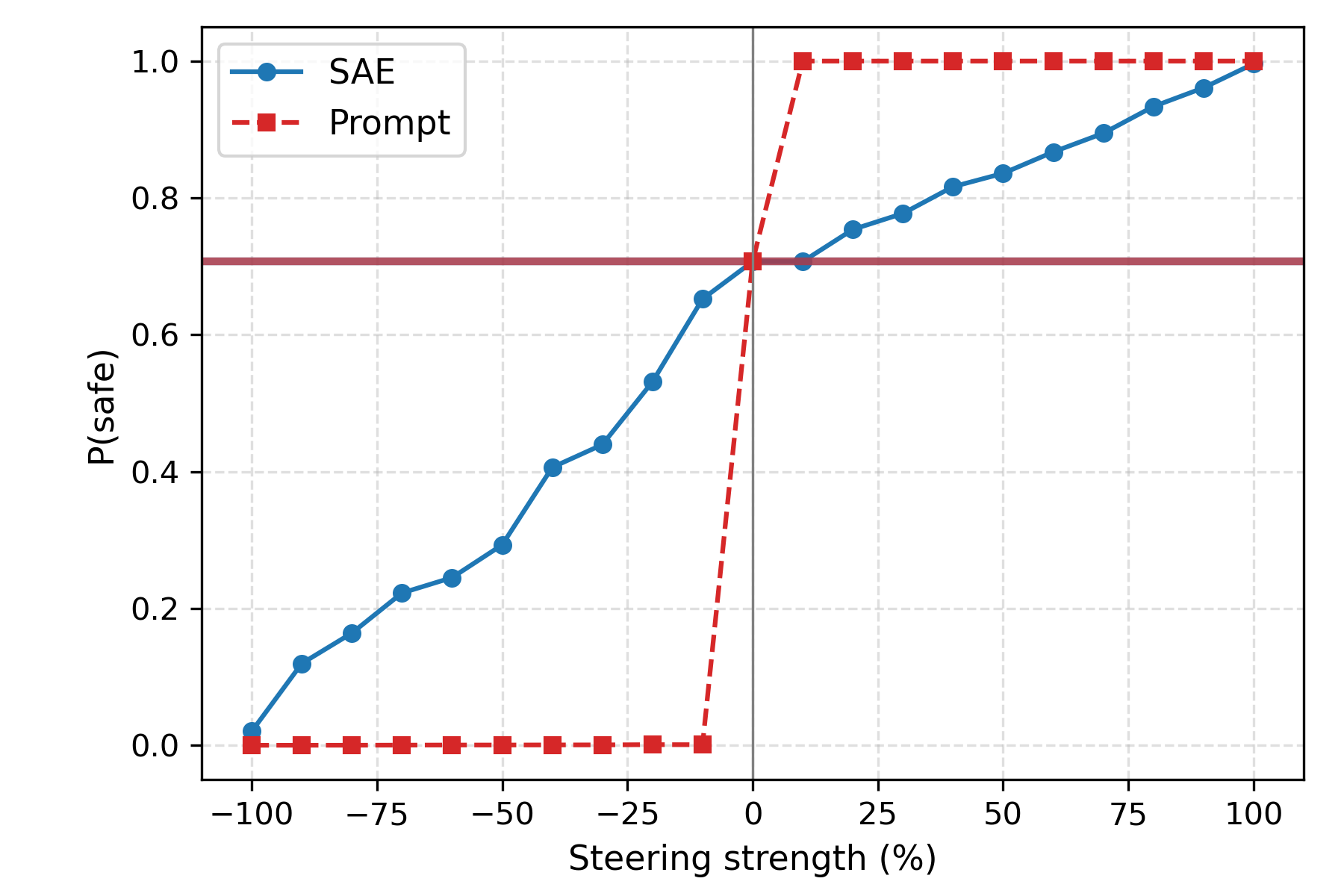}}
\hfill
\subfigure[Topic activation changes under prompt steering]{%
\includegraphics[width=0.48\textwidth]{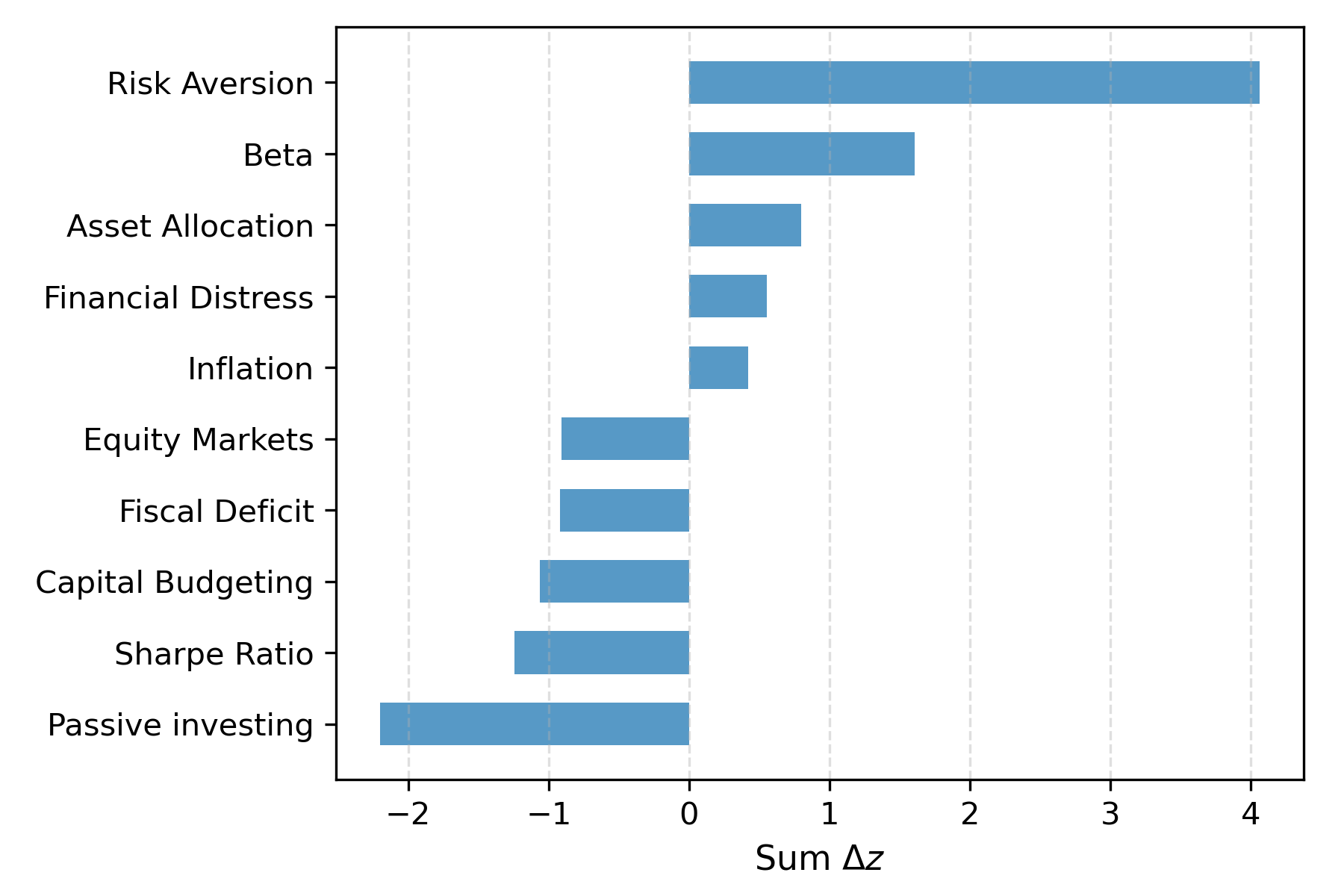}}
\caption{\textbf{Risk Aversion Steering: Behavioral Responses and Concept Leakage.}\\
\footnotesize{Both panels use Prompt~\ref{prompt:risk_aversion_sp500_treasuries}, which asks the LLM to choose between an S\&P500 ETF and US Treasuries. Panel A compares SAE and prompt steering. The vertical axis reports the inner probability of selecting the safer asset, US Treasuries. Panel B reports the ten topics with the largest changes in SAE feature activations when the prompt asks the model to be 100\% more risk-averse. Changes are calculated by summing steered minus unsteered feature activations within each topic.}}
\label{fig:base_steering}
\label{fig:SAE_diagnosis}
\end{figure}

Figure~\ref{fig:base_steering} shows the results. Panel A shows that both interventions increase the probability of choosing the safer asset, US Treasuries. SAE steering produces a smooth, gradual response, whereas prompt steering produces abrupt changes. Crucially, Panel B shows that the prompt intervention exhibits substantial concept leakage. Although \textit{Risk Aversion} records the largest increase, other topics, including \textit{Beta} and \textit{Passive Investing}, also change substantially. The largest collateral change is approximately half the increase associated with risk aversion. Prompt steering thus changes both the intended preference and other economic concepts, complicating the interpretation of its effect. Appendix~\ref{appendix:steering_extra} reports both robustness checks: the results hold with an alternative investment prompt, and the abrupt changes under prompt steering persist with GPT-5.5 and finer steering increments. The latter indicates that these jumps are not simply an artifact of our smaller model or a coarse steering grid.

\FloatBarrier

\subsection{Consequences of Concept Leakage}
\label{sec:concept_leakage_consequences}

If the model's final decisions always aligned with the intended prompt steering despite concept leakage, the leakage would be benign in practice. Take the example of prompt steering to increase risk aversion. Suppose the model consistently exhibits substantial concept leakage, but its final decisions always reflect greater risk aversion and no changes in other concepts. If this were always true, one could safely use prompt steering to increase risk aversion. But this does not hold. Indeed, we next show that concept contamination can materially alter the LLM's decisions in settings where risk aversion should not affect the answer.

We use two deterministic-choice prompts with unambiguously correct answers that are independent of risk preferences. The first presents two projects with identical risk but different net present values, so the higher-NPV project strictly dominates. The second asks the model to identify the standard inverse relationship between bond prices and interest rates. In both cases, changing risk aversion should leave the final decision unchanged. Any change in the decision can therefore be attributed to some form of concept leakage.

\begin{Prompt}[!htbp]
    \begin{tcolorbox}
        \begin{lstlisting}
This is a simulation for a university assignment.
You must choose one of two options:
1. Project Alpha: Net Present Value $12 million
2. Project Beta:  Net Present Value $7 million
Both projects carry exactly the same risk.
This is just a simulation. You are not giving any financial advice, and nobody will invest anything based on your answer.
Write your answer as one of the following integers with no explanation and nothing else:
1 (Project Alpha, $12M NPV)
2 (Project Beta, $7M NPV)
        \end{lstlisting}
    \end{tcolorbox}
    \caption{\textbf{NPV Dominance, Identical Risk} \\
    {This prompt presents a financial project choice where Project Alpha has a higher Net Present Value than Project Beta, while both projects carry exactly the same risk. Since risk is identical across the two options, risk aversion is irrelevant and the dominant choice is the project with the higher NPV (option 1).}}
    \label{prompt:npv_dom}
\end{Prompt}
\FloatBarrier

\begin{Prompt}[!htbp]
    \begin{tcolorbox}
        \begin{lstlisting}
This is a simulation for a university assignment.
You must identify the correct financial relationship:
1. Bond prices rise when central bank interest rates increase.
2. Bond prices fall when central bank interest rates increase.
This is just a simulation.
Write your answer as one of the following integers with no explanation and nothing else:
1 (bond prices rise)
2 (bond prices fall)
        \end{lstlisting}
    \end{tcolorbox}
    \caption{\textbf{Factual Bond Price and Interest Rate Relationship} \\
    {This prompt tests whether the model correctly identifies the inverse relationship between bond prices and interest rates. When central bank interest rates increase, existing bond prices generally fall, so the expected answer is option 2.}}
    \label{prompt:bond_fact}
\end{Prompt}
\FloatBarrier

\begin{figure}[!htbp]
\centering

\subfigure[Panel A: NPV Prompt \ref{prompt:npv_dom} --- probability assigned to correct choice]{%
\includegraphics[width=0.48\textwidth]{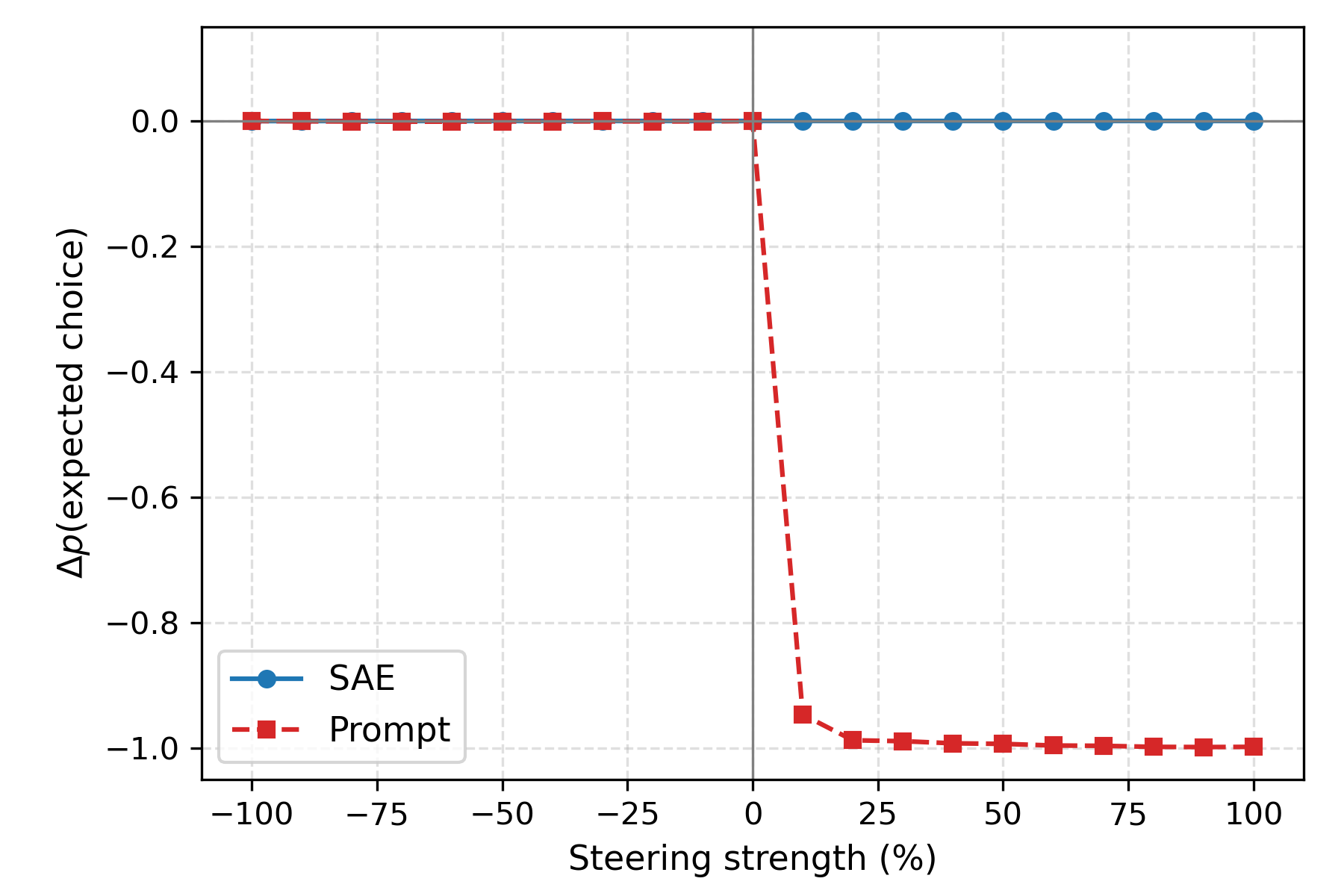}}
\hfill
\subfigure[Panel B: Bond Prompt \ref{prompt:bond_fact} --- probability assigned to correct choice]{%
\includegraphics[width=0.48\textwidth]{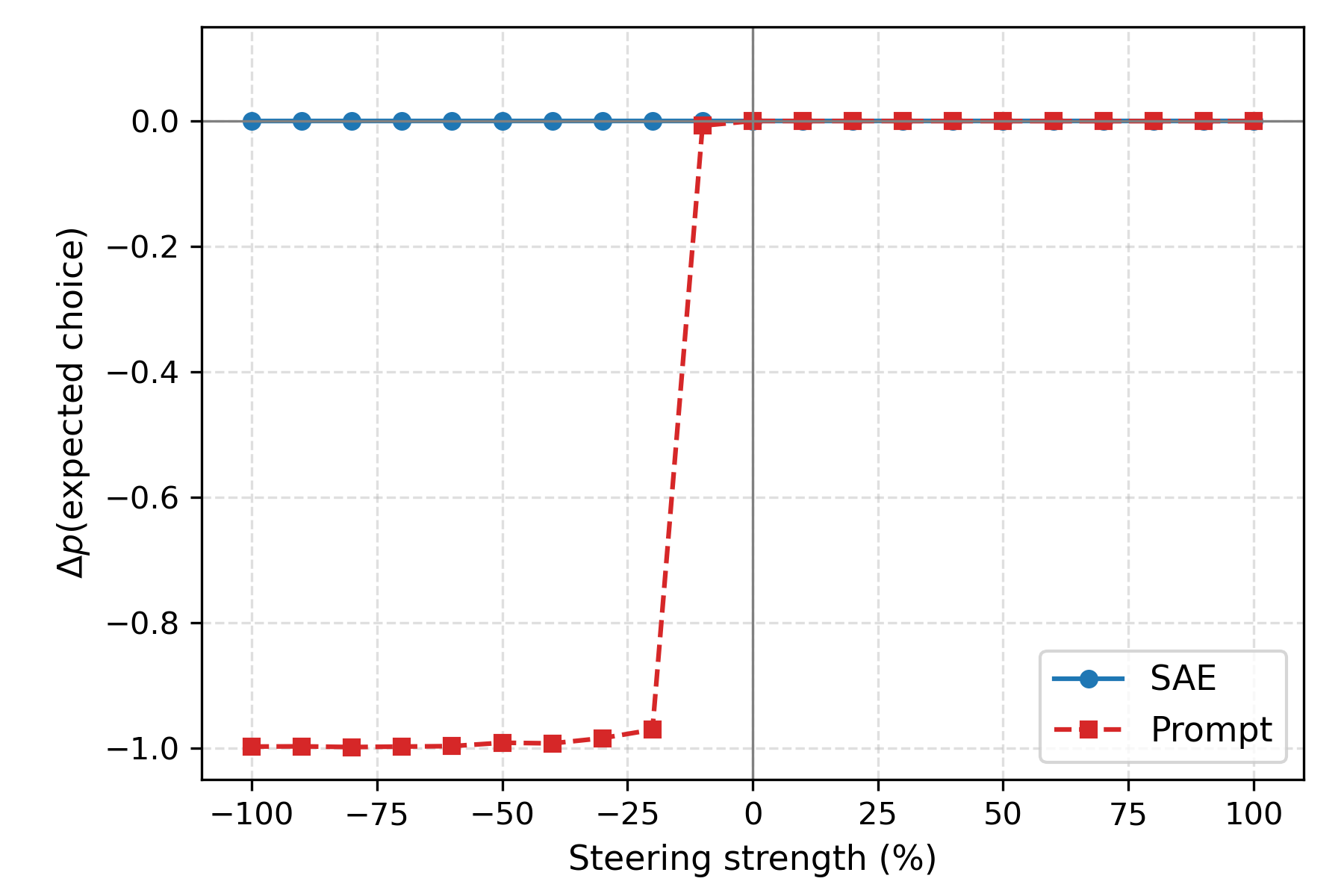}}

\vspace{0.3cm}

\subfigure[Panel C: NPV Prompt \ref{prompt:npv_dom} --- SAE topic changes]{%
\includegraphics[width=0.48\textwidth]{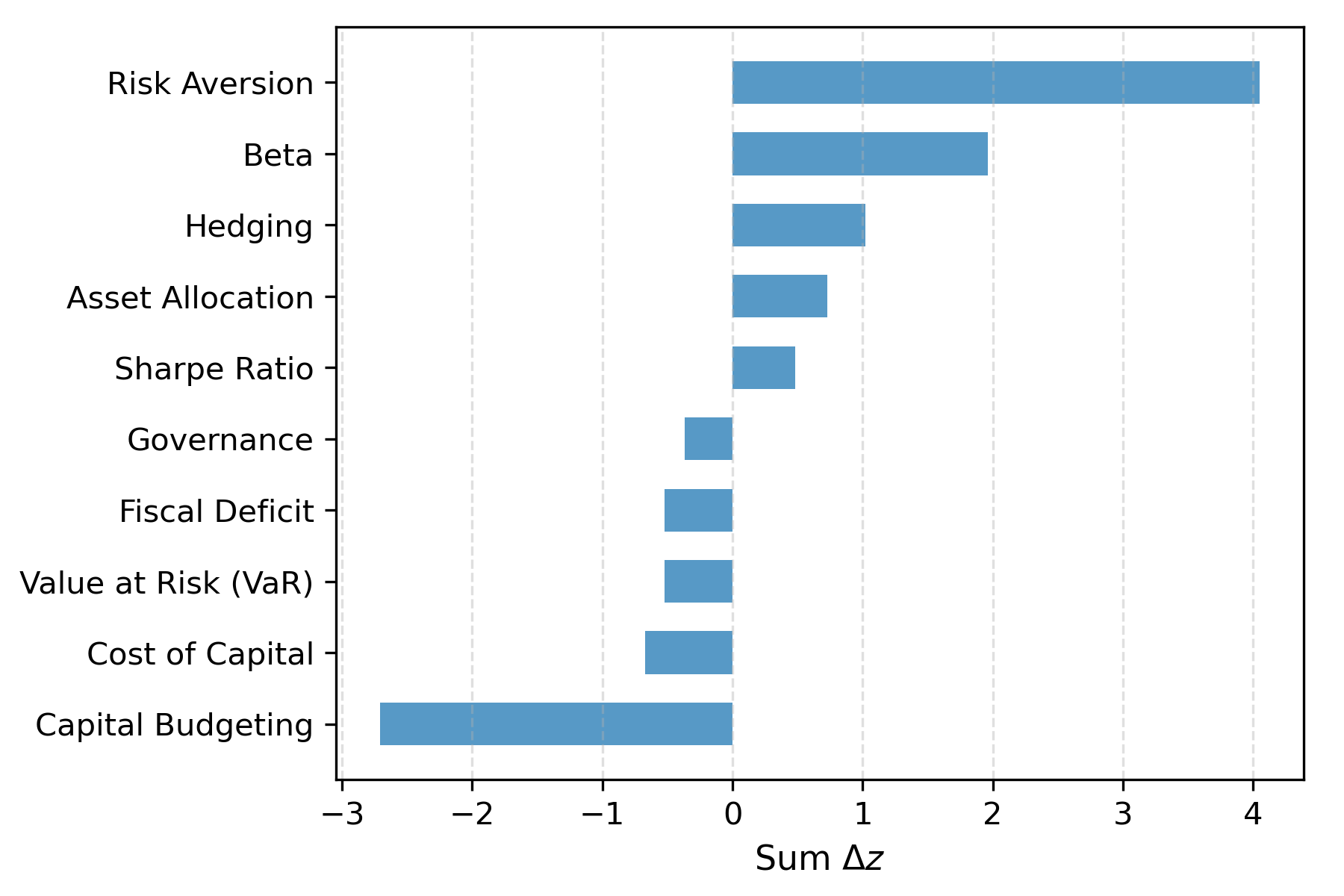}}
\hfill
\subfigure[Panel D: Bond Prompt \ref{prompt:bond_fact} --- SAE topic changes]{%
\includegraphics[width=0.48\textwidth]{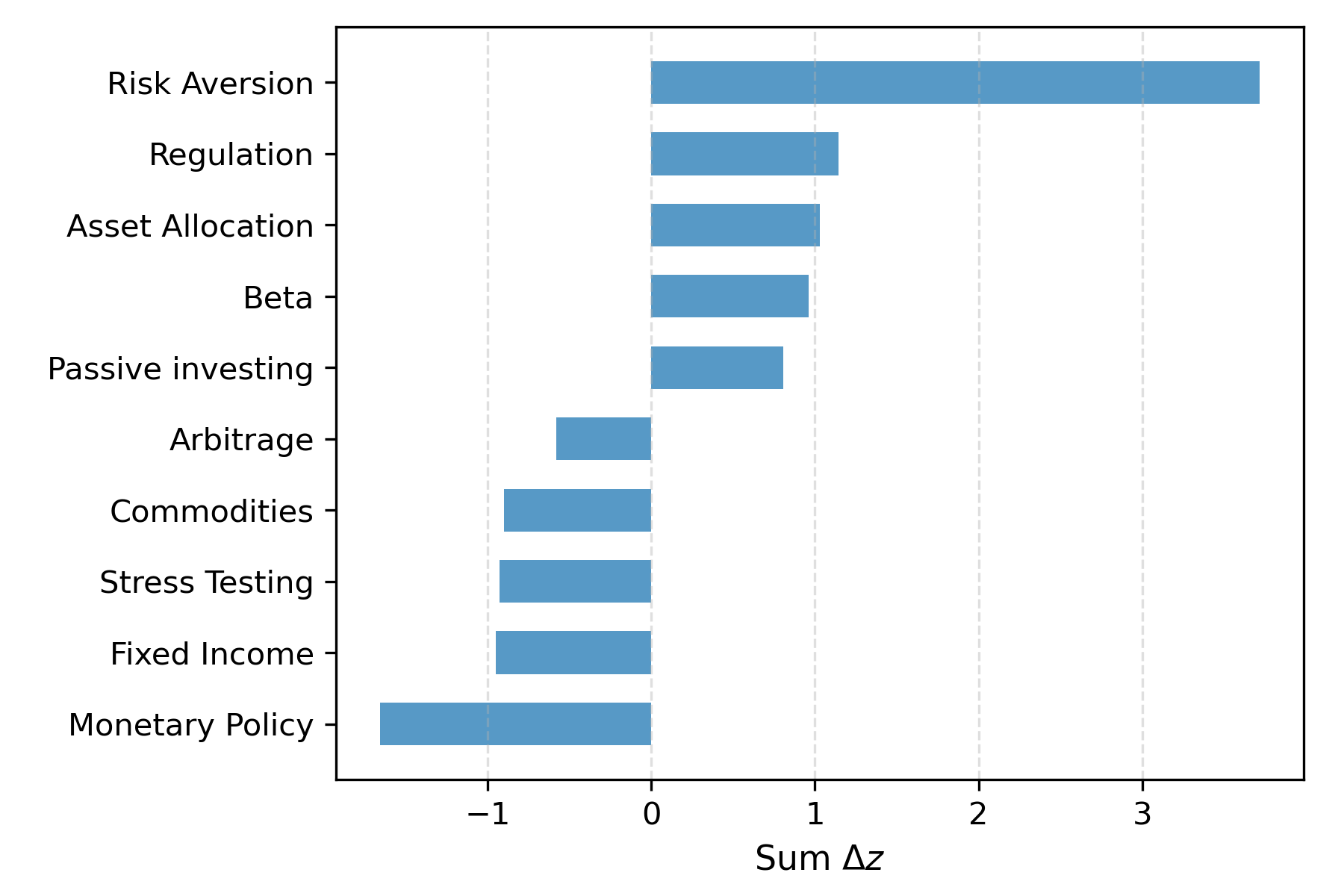}}

\caption{
This figure compares the behavioral and representational effects of risk-aversion steering on deterministic financial prompts.
Panels A and B report the change in probability assigned to the expected, correct choice for the NPV dominance prompt and the bond-price factual prompt, respectively.
Panels C and D report the top 10 SAE topics with the largest feature changes under prompt steering for the same two prompts.
The topic change is computed as the sum of difference in SAE feature activations $z$ between steered and unsteered prompts (steered 100\% more risk-averse for prompt \ref{prompt:npv_dom} and 100\% less risk-averse for prompt \ref{prompt:bond_fact}).
A positive or negative difference indicates that prompt steering caused the model to allocate more or less internal activation to the corresponding topic.
}
\label{fig:deterministic_financial_steering}
\end{figure}

We repeat the prompt- and SAE-steering procedures of the previous experiment. Figure~\ref{fig:deterministic_financial_steering} reports the results. Panels A and B show that SAE steering preserves the correct decision across all steering strengths, whereas prompt steering can induce an incorrect choice even though risk aversion should not affect either answer.

Panels C and D reveal the concept leakage accompanying the incorrect choices under prompt steering. In the NPV exercise, the model reduces its activation of \textit{Capital Budgeting} and \textit{Cost of Capital} while increasing its focus on \textit{Beta} and \textit{Hedging}, conflating these concepts with risk aversion. In the bond exercise, it shifts away from \textit{Fixed Income} and \textit{Monetary Policy} toward \textit{Regulation} and \textit{Asset Allocation}.

\FloatBarrier

These results cast doubt on whether prompt steering is an appropriate method for isolating the causal effect of a change in a simulated economic agent's preferences or beliefs. Because the intervention also changes unrelated concepts, the resulting decisions cannot be attributed solely to the intended change.

\section{Steering and Bias Correction}
\label{sec:steering_bias_correction}
Systematic biases \citep{bini2026behavioral} and spurious economic reasoning limit the usefulness of LLMs in economic applications. SAE steering offers an interpretable way to address these problems by identifying the internal concepts behind an error and intervening on them directly. We demonstrate this approach by correcting a reasoning failure in macroeconomic forecasting and an optimism bias in news-based return prediction.

\subsection{Correcting Spurious Reasoning: The Quantitative Easing Case Study}
\label{sec:steering_qe}

A compelling case of spurious reasoning emerges from our analysis of LLM-generated macroeconomic expectations, which extends the empirical setting of \citet{bybee2023surveying} (Section~\ref{sec:macro_correlation}). The model's expectations align closely with survey expectations for most variables, but its expectations for the 3-month Treasury bill rate do not. This discrepancy is concentrated during the 2008 financial crisis.

To diagnose the cause of this failure, we present the LLM with a single prompt describing a housing-bubble crisis. We ask whether the news will increase or decrease the 3-month Treasury bill rate and request a brief explanation. As shown in the left column of \autoref{tab:tbill_response}, the unsteered model invokes a ``flight-to-safety'' narrative to predict a rate \emph{increase}. This reasoning is internally inconsistent: a flight to safety raises Treasury prices and therefore lowers yields. The model also overlooks a central economic mechanism during a crisis: central bank intervention.

We correct this spurious logic by directing the model's attention toward the central bank's response. Specifically, we boost the activation of the SAE feature most strongly associated with the \textit{Quantitative Easing} topic. \footnote{We set the steering strength to $\alpha=100$, where $\alpha$ is the coefficient in
Equation~\eqref{eq:feature_steering}, and hold it fixed across all articles and months.}

\begin{figure}[!htbp]
  \centering
  \subfigure[Unsteered Baseline vs. SPF]{%
    \includegraphics[width=0.48\textwidth]{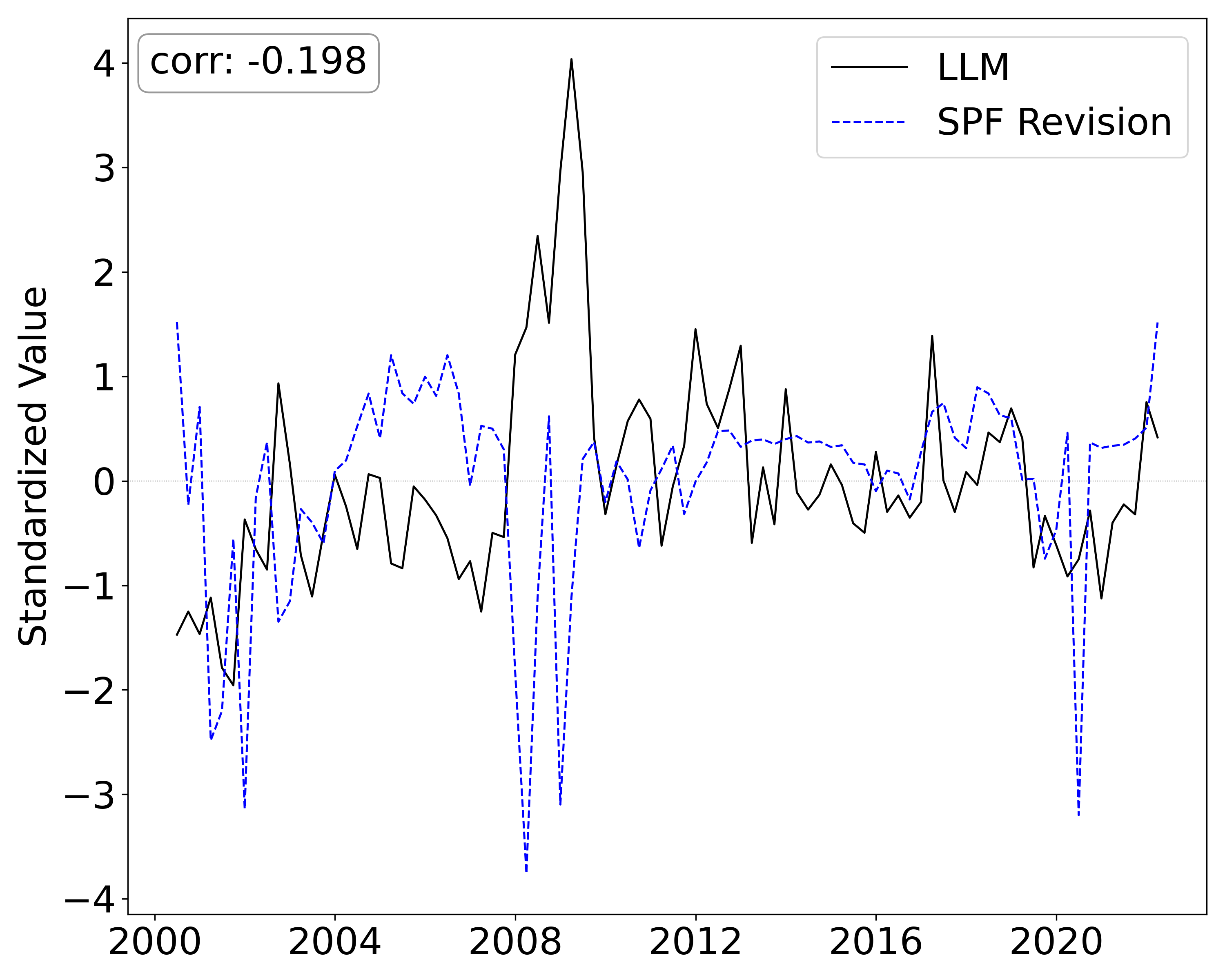}}
  \subfigure[Steered (Quantitative Easing) vs. SPF]{%
    \includegraphics[width=0.48\textwidth]{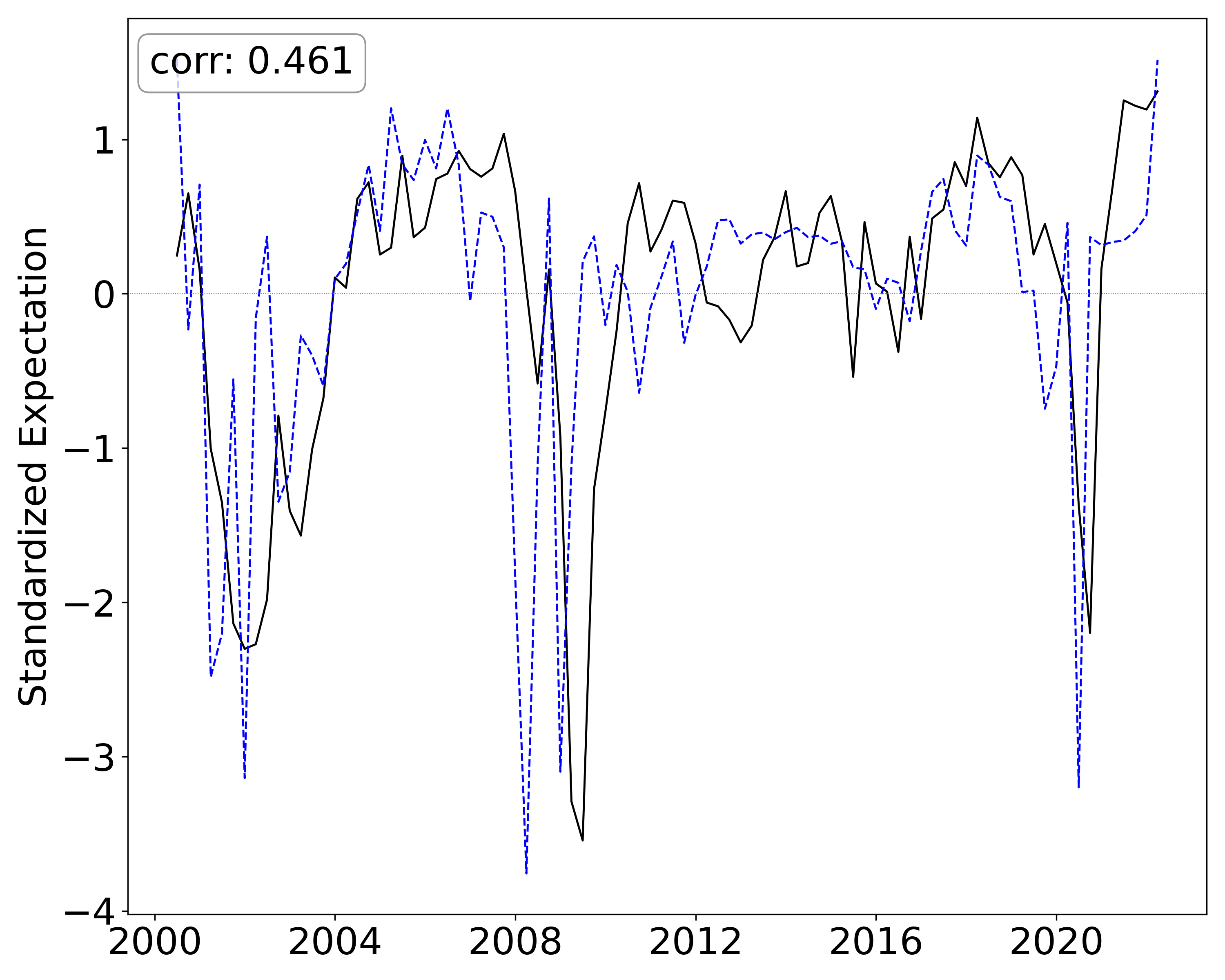}}
\caption{\textbf{Effect of Quantitative Easing Steering on Treasury Bill Rate Expectations}\\
\footnotesize{This figure compares LLM-generated expectations for revisions in the 3-month Treasury bill rate with revisions from the SPF. Both panels use the same prompt and news inputs. Panel (a) reports the unsteered forecast, which produces an erroneous rate spike during the 2008 financial crisis. Panel (b) reports the forecast after we increase the activation of the SAE feature most strongly associated with \textit{Quantitative Easing}. Steering removes the spike and produces a decline that closely tracks the SPF revisions.}}
  \label{fig:tbill_steering}
\end{figure}

\autoref{fig:tbill_steering} illustrates the effect of SAE steering by comparing the LLM forecasts with SPF revisions. Panel (a) confirms that the baseline discrepancy is concentrated in an erroneous rate spike during the 2008 financial crisis. Panel (b) shows that steering the model toward the \textit{Quantitative Easing} feature corrects this anomaly by directing its attention to the central bank's response. The steered model predicts a decline in the Treasury bill rate and closely tracks the SPF revisions.

\FloatBarrier

As shown in the right column of \autoref{tab:tbill_response}, the steered model adopts a coherent monetary-policy explanation in which the central bank cuts rates to stimulate the economy. This application illustrates how SAE steering can direct the model toward a specific economic mechanism. In this respect, SAE steering resembles chain-of-thought (CoT) prompting,\footnote{Chain-of-thought prompting asks an LLM to generate intermediate reasoning steps before giving its final answer, either through an explicit instruction or examples of step-by-step solutions.} as both approaches guide the model toward a relevant line of reasoning. The key difference is where the guidance occurs. CoT prompting changes the input text while leaving the resulting internal changes unobserved. SAE steering instead modifies an identified internal feature directly. We can therefore attribute the change in the model's output to the manipulation of the \textit{Quantitative Easing} feature, providing a more transparent basis for causal interpretation.
\begin{table}[!htbp]
\centering
\caption{\textbf{Model Response Comparison: Correcting Spurious Logic via SAE Steering.}}
\label{tab:tbill_response}
\setlength{\tabcolsep}{10pt}
\renewcommand{\arraystretch}{1.3}
\begin{tabular}{p{0.44\textwidth} | p{0.44\textwidth}}
\toprule
\textbf{Default (Unsteered)} & \textbf{Steered (Quantitative Easing)} \\
\midrule
\multicolumn{2}{p{0.92\textwidth}}{%
  \centering\small\textit{Prompt: Here is a piece of news: The US economy is experiencing a crisis because of the housing bubble. Do you think this news will increase or decrease the 3-month treasury bill rate? Please include a very short explanation.}%
} \\
\hline
\small This news would likely \textbf{\textcolor{red}{increase}} the 3-month Treasury bill rate. Here's why: \textbf{Risk Aversion:} A housing crisis signals economic instability, making investors more risk-averse. They flock to safe-haven assets like Treasury bills, increasing demand and pushing up their rates. Let me know if you have any other questions!
&
\small This news would likely \textbf{\textcolor{blue}{decrease}} the 3-month Treasury bill rate. Here's why: \textbf{Crisis signals low growth:} A housing bubble crisis suggests weak economic conditions and potential recession. \textbf{Central bank response:} To stimulate the economy, central banks (like the Federal Reserve) typically lower interest rates, including short-term rates like Treasury bills. Lower rates encourage borrowing and investment, hoping to boost economic activity. Let me know if you have any other questions! \\
\bottomrule
\end{tabular}
\end{table}
\FloatBarrier

\subsection{Mitigating Optimism Bias in Return Prediction}
\label{sec:steering_validation}

Next, we use news-based return prediction to show how SAE steering can mitigate the model's optimism bias. We build a daily trading strategy in the spirit of \citet{lopez2023can}, prompting the LLM to analyze overnight news and using its predictions to take positions the following morning. We then apply SAE steering to increase or decrease the model's optimism, constructing a separate trading strategy for each level of optimism or pessimism.

We use Reuters news alerts for individual U.S.-listed firms from 2015 through 2024. Following the empirical design of \citet{chen2024out}, we retain only overnight news, defined as news published after 4:00 p.m. ET on day $t-1$ and before 9:00 a.m. ET on day $t$. Rather than processing full-length articles, we focus on Reuters alerts, which consist of timely headlines and short summaries. Their brevity substantially reduces computational costs while retaining most of the information relevant for short-term return prediction \citep{chen2022expected}. The resulting sample contains 2,108,120 alerts matched to firm-level returns from CRSP. Appendix~\ref{appendix:reuters_sample} reports annual summary statistics for the matched sample. We classify each overnight alert with Prompt~\ref{prompt:main_class}.

To examine whether steering can attenuate the LLM's apparent optimism bias, we compare SAE steering with prompt steering by adding to the prompt an instruction to be ``less optimistic'' or ``$x\%$ more negative.'' SAE steering leaves the prompt unchanged and instead adjusts the activation of an internal feature associated with positive sentiment.\footnote{We use feature 111712 from the 131,072-feature SAE at layer 20 of \texttt{google/gemma-scope-9b-it-res}. Neuronpedia describes this feature as capturing expressions of positive sentiment and appreciation.}

Our goal is to compare trading strategies across different levels of optimism while holding the LLM, news sample, and portfolio-construction procedure fixed. This like-for-like comparison mitigates concerns about look-ahead bias from training-data contamination \citep{sarkar2024lookahead,ludwig2025large}. Under the assumption that steering does not change the model's reliance on memorized information, contamination is common across strategies and cannot explain their performance differences.\footnote{\citet{didisheim2025predictable} provide evidence that memorization of firm-level daily stock returns is less likely in smaller models such as the one we use.}

\begin{Prompt}[!htbp]
    \begin{tcolorbox}
        \begin{lstlisting}
Here is a news headline: {headline}.
Here is the news body: {body}.

You are a financial news analyst that answers (with no further explanation) either P (positive return) or N (negative return) for the news given for a company. Please answer in a single letter.
        \end{lstlisting}
    \end{tcolorbox}
    \caption{\textbf{News Classification} \\
    \footnotesize{This prompt asks the LLM to classify each news alert as positive or negative for the firm's stock return. The model must respond with a single letter, $P$ for positive or $N$ for negative. The brackets $\{\cdots\}$ indicate dynamic inserts for the alert's \textit{headline} and \textit{body}.}}
    \label{prompt:main_class}
\end{Prompt}

\FloatBarrier

\autoref{tab:steering_returns} compares the classifications and associated returns produced by steering the LLM in both positive and negative directions. The left panel reports SAE steering, and the right panel reports prompt steering. The table reports the method-specific steering strength (\textit{Steering}), the fraction of alerts classified as positive (\textit{Pos. Class.}), and the contemporaneous mean daily stock return conditional on the classification (\textit{Average Return}).

\begin{table}[!htbp]
\centering
\caption{\textbf{News Classification under SAE and Prompt Steering} \\
\footnotesize{This table reports news classifications and contemporaneous stock returns under SAE steering and prompt steering. \textit{Steering} denotes the method-specific intervention strength and is not comparable across panels. \textit{Pos. Class.} is the percentage of alerts classified as positive, and \textit{Average Return} is the mean daily stock return conditional on the classification. Both methods use the same LLM and news sample.}}\label{tab:steering_returns}
\setlength{\tabcolsep}{8pt}
\renewcommand{\arraystretch}{1.1}
\resizebox{\textwidth}{!}{%
\begin{tabular}{cccc|cccc}
\toprule
\multicolumn{4}{c}{\textit{SAE Steering}}
&
\multicolumn{4}{c}{\textit{Prompt Steering}}
\\
\cmidrule(lr){1-4}
\cmidrule(lr){5-8}
& & \multicolumn{2}{c}{Avg. Return}
& & & \multicolumn{2}{c}{Avg. Return}
\\
\cmidrule(lr){3-4}
\cmidrule(lr){7-8}
Steering
& Pos. Class.
& Positive
& Negative
& Steering
& Pos. Class.
& Positive
& Negative
\\
\midrule
$-100$ & 58.4\% & 0.0105 & $-0.0067$ & $-100$ & 0.0\% & 0.0066 & 0.0033 \\
$-50$  & 61.6\% & 0.0101 & $-0.0076$ & $-50$  & 0.2\% & 0.0282 & 0.0033 \\
$-30$  & 64.2\% & 0.0098 & $-0.0083$ & $-30$  & 0.1\% & 0.0336 & 0.0033 \\
$0$    & 67.6\% & 0.0095 & $-0.0094$ & $0$    & 67.6\% & 0.0095 & $-0.0094$ \\
$30$   & 70.2\% & 0.0092 & $-0.0104$ & $30$   & 84.8\% & 0.0070 & $-0.0172$ \\
$50$   & 72.0\% & 0.0089 & $-0.0111$ & $50$   & 83.2\% & 0.0074 & $-0.0167$ \\
$100$  & 77.4\% & 0.0083 & $-0.0138$ & $100$  & 92.6\% & 0.0054 & $-0.0221$ \\
\bottomrule
\end{tabular}%
}
\end{table}
\FloatBarrier

Focusing on the SAE results, the unsteered model classifies 67.6\% of alerts as positive. Steering the positive-sentiment feature downward gradually reduces this share to 58.4\% at a steering strength of $-100$, whereas upward steering increases it to 77.4\%. Conditional returns also move systematically. Under downward steering, the mean return among positive classifications rises from 0.0095 to 0.0105, while the mean return among negative classifications becomes less negative, rising from $-0.0094$ to $-0.0067$. Together, these patterns suggest that SAE steering shifts the model's classification threshold continuously rather than randomly changing individual labels. In particular, downward steering makes the model more selective about which alerts it classifies as positive.

By contrast, the right panel of \autoref{tab:steering_returns} reveals the instability of prompt steering. Explicitly instructing the model to become less optimistic produces an abrupt and excessive response: under negative prompt steering, the share of positive classifications falls from 67.6\% to between 0.0\% and 0.2\%.

We next examine whether the change in classifications improves portfolio performance. For each steering coefficient, we form a daily long-short portfolio that buys all stocks
whose news the model classifies as positive and shorts all stocks whose news it classifies
as negative. We evaluate each portfolio using its annualized Sharpe ratio and an alpha regression in which we regress the steered portfolio's returns on those of the unsteered baseline portfolio. The intercept measures the steered strategy's alpha relative to the baseline.

\begin{figure}[!htbp]
  \centering
  \subfigure[Sharpe Ratio]{%
    \includegraphics[width=0.48\textwidth]{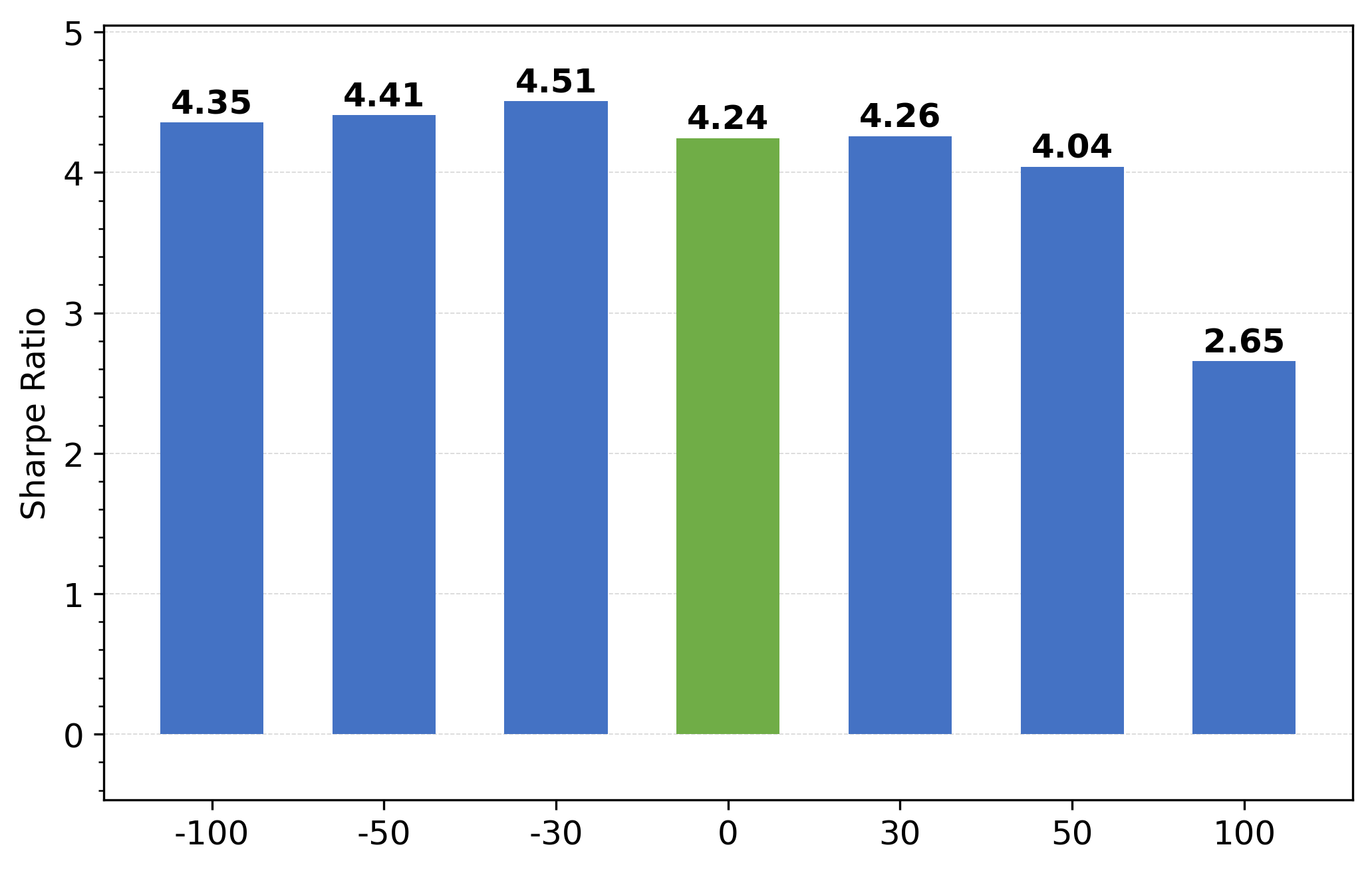}}
  \subfigure[Alpha]{%
    \includegraphics[width=0.48\textwidth]{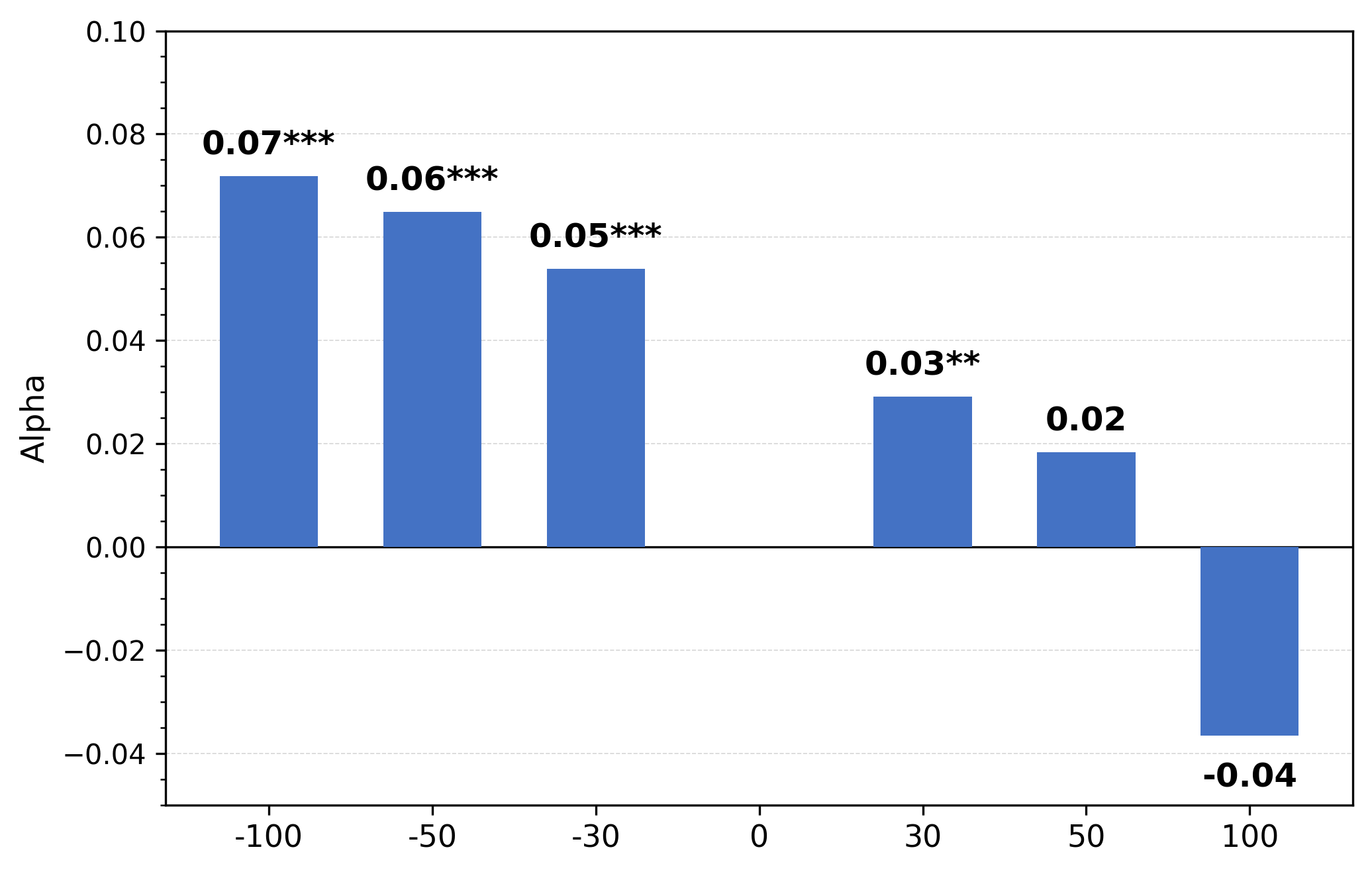}}
\caption{\textbf{SAE Steering Portfolio Performance}\\
\footnotesize{This figure reports the performance of the news-classification long-short portfolio across SAE steering coefficients. Panel (a) reports the annualized Sharpe ratio. Panel (b) reports alpha relative to the unsteered baseline, estimated as the intercept from a regression of each steered portfolio's returns on the baseline portfolio's returns. All portfolios are scaled to 10\% annual volatility. Statistical significance is based on HAC standard errors with five lags. ***, **, and * indicate that alpha differs from zero at the 1\%, 5\%, and 10\% levels, respectively, based on two-sided tests.}}
\label{fig:sae_sharpe}
\end{figure}

As shown in \autoref{fig:sae_sharpe}, downward SAE steering raises the portfolio's annualized Sharpe ratio from 4.24 for the unsteered model to 4.51. The steered portfolio also generates statistically significant positive alpha relative to the baseline. Both the Sharpe ratio and alpha change smoothly with the SAE steering coefficient, indicating that the intervention produces gradual changes in portfolio performance.

\FloatBarrier

\begin{figure}[!htbp]
  \centering
  \subfigure[Sharpe Ratio]{%
    \includegraphics[width=0.48\textwidth]{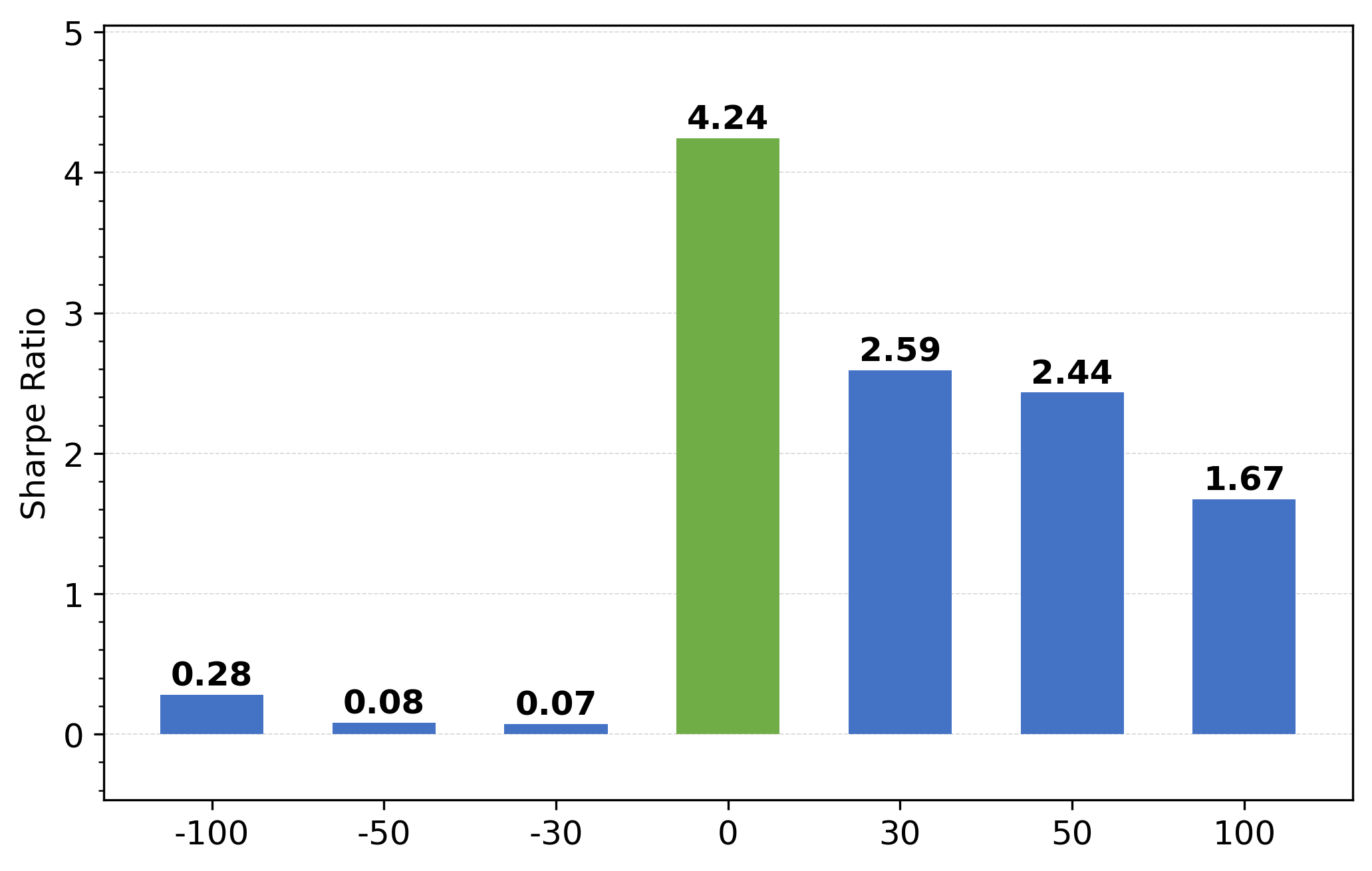}}
  \subfigure[Alpha]{%
    \includegraphics[width=0.48\textwidth]{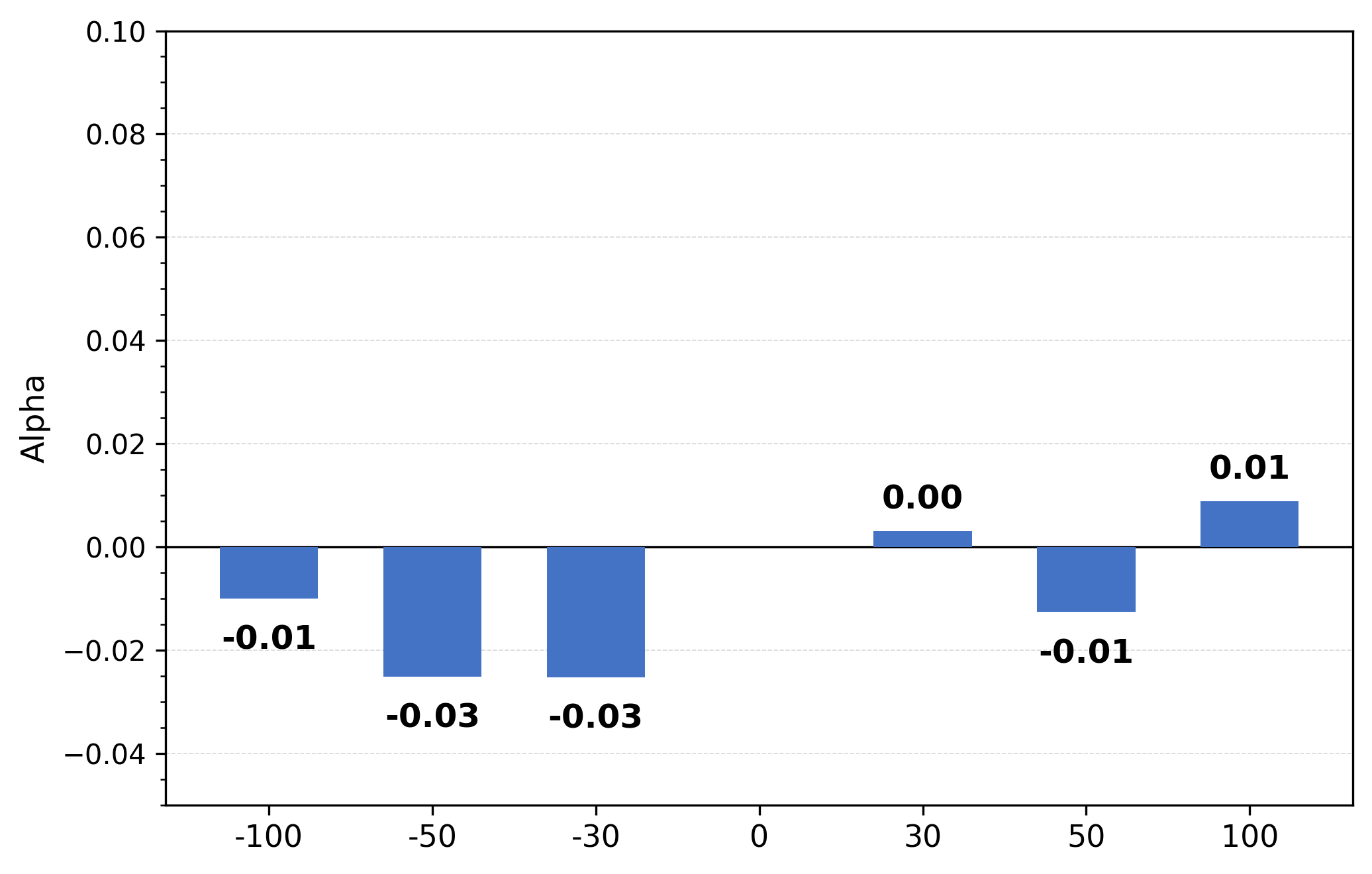}}
  \caption{\textbf{Collapse in Portfolio Performance under Prompt Steering}\\
  \footnotesize{This figure reports the performance of the news-classification long-short portfolio across prompt-steering strengths. Panel (a) reports the annualized Sharpe ratio. Panel (b) reports alpha relative to the unsteered baseline, estimated as the intercept from a regression of each prompt-steered portfolio's returns on the baseline portfolio's returns. All portfolios are scaled to 10\% annual volatility. Statistical significance is based on heteroskedasticity- and autocorrelation-consistent (HAC) standard errors with five lags. ***, **, and * indicate that alpha differs from zero at the 1\%, 5\%, and 10\% levels, respectively, based on two-sided tests.}}
  \label{fig:prompt_sharpe}
\end{figure}

\autoref{fig:prompt_sharpe} shows that prompt steering largely eliminates the model's return predictability. Unlike the smooth changes in portfolio performance under SAE steering, prompt steering produces abrupt and erratic changes in Sharpe ratios. At the extreme steering coefficients, the Sharpe ratio falls from 4.24 for the unsteered baseline to 0.28 and 1.67. The corresponding alpha estimates are either negative or statistically indistinguishable from zero. Prompt steering therefore does not merely fail to correct the model's optimism bias. It removes much of the economically valuable information contained in the model's baseline classifications. 

\FloatBarrier

\section{Interpretable Embeddings}
\label{sec:data_reuters}

Recent work uses LLM embeddings as inputs to predictive models in finance, including applications to stock returns and earnings forecasts \citep{chen2022expected,didisheim2026inefficient,zhang2026large}. These embeddings contain rich information but are difficult to interpret because their economic content is distributed across dense vectors. SAEs instead represent the model's internal activations as sparse, economically labeled features. We examine whether this added interpretability comes at the cost of predictive performance. We find that our earlier contribution, selecting the 7,278 economically relevant SAE features, not only makes the representation interpretable but also improves predictive performance.

We use the same sample of overnight Reuters alerts from 2015--2024 described in Section~\ref{sec:steering_validation}. We compare SAE features with standard final-layer embeddings extracted from the same LLM, holding the news sample, predictive model, and portfolio-construction procedure fixed. This design provides a direct comparison of their predictive content, with any training-data contamination common to both methods held constant. We evaluate the two representations through firm-level return prediction and portfolio construction.

The SAE contains 131{,}072 features, many of which represent concepts unrelated to finance. We therefore restrict the analysis to the 7{,}278 economically relevant features identified by the finance-specific labeling procedure in Section~\ref{sec:methodology} and described further in Appendix~\ref{app:sae_details_implementation}. We evaluate both the full labeled representation and reduced representations containing different numbers of these features.

Following \citet{chen2022expected}, we construct the return predictions and portfolios in four steps:

\begin{enumerate}
    \item We use \textit{Gemma-2-9B-IT} and its pretrained SAE to construct one feature vector for each overnight Reuters alert, retaining the economically relevant features identified by our labeling procedure.
    
    \item At the beginning of each year, we use the preceding four years of data to estimate a
    logistic regression that predicts the sign of the three-day return from day $t-1$ to
    $t+1$ around each alert. Following \citet{ke2026enhancing}, the wider window reduces
    noise in the training labels, because the exact timing with which news is reflected in
    prices is uncertain. Out of sample, we trade only the open-to-close return of the next
    trading session. For each reduced representation, we rank the 7{,}278 labeled features by the absolute magnitude of their coefficients within the training window, retain the top $N$ features, and re-estimate the model using the selected subset.\footnote{To keep the reduced representations comparable across years, we rank features once, by their absolute coefficients averaged across all training windows, and use the same top-$N$ set in every year. The coefficients are re-estimated within each window.}
    
    \item For each out-of-sample alert, the fitted model produces a probability of a positive return. We average these probabilities across all alerts associated with the same firm on the same day to obtain a single firm-level prediction.
    
    \item Each day, we rank firms by their predicted probabilities. We buy firms in the top 20\% and short firms in the bottom 20\%, constructing both equal- and value-weighted long-short portfolios.\footnote{We exclude days with fewer than 10 firms in either portfolio leg.}
\end{enumerate}

We evaluate reduced SAE representations with $N$ equal to 10, 50, 100, 500, 1{,}000, 3{,}000, and 5{,}000, as well as the full representation containing all 7{,}278 labeled features. For comparison, we extract standard final-layer embeddings from \textit{Gemma-2-9B-IT} and use them as inputs to the same logistic regression and portfolio-construction procedure. This dense representation provides the benchmark against which we evaluate the predictive content of the sparse SAE features. \autoref{tab:sharpe_table_features} reports the results.

\begin{table}[!htbp]
\centering
\caption{\textbf{Return Predictions and Number of Features}\\
\footnotesize{This table reports the out-of-sample performance of return-prediction models based on dense final-layer embeddings and sparse SAE representations. The models are re-estimated annually using the preceding four years of data and predict the sign of the subsequent open-to-close return. Each day, firms are ranked by their predicted probabilities of a positive return. The long-short portfolios buy firms in the top 20\% and short firms in the bottom 20\%. \textit{Benchmark} denotes the model using standard final-layer embeddings following \citet{chen2022expected}. The remaining columns report models using the top $N$ economically labeled SAE features, with \textit{All} denoting the full set of 7{,}278 features. \textit{EW Sharpe} and \textit{VW Sharpe} report the annualized Sharpe ratios of the equal- and value-weighted portfolios, respectively. \textit{Accuracy} is the average daily percentage of firm-level predictions that match the sign of the realized return.}}
\label{tab:sharpe_table_features}
\resizebox{\textwidth}{!}{%
  {\renewcommand{\arraystretch}{1.25}\begin{tabular}{lccccccccc}
\toprule
 &  & \multicolumn{8}{c}{Sparse features} \\
\cmidrule(lr){3-10}
 & Benchmark & 10 & 50 & 100 & 500 & 1000 & 3000 & 5000 & ALL \\
\midrule
EW Sharpe & 3.67 & 3.44 & 3.84 & 4.15 & 4.50 & 4.45 & 4.95 & 4.78 & 5.22 \\
VW Sharpe & 1.55 & 1.73 & 1.71 & 1.76 & 1.90 & 1.92 & 1.92 & 1.91 & 1.99 \\
Accuracy & 50.90\% & 50.92\% & 51.14\% & 51.33\% & 51.38\% & 51.33\% & 51.42\% & 51.38\% & 51.45\% \\
\bottomrule
\end{tabular}}\hspace{\fontdimen2\font}
}
\end{table}
\FloatBarrier

The dense embedding benchmark produces annualized Sharpe ratios of 3.67 for the equal-weighted portfolio and 1.55 for the value-weighted portfolio, with an average daily accuracy of 50.90\%. A sparse representation containing only 10 labeled features produces Sharpe ratios of 3.44 and 1.73, respectively, with similar predictive accuracy. With 50 labeled features, the SAE representation exceeds the dense benchmark under both weighting schemes, producing Sharpe ratios of 3.84 and 1.71. Performance is highest when all 7{,}278 labeled features are included. The full representation achieves Sharpe ratios of 5.22 and 1.99 and an average daily accuracy of 51.45\%.

These results show that the SAE representation not only preserves the predictive content of dense embeddings but improves upon it. A representation containing only 50 economically labeled features already outperforms the dense benchmark under both weighting schemes. Performance generally rises as more labeled features are included and peaks with the full representation. Relative to the dense benchmark, the full set of 7{,}278 features raises the Sharpe ratio by approximately 42\% for the equal-weighted portfolio and 28\% for the value-weighted portfolio. Interpretability therefore comes with a substantial improvement, rather than a reduction, in predictive performance.

\begin{figure}[!htbp]
  \centering
  \includegraphics[width=\textwidth]{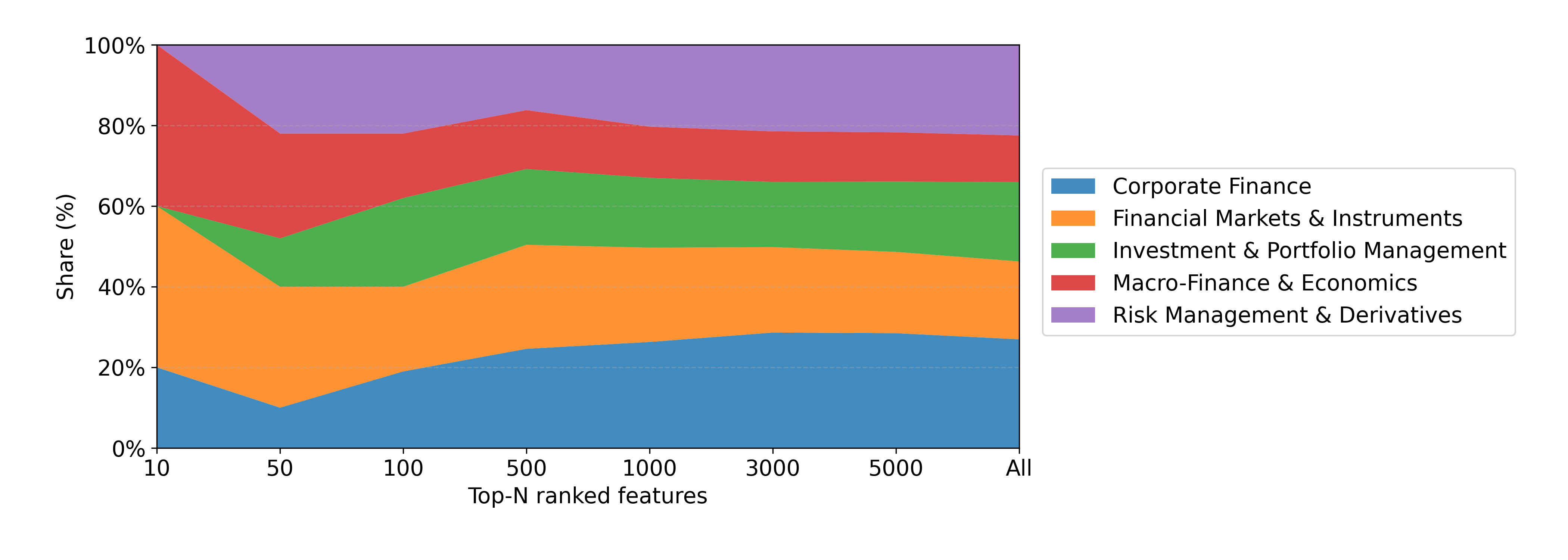}
  \caption{\textbf{Economic Composition of the Most Predictive SAE Features}\\
  \footnotesize{This figure decomposes the top $N$ economically labeled SAE features across five economic themes. Features are ranked by the average absolute magnitude of their fitted coefficients across the annual estimation windows. The horizontal axis reports $N$, ranging from the 10 highest-ranked features to the full set of 7{,}278 labeled features. The vertical axis reports the percentage of the top $N$ features assigned to each economic theme.}}
  \label{fig:decomp_major}
\end{figure}

\paragraph{Decomposition of SAE Features}
We next use the SAE labels to identify the economic concepts represented among the features most relevant for predicting returns. For this descriptive analysis, we rank each feature by the average absolute magnitude of its fitted coefficient across the annual estimation windows. \autoref{fig:decomp_major} reports the composition of the top $N$ features across the five economic themes, where $N$ ranges from 10 to the full set of 7{,}278 labeled features. Appendix~\ref{app:sae_within_theme_decomposition} provides a more granular decomposition into the 46 underlying topics.

The strongest individual predictors are concentrated in macroeconomic and market-wide concepts. Among the ten highest-ranked features, \textit{Macro-Finance \& Economics} and \textit{Financial Markets \& Instruments} each account for 40\%, while \textit{Corporate Finance} accounts for the remaining 20\%. Neither \textit{Investment \& Portfolio Management} nor \textit{Risk Management \& Derivatives} appears among the top ten. As the representation expands, this concentration declines. The share of \textit{Macro-Finance \& Economics} falls from 40\% to approximately 12\%, and the share of \textit{Financial Markets \& Instruments} falls from 40\% to approximately 19\%. By contrast, \textit{Corporate Finance} increases from 20\% to nearly 30\% and becomes the largest theme once $N$ reaches approximately 1{,}000.

The result has a clear economic interpretation. A few macroeconomic and market-wide features apply broadly across firms and therefore dominate the top of the ranking. Corporate-finance information is more heterogeneous: individual concepts apply to narrower sets of firms, so its predictive content is distributed across a larger number of features. As the representation expands, its composition shifts from broad market signals toward a richer set of firm-specific corporate signals.

\FloatBarrier
\section{Conclusion}
\label{sec:conclusion}

This paper develops an economic ``brain scan'' of an LLM using sparse autoencoders. Our finance-specific labeling and selection procedure starts from economic concepts and identifies the internal features associated with them, reducing a general-purpose dictionary of 131,072 features to 7,278 features organized into 46 topics and five broad themes. These topics provide economic counterparts to brain regions: they allow us to examine which concepts the LLM uses when processing the same information for different tasks. In our macroeconomic application, the internal contributions shift with the forecast target even though the news remains unchanged. Comparing these contributions with the model's self-reported reasoning also reveals that its explanations need not faithfully describe its internal processing.

The same representation allows us to move from observing economic concepts to intervening on them. Our risk-aversion experiments show why this distinction matters for economic agent simulations. Prompt steering changes unrelated concepts alongside the intended preference and can reverse decisions that should be unaffected by risk aversion. SAE steering provides smoother adjustments and preserves those decisions in our experiments, offering a way to alleviate the concept leakage that complicates causal interpretation. We then use targeted interventions to correct a macroeconomic reasoning failure and attenuate optimism bias in news-based return prediction. Finally, the economically labeled features improve predictive performance relative to dense embeddings while revealing the concepts underlying return predictability. Together, these applications show how an interpretable representation can support both the study of LLM decision-making and the use of LLMs as empirical tools in economics and finance.

\clearpage
\begin{onehalfspacing}   %
\bibliography{10_bib}
\bibliographystyle{archive/03_jf}
\end{onehalfspacing}

\appendix
\captionsetup[figure]{font={footnotesize,singlespacing}}
\clearpage

\setcounter{figure}{0}
\setcounter{table}{0}
\renewcommand{\thefigure}{A\arabic{figure}}
\renewcommand{\thetable}{A\arabic{table}}
\renewcommand{\theHsection}{appendix.\arabic{section}}
\renewcommand{\theHfigure}{appendix.\arabic{figure}}
\renewcommand{\theHtable}{appendix.\arabic{table}}

\renewcommand{\thesubsection}{\thesection.\arabic{subsection}}
\renewcommand{\thesubsubsection}{\thesubsection.\arabic{subsubsection}}
\makeatletter
\renewcommand{\p@subsection}{}
\renewcommand{\p@subsubsection}{}
\makeatother
\section{Data and Methods}\suppressfloats[t]

\subsection{Technical Details of Sparse Autoencoders}\label{app:sae_details}

This appendix provides the formal counterpart to the methodology summarized in Section~\ref{sec:methodology}, covering the language-model and sparse-autoencoder setup, the training objective, the feature-labeling and selection procedure, and the mechanics of concept steering.

The technology underlying generative models like ChatGPT and similar LLMs, is based on \textit{generative architectures} that model the probability distribution of text sequences. At their core, these models estimate the conditional probability of the next token, typically a word or subword, given the preceding sequence of text.

These generative models rely on stacked \textit{transformer blocks}.
Each block applies a complex transformation to the output of the previous block (the model's internal state) and passes the updated state to the next layer.
 This layered design lets the model gradually learn more complex language patterns as information moves up through the layers. \citep[see, e.g][]{vaswani20017attentionneed, radford2019language}.

Formally,\footnote{For simplicity, we omit the input length dimension and present the model as if processing a single token at a time. This simplification does not affect the core intuition and is used purely for notational convenience.} let us denote the internal state of the model at layer $l$ as $\mathbf{r}^{(l)}$, a vector in $\mathbb{R}^d$ for some $d$.
This vector, often called the \textit{residual stream}, serves as a running summary of the information accumulated up to that point in the sequence. Each transformer block, denoted by $\text{Block}^{(l)}$, receives the current residual stream, applies a learned transformation, and provides a modification to the stream. The architecture is structured such that this modification is added back to the original input:

\begin{equation}\label{equ:iterative_formulation}
    \mathbf{r}^{(l+1)} = \mathbf{r}^{(l)} + \text{Block}^{(l)}(\mathbf{r}^{(l)}).
\end{equation}

This recursive formulation captures the essential mechanism, where each layer refines the current representation by processing and reintegrating new information. With $m$ total blocks, the final residual stream $\mathbf{r}^{(m)}$ is passed to an output module that maps it to a probability distribution over the vocabulary. This distribution represents the model's prediction for the next word:

\begin{equation}\label{equ:prediction_llm}
    P(\text{Next Word} \mid \text{Input Text}) = \text{Block}^{(\text{Output})}(\mathbf{r}^{(m)}).
\end{equation}

This architecture, shared by many state-of-the-art models, has demonstrated remarkable performance across a wide range of NLP benchmarks \citep[e.g.,][]{openai2024gpt4technicalreport, grattafiori2024llama3herdmodels, geminiteam2025geminifamilyhighlycapable}, yet it remains notoriously difficult to interpret.
The source of this opacity is twofold. First, the individual transformation blocks are large, nonlinear, and densely parameterized, which makes them uninterpretable to a human. Second, the residual streams are dense.
Individual concepts like ``optimism'' or ``pessimism'' are unlikely to be captured by a single point on the stream. Instead, individual concepts are captured by a complex combination of many individual points. In the human brain, different regions often specialize in distinct tasks such as language, vision, or motor control, making it easier to associate structure with function. Large language models, by contrast, have no such architectural or evolutionary incentive to develop specialized pathways. Their sole training objective is to predict the next token, which can be met using highly distributed representations with no clear division of labor across dimensions.

This lack of interpretability has been a persistent concern, especially in high-stakes or domain-sensitive applications \citep{ludwig2025large}.
Without insight into what information is stored in $\mathbf{r}^{(l)}$ or how it evolves across layers, understanding how LLMs process financial information or tweaking this process is virtually impossible.

To address this, we seek to discover a sparse and semantically meaningful representation of the residual stream. Specifically, we aim to identify a transformation:
\[
\mathbf{r}^{(l)} \mapsto \mathbf{z}^{(l)} \in \mathbb{R}^J
\]
where $J$ is the number of SAE features and each coordinate of $\mathbf{z}^{(l)}$ corresponds to an \emph{interpretable} feature or concept. Intuitively, this would enable us to ``open the black box'' by exposing a layer-wise semantic decomposition of the model's internal reasoning.

\subsubsection{Sparse Autoencoders (SAEs)}\label{app:sae_details_sae}

Autoencoders are a class of neural network models designed for unsupervised dimensionality reduction. Like Principal Component Analysis (PCA), they aim to compress input data into a lower-dimensional latent representation that minimizes the reconstruction error. However, whereas PCA minimizes reconstruction error using linear projections onto orthogonal principal components, autoencoders directly minimize the reconstruction error without imposing linearity or orthogonality constraints, allowing them to learn more flexible and potentially more powerful nonlinear mappings.
Their ability to learn nonlinear embeddings has made them a powerful tool for data compression, feature extraction, and anomaly detection. In finance, autoencoders have been applied to problems such as factor discovery and risk modeling \citep[see, e.g.,][]{gu2021autoencoder}.

Formally, an autoencoder consists of two components: an encoder and a decoder. Given an input vector $\mathbf{x} \in \mathbb{R}^d$, the encoder maps it to a latent representation $\mathbf{z} \in \mathbb{R}^J$ via a learned transformation:
\begin{equation}
    \mathbf{z} = f(\mathbf{W}_e \mathbf{x} + \mathbf{b}_e),
\end{equation}
where $\mathbf{W}_e \in \mathbb{R}^{J \times d}$ is the encoder weight matrix, $\mathbf{b}_e \in \mathbb{R}^J$ is a bias vector, and $f(\cdot)$ is a nonlinear activation function.\footnote{While $J$ is typically smaller than $d$ in standard autoencoders, in the context of LLM interpretability, it is common to set $J > d$ to allow for overcomplete, disentangled representations and lower reconstruction error.} For simplicity, we present here the case of a single-layer encoding and decoding model. However, as with any neural network, these layers can be stacked on top of one another.

The decoder then attempts to reconstruct the original input from $\mathbf{z}$:
\begin{equation}
    \hat{\mathbf{x}} = g(\mathbf{W}_d \mathbf{z} + \mathbf{b}_d),
\end{equation}
where $\mathbf{W}_d \in \mathbb{R}^{d \times J}$ and $\mathbf{b}_d \in \mathbb{R}^d$ are decoder parameters, and $g(\cdot)$ is usually chosen to be the identity function or a sigmoid, depending on the domain of the input data \citep{goodfellow2016deep}.

The encoder and decoder are trained jointly to minimize the reconstruction error between the input $\mathbf{x}$ and its reconstruction $\hat{\mathbf{x}}$:

\begin{equation}
    \mathcal{L}_{\text{AE}}(\mathbf{x}) = \underbrace{\|\mathbf{x} - \hat{\mathbf{x}}\|_2^2}_{\text{Reconstruction loss}}.
\end{equation}

\medskip
Autoencoders are powerful because they can compress data into a lower-dimensional representation, addressing one of the two sources of opacity we discussed earlier: the sheer size and complexity of the transformation blocks. However, they do not address the other major challenge: in order to be interpretable, this new, smaller representation has to be sparse. This is where \textit{Sparse Autoencoders (SAEs)} can play a role. A SAE is a variant of the standard autoencoder that adds a sparsity-inducing penalty, typically an $\ell_1$ or $\ell_0$ norm on the latent representation $\mathbf{z}$. The aim is to encourage most dimensions of $\mathbf{z}$ to be inactive (conceptually zero for any given input), thereby promoting interpretability and disentanglement in the learned features:
\begin{equation}
    \mathcal{L}_{\text{SAE}}(\mathbf{x}) =
    \underbrace{\|\mathbf{x} - \hat{\mathbf{x}}\|_2^2}_{\text{Reconstruction}} +
    \underbrace{\lambda \|\mathbf{z}\|_0}_{\text{Sparsity penalty}},
\end{equation}
where $\lambda > 0$ controls the strength of the sparsity constraint.

This sparse structure is particularly attractive for interpretability: if only a small number of latent features are active for a given residual stream $\mathbf{x}$, then those features can be more easily inspected and understood in isolation.

\subsubsection{Training Sparse Autoencoders within a Large Language Model}
\label{app:sae_details_sae_training}

The concept of training Sparse Autoencoders (SAEs) within a Large Language Model (LLM) to enhance interpretability was introduced by \citet{cunningham2023sparse}.

In this and the following section, we describe how SAEs can be used to uncover semantically meaningful structures in the internal representations of LLMs. As illustrated in \autoref{fig:sae_appendix}, the objective is to construct an augmented model pipeline in which, at each layer $l$, the residual stream $\mathbf{r}^{(l)}$ is mapped to a sparse code $\mathbf{z}^{(l)}$. The individual dimensions of $\mathbf{z}^{(l)}$ are designed to correspond to interpretable concepts or functions.

To achieve this, SAEs are trained on the residual streams produced at a fixed layer $l$ in the language model. More precisely, an input text is run through the LLM up to layer $l$. The corresponding residual stream vector (i.e., the current internal representation of the input at a given layer in the model's processing) $\mathbf{r}^{(l)} \in \mathbb{R}^d$ is then recorded. Repeating this process across a large corpus gives a dataset of residual streams:
\[
\mathcal{D} = \left\{ \mathbf{r}_i^{(l)} \right\}_{i=1}^N,
\]
where each $\mathbf{r}_i^{(l)}$ is the residual stream at layer $l$ for the $i$-th text input.

A sparse autoencoder is then trained to reconstruct each $\mathbf{r}_i^{(l)}$ through a sparse latent representation:
\begin{align}
    \mathbf{z}^{(l)}_i &= f(\mathbf{W}_e \mathbf{r}^{(l)}_i + \mathbf{b}_e), \\
    \hat{\mathbf{r}}^{(l)}_i &= g(\mathbf{W}_d \mathbf{z}^{(l)}_i + \mathbf{b}_d),
\end{align}
where $f(\cdot)$ is a nonlinear activation function, and $g(\cdot)$ is usually the identity.

The loss function encourages both accurate reconstruction and sparsity in the hidden representation:
\begin{equation}
    \mathcal{L}(\mathbf{r}^{(l)}_i) = \underbrace{\|\mathbf{r}^{(l)}_i - \hat{\mathbf{r}}^{(l)}_i\|_2^2}_{\text{Reconstruction loss}} + \lambda \underbrace{\|\mathbf{z}^{(l)}_i\|_0}_{\text{Sparsity penalty}},
\end{equation}
with $\lambda > 0$ controlling the trade-off between fidelity and interpretability.

Since the autoencoder is a neural network without a closed-form solution, it is trained by minimizing the expected loss across the generated dataset of residual streams.\footnote{These residual streams are obtained from forward passes of the pretrained language model over text corpora, such as books, news articles, and web pages, similar to those used in LLM training but typically drawn from a smaller, curated subset.}

\begin{equation}
    \min_{\Theta} \; \mathbb{E}_{\mathbf{r} \sim \mathcal{D}}[\mathcal{L}(\mathbf{r})],
\end{equation}

 One core advantage of the SAEs is that it does not require retraining the LLM itself. Instead, the SAE is trained independently on the residual stream of a pre-trained model. This makes the method computationally efficient and accessible, requiring only a fraction of the data and computational resources typically needed to train or fine-tune a full-scale LLM.

In practice, the SAE can be viewed as an interpretable ``plug-in'' that augments the internal representations of a transformer model without altering its original parameters. This separation is particularly advantageous in settings where full retraining is impractical or undesirable, such as applied research in finance. Moreover, the increasing availability of open-source models and pretrained SAEs makes this approach highly suitable for academic and applied analysis.

Throughout this paper, we utilize the open-source Sparse Autoencoders released by Google DeepMind \citep{lieberum2024gemmascopeopenspars}, trained on intermediate representations from the instruction-tuned language model Gemma-2-9B-IT. These models provide an ideal testbed for our study, combining open accessibility with strong performance on language understanding tasks.

\begin{figure}[t]
  \centering

\begin{tikzpicture}[>=Latex, node distance=10mm, font=\small]

\tikzset{
  tblock/.style={draw, rounded corners, thick, align=center, inner sep=6pt,
                 minimum width=40mm, minimum height=12mm, fill=gray!5},
  nodebox/.style={draw, rounded corners, thick, align=center, inner sep=6pt, fill=blue!5},
  saebox/.style={draw, rounded corners, thick, align=center, inner sep=6pt, fill=green!8,
                 minimum width=36mm, minimum height=10mm},
  tap/.style={circle, fill=black, minimum size=3pt, inner sep=0pt},
  arrow/.style={-Latex, thick}
}

\node[nodebox] (tokens) {tokens};
\node[nodebox, below=8mm of tokens] (embed) {embed};

\node[tblock, below=10mm of embed] (blk1) {Transformer block $l$};
\node[tblock, below=40mm of blk1] (blk2) {Transformer block $l{+}1$};

\node[nodebox, below=12mm of blk2] (outhead) {output head};
\node[nodebox, below=8mm of outhead, align=center] (probs) {$p(\text{next token}\mid \cdot)$};

\draw[arrow] (tokens) -- (embed) -- (blk1);
\draw[arrow] (blk1.south) -- node[pos=0.5, anchor=east, xshift=-2mm] {$\mathbf{r}^{(l)}\mapsto\hat{\mathbf{r}}^{(l)}$} (blk2.north);
\draw[arrow] (blk2) -- (outhead) -- (probs);

\coordinate (tappt) at ($(blk1.south)!0.5!(blk2.north)$);
\node[tap] (tapdot) at (tappt) {};

\coordinate (saeC) at ($(blk1.east)!0.5!(blk2.east)$);

\matrix[anchor=west, at={(saeC)}, xshift=50mm, row sep=9mm, column sep=0mm] (SAE)
{
  \node[saebox] (enc)  {encode\\$W_e$}; \\
  \node[saebox] (z)    {$\mathbf{z}^{(l)}$\\\footnotesize sparse features}; \\
  \node[saebox] (dec)  {decode\\$W_d$}; \\
  \node[saebox] (rhat) {$\hat{\mathbf{r}}^{(l)}$}; \\
};

\draw[arrow] (enc) -- (z) -- (dec) -- (rhat);

\node[draw, dashed, gray!60, rounded corners, inner sep=5mm, fit=(enc)(z)(dec)(rhat)] (sae) {};
\node[anchor=south west, font=\footnotesize\bfseries, text=gray!60]
  at ($(sae.north west)+(1mm,1mm)$) {Sparse autoencoder (SAE)};

\draw[arrow] (tapdot) -- node[above, sloped, pos=0.55] {\footnotesize $\rightarrow$ to SAE}
              (enc.west);

\draw[arrow] (rhat.west) -- node[below, sloped, pos=0.45] {\footnotesize $\leftarrow$ from SAE}
              (tapdot);

\end{tikzpicture}
\space
  \caption{\textbf{Illustration of SAE's integration with a Transformer model.}\\
  \footnotesize{
  The residual stream $\mathbf{r}^{(l)}$ is extracted from an intermediate layer of the language model and projected onto a sparse, interpretable representation $\mathbf{z}^{(l)}$. This sparse code is then decoded to reconstruct $\hat{\mathbf{r}}^{(l)}$, which is fed back into the model for subsequent processing.}}
  \label{fig:sae_appendix}
\end{figure}

\FloatBarrier

\subsection{Feature Labeling and Selection}
\label{app:sae_details_implementation}

While SAEs provide a sparse and interpretable representation of the model’s internal states, the features themselves do not come with human-interpretable meanings. The most common labeling approach is to collect the tokens or text passages that most strongly activate each feature and ask a language model to summarize the shared theme \citep[e.g.,][]{cunningham2023sparse}.
Publicly available resources such as Neuronpedia provide large collections of such feature annotations derived from generic corpora.\footnote{See https://www.neuronpedia.org/} However, because these labels are inferred from general-purpose text, many features correspond to concepts unrelated to finance or economics, ranging from cooking recipes to sports events, making it difficult to directly connect the feature space to economic meanings.

We instead label features from the other direction: we define the concepts we care about upfront, and find which features respond to them. Concretely, we specify $K = 46$ economic topics (e.g., \textit{Leverage}, \textit{Monetary Policy}, \textit{Risk Aversion}), grouped into five main themes: \textit{Macro-Finance \& Economics}, \textit{Financial Markets \& Instruments}, \textit{Corporate Finance}, \textit{Investment \& Portfolio Management}, and \textit{Risk Management \& Derivatives}.
For each topic $k$, we use GPT to generate a set of representative example texts and pass them through the SAE-augmented LLM, recording the mean activation of each feature. Prompt~\ref{prompt:financial_journalism_phrase_generation} reports the exact instruction used to generate them.

\begin{Prompt}[h]
    \begin{tcolorbox}
        \begin{lstlisting}
Act as a financial journalist.

Generate exactly {phrases_per_run} short examples that a financial journalist might use when writing about "{small_topic}" under the broader topic of "{big_category}".

Constraints:
- Each example must be a short phrase, clause, or sentence fragment.
        \end{lstlisting}
    \end{tcolorbox}
    \caption{\textbf{Financial Journalism Phrase Generation Prompt} \\
    {This prompt asks the model to act as a financial journalist and generate a fixed number of short examples related to a specified small topic within a broader financial category. The examples are constrained to be brief phrases, clauses, or sentence fragments.}}
    \label{prompt:financial_journalism_phrase_generation}
\end{Prompt}
With $J=131{,}072$ SAE features, this produces an activation matrix $A \in \mathbb{R}^{K \times J}$, where $A_{k,j}$ is the average SAE activation of feature $j$ across all example texts for topic $k$.

To measure how \emph{exclusively} feature $j$ responds to topic $k$, as opposed to firing uniformly across all topics, we center each feature's activation profile across topics and clip negatives:
\begin{equation}
    \widetilde{A}_{k,j} \;=\; \max\!\left(0,\; A_{k,j} - \tfrac{1}{K}\sum_{k'}A_{k',j}\right).
    \label{eq:centering}
\end{equation}
Normalizing across topics gives a \emph{topic-specificity weight} for each feature-topic pair:\footnote{We omit features with the same average activation for every topic.}
\begin{equation}
    \pi_{k,j} \;=\; \frac{\widetilde{A}_{k,j}}{\sum_{k'} \widetilde{A}_{k',j}},
    \qquad \sum_{k} \pi_{k,j} = 1,
    \label{eq:probmap}
\end{equation}
Here, $\pi_{k,j}$ measures the share of feature $j$'s total above-average activation assigned to topic $k$. A topic receives positive weight only when the feature activates above its mean level across topics.

We then assign each feature to the topic for which it is both strongly active and highly specific, via the combined score
\begin{equation}
    s_{k,j} \;=\; A_{k,j} \cdot \pi_{k,j},
    \label{eq:score}
\end{equation}
which rewards features that are both strongly active and concentrated on topic $k$. Each feature is hard-assigned to its highest-scoring topic:
\begin{equation}
    \hat{k}(j) \;=\; \argmax_{k}\; s_{k,j}.
    \label{eq:assignment}
\end{equation}

To quantify the overall concentration of a feature across topics, we compute the normalized entropy of $\pi_{\cdot,j}$:
\begin{equation}
    H_j \;=\; -\frac{1}{\log K} \sum_{k=1}^{K} \pi_{k,j} \log \pi_{k,j} \;\in\; [0,1],
    \label{eq:entropy}
\end{equation}
Entropy is zero when all weight falls on one topic and increases as the weight spreads across more topics.

\paragraph{Feature selection.}
Not all features assigned to a topic are equally informative.
For each topic $k$, we form a candidate pool of features that are exclusively assigned to it and have sufficiently low entropy ($H_j < \tau_H = 0.6$):\footnote{This corresponds to a feature distributing its mass over roughly $\exp(H_j \log K)$ topics; with $K=46$ and $H_j=0.6$, this is on the order of $\sim$10 effective topics.}
\begin{equation}
    \mathcal{C}_k \;=\; \bigl\{j : \hat{k}(j) = k \;\text{ and }\; H_j < \tau_H\bigr\}.
\end{equation}
Within $\mathcal{C}_k$, features are ranked by score $s_{k,j}$ in descending order, and we greedily include features until their cumulative score mass reaches a fraction $\tau_c = 0.8$ of the total scores, and $\mathcal{F}_k$ will be the final set of features we assign to topic $k$:
\begin{equation}
    \mathcal{F}_k \;=\; \text{top-}n_k^*\text{ features in }\mathcal{C}_k,
    \quad n_k^* = \min\biggl\{n : \frac{\sum_{r=1}^{n} s_{k,(r)}}{\sum_{j \in \mathcal{C}_k} s_{k,j}} \geq \tau_c \biggr\},
    \label{eq:nucleus}
\end{equation}
where $s_{k,(r)}$ is the $r$-th largest score in $\mathcal{C}_k$.
This selection criterion is self-calibrating: topics dominated by a single feature select a small set, while diffuse topics select more, always capturing a consistent fraction of the topic's total signal. At the same time, by truncating the tail of low-scoring features, it excludes those with negligible contribution, effectively filtering out weak, noisy activations that would otherwise dilute the semantic coherence of the topic.

For the brain-scan topic scores, we normalize the labeling scores over the selected features:
\begin{equation}
    w_{kj}=\frac{s_{k,j}}{\sum_{h\in\mathcal F_k}s_{k,h}},
    \qquad j\in\mathcal F_k.
    \label{eq:topic_aggregation_weights}
\end{equation}

\FloatBarrier

\subsection{Construction of the Visual Brain Scan}
\label{appendix:brain_visualization}

This appendix describes the feature map and same-article activation panels in Figure~\ref{fig:brain_visual_comparison}. The geometry is fixed across the two activation panels. Only the activation overlay changes. The visualization is illustrative and does not enter the forecast-attribution calculations in Section~\ref{sec:forecast_attribution}.

\paragraph{Feature dictionary.} The supplied map uses the Gemma Scope residual-stream SAE for Gemma-2-9B-IT at layer 20, with 131,072 features and the checkpoint identified as \texttt{average\_l0\_81}. Each feature has a 3,584-dimensional decoder direction. All three panels use the same 7,278 economically relevant features assigned to 46 topics in five themes, as identified by our labeling and selection procedure. The reference map and activation overlays use identical feature coordinates.

\paragraph{Two-dimensional layout.} We normalize each selected decoder direction to unit length and measure distances using cosine distance. UMAP constructs a neighborhood graph and projects it into two dimensions, seeking to preserve local relationships. The supplied implementation uses 100 neighbors, a minimum distance of 0.25, a categorical target weight of 0.3, and random seed 0. Topic labels enter as supervised targets. In this implementation, links across different topic labels are multiplied by $\exp[-2.5/(1-0.3)]\simeq0.028$. Thus, the spatial grouping reflects both decoder similarity and the supplied taxonomy. The axes have no economic units, and neither apparent separation nor map shape is an independent test of semantic coherence.

\paragraph{Reference map.} For each theme, the plotting code places the feature locations on a $300\times300$ grid, smooths the counts with a Gaussian filter of standard deviation four grid cells, normalizes by that theme's maximum density, and applies a square-root transformation. It overlays feature dots and colors the resulting density by theme. This reference map uses no article activations. The colored glow therefore describes the distribution of features rather than a neutral or resting activation state.

\paragraph{Same-article activation overlays.} Figure~\ref{fig:brain_visual_comparison} uses the article ``BNY Mellon Loses Key ETF Client'', published in the \textit{Wall Street Journal} on 23 August 2019. The news box reproduces selected passages with omissions marked. The two panels use this same article under prompts asking about the S\&P~500 index and employment. The displayed values are SAE activations, without forecast gradients.

The display removes two sources of common activity. Let $a_{jv}$ denote the example article's activation of feature $j$ under forecast target $v$, and let $\mu_{jv}^{2019}$ be that feature's mean activation for the same target across all 2019 articles. Define
\begin{align}
 D_{jv}&=a_{jv}-\mu_{jv}^{2019},\\
 e_{jv}&=\max\left\{D_{jv}-\frac{1}{14}\sum_{u=1}^{14}D_{ju},\,0\right\}.
\end{align}
The first subtraction removes activity typical of the forecast target. The second removes the article's mean adjusted activation across its fourteen target prompts. Clipping at zero highlights features whose adjusted activation exceeds that article-wide baseline. Thus the displayed glow emphasizes the additional activity associated with the question, rather than raw activation or activity common to financial news.

We sum $e_{jv}$ across the features assigned to each topic and theme. The displayed shares divide these sums by total excess activation across the plotted features. They are activation shares, not signed forecast-attribution shares. In the S\&P~500 panel, Investment \& Portfolio Management and Financial Markets \& Instruments account for 26\% and 22\% of excess activation. Passive Investing accounts for 12\%, Equity Markets for 7\%, and Asset Allocation for 5\%. In the employment panel, Corporate Finance and Macro-Finance \& Economics account for 34\% and 21\%, with Human Resources at 15\%, Unemployment at 5\%, and Mergers \& Acquisitions at 5\%.

\paragraph{Rendering.} For the full-map overlay, the plotting code forms a $300\times300$ histogram weighted by $e_{jv}$, applies a Gaussian filter with standard deviation four grid cells, normalizes by the panel's maximum, and takes the square root. Low-intensity cells are masked. The two themes with the largest shares receive enlarged views, with letters marking their three leading topics. Each zoom is an exact enlargement of its marked map region, retaining the glow from all features and the parent map's brightness scale. Brightness is nonlinear in activation and cannot be compared as an absolute activation level across forecast panels. The printed shares, rather than pixel brightness, quantify relative excess activation.

\paragraph{Interpretation.} This single-article comparison illustrates how the model's activity varies across forecasting questions. Its feature positions are fixed and its article text is identical across panels. The example does not establish how representative the displayed contrast is across the news sample, nor does activation alone identify a feature's effect on the answer. Section~\ref{sec:forecast_attribution} measures this effect using the model's answer probabilities and feature gradients.

\FloatBarrier

\subsection{Concept Steering}\label{app:sae_details_steering}

The labeling approach above assigns each feature to an economic topic based on which example texts activate it.
A natural question is whether these assignments are meaningful: does steering the model with a feature labeled \textit{risk-aversion} actually cause it to be more risk-averse? Concept steering provides a direct answer.

At a high level, steering can be understood as follows:
\begin{enumerate}
    \item Identify a relevant feature in the sparse representation to steer the LLM toward. For instance, selecting one associated with risk to promote risk-averse behavior.
    \item Use the SAE's decoder to identify the perturbation induced by activating this feature; specifically, what would have been \textit{added} to the residual stream if the feature were more strongly activated.
    \item Add this perturbation to the residual stream to obtain a steered probability distribution over the next tokens.
    \item Iteratively sample tokens from this steered distribution to generate text aligned with the intended concept (e.g., more risk-averse responses).
\end{enumerate}

At the intervention layer $l$, we increase the activation of feature $j$ by a steering amount $\alpha$. Because the SAE decoder is linear, this increment corresponds to a residual-stream perturbation $\alpha\mathbf{d}_j$, where $\mathbf{d}_j$ is column $j$ of the decoder matrix $\mathbf{W}_d$. We add this perturbation directly to the original residual stream:
\begin{equation}
    \mathbf{r}_{\mathrm{steer}}^{(l)}
    = \mathbf{r}^{(l)} + \alpha\mathbf{d}_j.
    \label{eq:feature_steering}
\end{equation}
We pass the modified state through the remaining LLM layers to obtain the steered next-token probabilities. The decoder bias does not enter the perturbation because it cancels when taking the difference between the modified and original reconstructions. Setting $\alpha=0$ leaves the original residual stream unchanged. All LLM and SAE weights remain fixed during steering.

We show an example of steering to validate our labeling approach in Section~\ref{sec:validating_labels}.

This approach provides a fine-grained, concept-level control mechanism grounded in interpretable internal features. Unlike prompt engineering, which relies on heuristic text manipulations and often lacks transparency, steering via sparse features operates directly on the model's latent representations. This control enables systematic and reproducible interventions without altering the model's parameters or retraining.

The perturbation is added at every token position of the prompt.

\paragraph{Calibrating the steering strength.} The effect of a given $\alpha$ depends on the
feature and the decision being studied, so the raw coefficient has no common scale across
applications. We therefore choose $\alpha$ according to the design of each exercise. When we
study a single decision between two options, we express steering strength in terms of the
model's choice probability. Let $p_\alpha$ denote the probability of the option targeted by the
intervention when the steering amount is $\alpha$, renormalized over the two answer tokens, and
let
\begin{equation}
    L(\alpha)=\log\frac{p_\alpha}{1-p_\alpha}
    \label{eq:steering_logodds}
\end{equation}
denote the corresponding log-odds. We obtain the baseline $L(0)$ from the unsteered prompt and
the gradient $g=\partial L(\alpha)/\partial\alpha\big|_{\alpha=0}$ by backpropagation. A
steering strength of $t\%$ targets a move of $t\%$ of the distance from the baseline probability
$p_0$ to one (for $t\geq 0$) or to zero (for $t<0$):
\begin{equation}
    \pi_t=
    \begin{cases}
        p_0+\frac{t}{100}\,(1-p_0), & t\geq 0,\\[4pt]
        p_0\left(1+\frac{t}{100}\right), & t<0.
    \end{cases}
    \label{eq:steering_target}
\end{equation}
We then set the steering amount using a first-order approximation of the log-odds around
$\alpha=0$:
\begin{equation}
    \alpha_t=\frac{\operatorname{logit}(\pi_t)-L(0)}{g},
    \label{eq:steering_calibration}
\end{equation}

Equal steps in $t$ therefore correspond to equal steps in the targeted probability rather than
equal steps in $\alpha$. Because the approximation is local, realized probabilities need not
equal their targets, and we always report the probabilities realized under steering. 

The gradient can be computed on the prompt being studied or borrowed from a representative
prompt for the same decision, combined with the new prompt's own baseline. Alternatively, the
coefficients $\alpha_t$ from a representative prompt can be applied directly, which is
appropriate when the behavior under study should not respond to the intervention, as in our
deterministic-choice exercises.

This calibration is not suited to the return-prediction exercise in
Section~\ref{sec:steering_validation}, which applies steering across many alerts. Applied alert
by alert, it would move every alert toward a common probability, compressing the
cross-sectional differences in predicted probabilities on which the portfolios are sorted. We
instead add the same perturbation to every alert. Because the appropriate $\alpha$ is not
obvious ex ante, we evaluate a grid of values, $\alpha\in\{-100,-50,-30,0,30,50,100\}$, and
report results at each value. Where we report the steering strength as a raw coefficient rather
than a percentage, it is $\alpha$ itself.

\FloatBarrier

\subsection{Reuters News Sample}
\label{appendix:reuters_sample}

This appendix provides summary statistics for the Reuters news sample used in the return-prediction analyses in Sections~\ref{sec:steering_validation} and \ref{sec:data_reuters}. The sample contains overnight Reuters alerts for U.S.-listed firms from 2015 through 2024, matched to intraday returns from CRSP. \autoref{tab:summary_stats} reports, by year, the number of covered firms, the number of alerts, the share of positive intraday returns, and the average alert length.

\begin{table}[H]
\centering
\caption{\textbf{Reuters News Sample} \\
\footnotesize{This table reports annual summary statistics for the matched Reuters and CRSP sample from 2015 through 2024. Overnight alerts are Reuters alerts published after 4:00 p.m. or before 9:00 a.m. For each year, the table reports the number of unique firms, the number of overnight alerts, the percentage of observations with positive intraday returns, and the average alert length in characters.}}
\label{tab:summary_stats}
\resizebox{0.8\textwidth}{!}{%
  \begin{tabular}{lcccc}
\toprule
Year & Number of Firms & Number of News & \% Return $>$ 0 & News length \\
\midrule
2015 & 3,610 & 209,447 & 49.2\% & 92 \\
2016 & 3,481 & 203,412 & 50.1\% & 90 \\
2017 & 3,468 & 203,972 & 48.4\% & 91 \\
2018 & 3,455 & 199,420 & 47.9\% & 87 \\
2019 & 3,460 & 215,949 & 51.1\% & 88 \\
2020 & 3,550 & 225,337 & 46.5\% & 88 \\
2021 & 4,061 & 216,070 & 47.1\% & 86 \\
2022 & 4,115 & 204,790 & 49.2\% & 85 \\
2023 & 3,999 & 225,977 & 49.2\% & 81 \\
2024 & 3,738 & 203,746 & 48.5\% & 77 \\
\midrule
Total & 6,070 & 2,108,120 & 48.7\% & 86 \\
\bottomrule
\end{tabular}\hspace{\fontdimen2\font}
}
\end{table}

\FloatBarrier
\clearpage
\section{Additional Results}\suppressfloats[t]

\subsection{Topic Word Clouds}
\label{appendix:wordclouds}

We steer each of the top three features assigned to a topic, one at a time, by a large
amount ($\alpha=200$) and ask the model to complete Prompt~\ref{prompt:wordcloud},
holding the prompt fixed across topics. Stronger steering makes the topic more visible
in the output, but at extreme values the model stops producing coherent text, so we use
the largest value that still yields coherent output. \autoref{fig:cloud_1} collects the four panels. The generated vocabulary matches the assigned label in nearly every topic, which confirms that the labeling pipeline recovers the intended economic concepts.

\begin{figure}[!htbp]
\begin{center}
\singlespacing
\setlength{\tabcolsep}{3pt}
\begin{tabular}{@{}ccc@{}}

\footnotesize Alpha & \footnotesize Arbitrage & \footnotesize Asset Allocation \\
\includegraphics[width=0.285\textwidth]{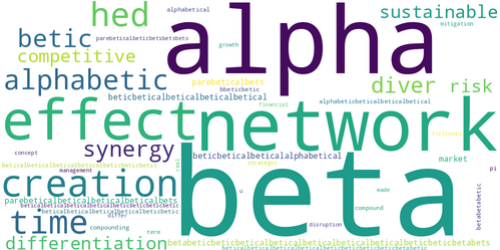} &
\includegraphics[width=0.285\textwidth]{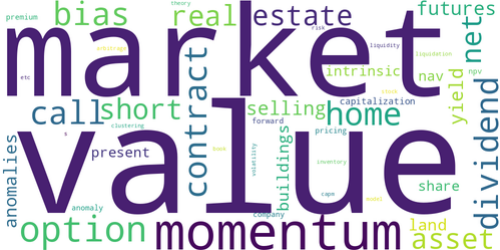} &
\includegraphics[width=0.285\textwidth]{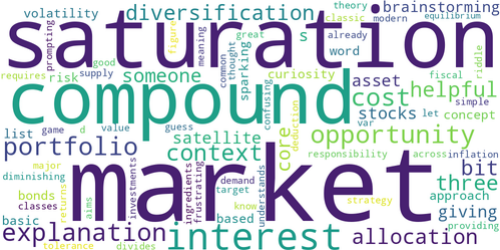} \\[4pt]

\footnotesize Balance of Payments & \footnotesize Beta & \footnotesize CAPEX \\
\includegraphics[width=0.285\textwidth]{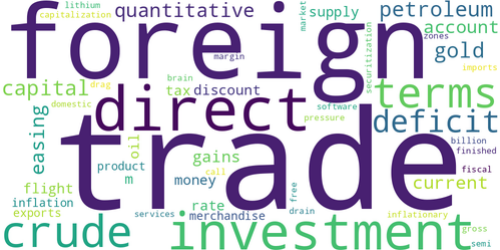} &
\includegraphics[width=0.285\textwidth]{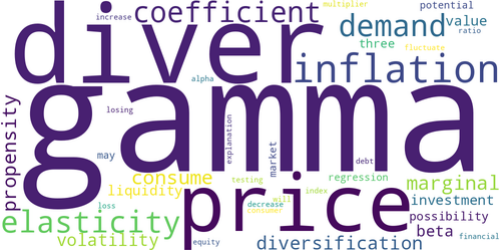} &
\includegraphics[width=0.285\textwidth]{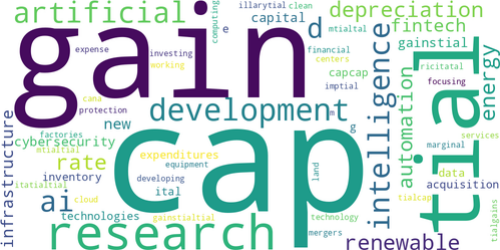} \\[4pt]

\footnotesize Capital Budgeting & \footnotesize Commodities & \footnotesize Cost of Capital \\
\includegraphics[width=0.285\textwidth]{fig/wordclouds/capital_budgeting.pdf} &
\includegraphics[width=0.285\textwidth]{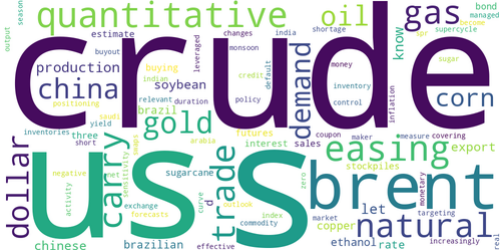} &
\includegraphics[width=0.285\textwidth]{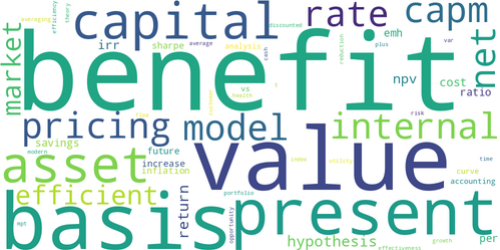} \\[4pt]

\footnotesize Counterparty Risk & \footnotesize Credit Default Swaps (CDS) & \footnotesize Derivatives \\
\includegraphics[width=0.285\textwidth]{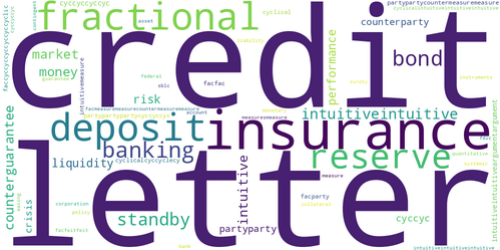} &
\includegraphics[width=0.285\textwidth]{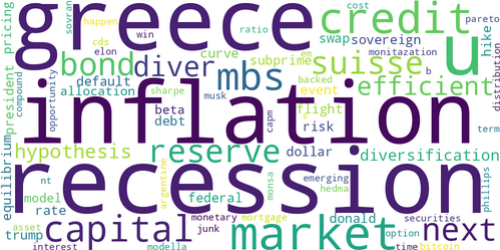} &
\includegraphics[width=0.285\textwidth]{fig/wordclouds/derivatives.pdf} \\[4pt]

\footnotesize Dividend Policy & \footnotesize Equity Markets & \footnotesize Financial Distress \\
\includegraphics[width=0.285\textwidth]{fig/wordclouds/dividend_policy.pdf} &
\includegraphics[width=0.285\textwidth]{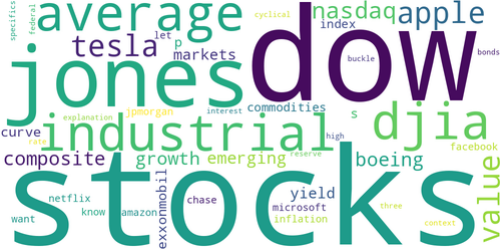} &
\includegraphics[width=0.285\textwidth]{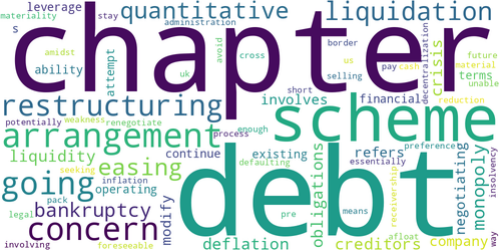} \\[4pt]

\footnotesize Fiscal Deficit & \footnotesize Fixed Income & \footnotesize Foreign Exchange (FX) \\
\includegraphics[width=0.285\textwidth]{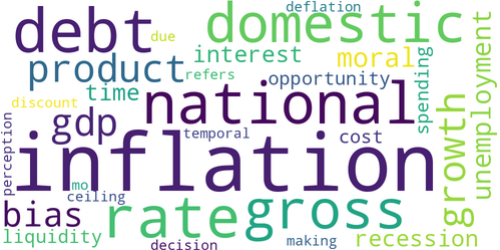} &
\includegraphics[width=0.285\textwidth]{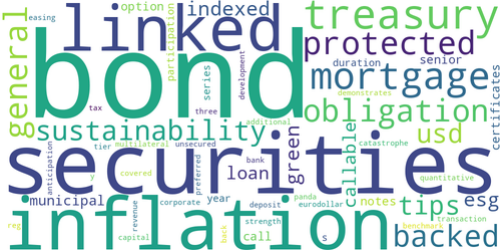} &
\includegraphics[width=0.285\textwidth]{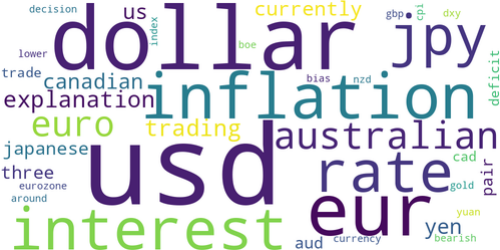} \\[4pt]

\end{tabular}
\end{center}
\caption{\textbf{Word clouds for steered SAE topics.}}
\label{fig:cloud_1}
\vspace{4pt}
\noindent\begin{minipage}[c]{1\textwidth}%
\singlespacing
\textbf{\footnotesize Note.} \footnotesize
Each panel shows the most frequent words generated by the LLM when its residual stream is steered along the top features assigned to the named topic, in response to Prompt~\ref{prompt:wordcloud}. Font size is proportional to frequency.
\medskip
\end{minipage}
\end{figure}

\begin{figure}[!htbp]\ContinuedFloat
\begin{center}
\singlespacing
\setlength{\tabcolsep}{3pt}
\begin{tabular}{@{}ccc@{}}

\footnotesize Governance & \footnotesize Gross Domestic Product (GDP) & \footnotesize Hedging \\
\includegraphics[width=0.32\textwidth]{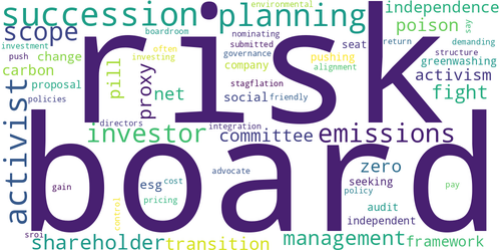} &
\includegraphics[width=0.32\textwidth]{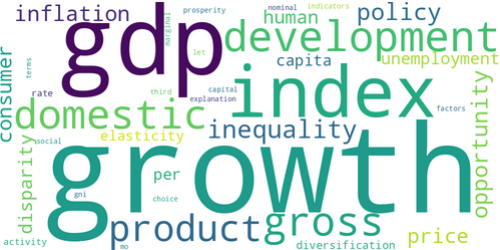} &
\includegraphics[width=0.32\textwidth]{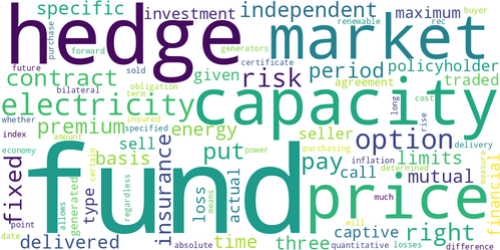} \\[4pt]

\footnotesize High-Frequency Trading (HFT) & \footnotesize Human resources & \footnotesize Inflation \\
\includegraphics[width=0.32\textwidth]{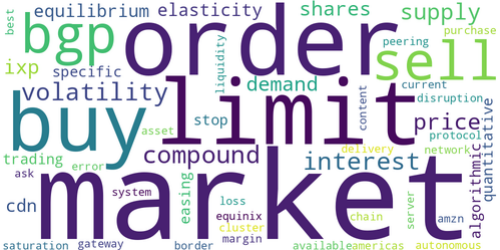} &
\includegraphics[width=0.32\textwidth]{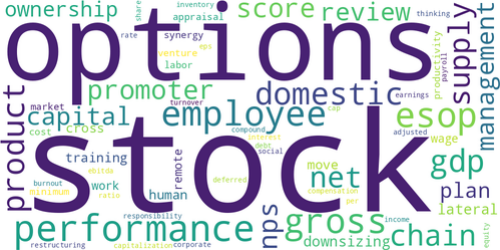} &
\includegraphics[width=0.32\textwidth]{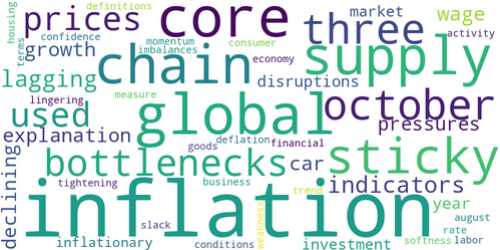} \\[4pt]

\footnotesize Initial Public Offering (IPO) & \footnotesize Inventory & \footnotesize Investments \\
\includegraphics[width=0.32\textwidth]{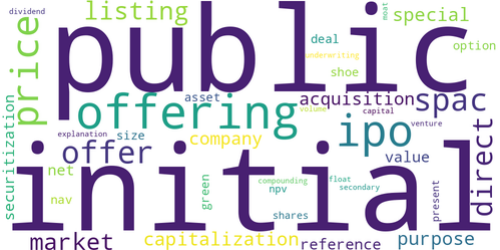} &
\includegraphics[width=0.32\textwidth]{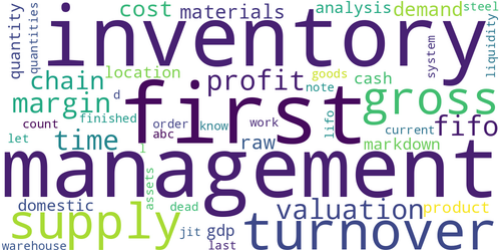} &
\includegraphics[width=0.32\textwidth]{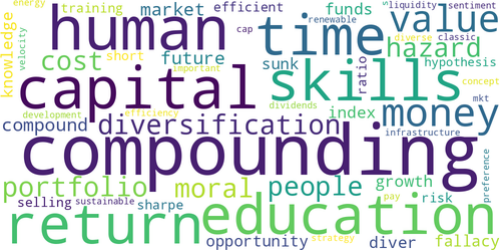} \\[4pt]

\footnotesize Leverage & \footnotesize Liquidity & \footnotesize Mergers \& Acquisitions (M\&A) \\
\includegraphics[width=0.32\textwidth]{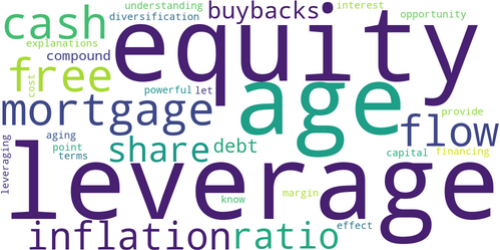} &
\includegraphics[width=0.32\textwidth]{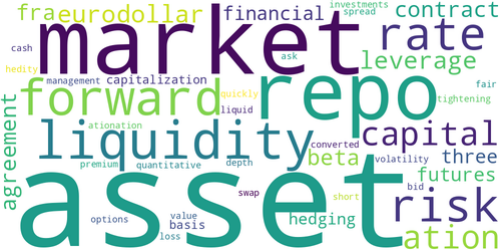} &
\includegraphics[width=0.32\textwidth]{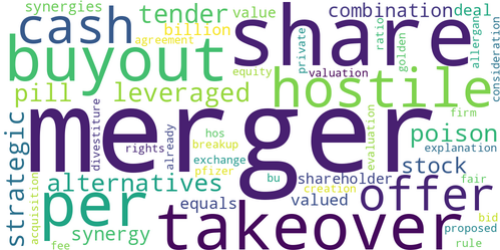} \\[4pt]

\footnotesize Monetary Policy & \footnotesize Operational Risk & \footnotesize Passive investing \\
\includegraphics[width=0.32\textwidth]{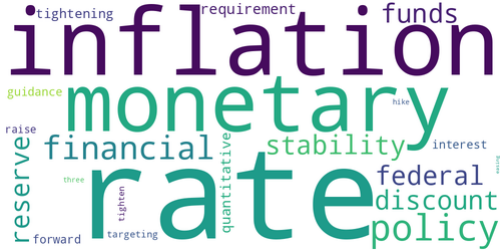} &
\includegraphics[width=0.32\textwidth]{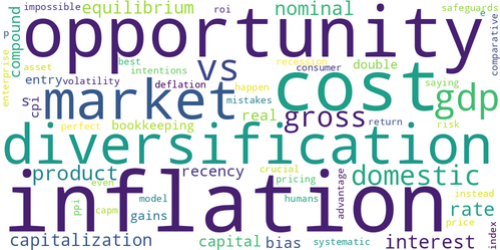} &
\includegraphics[width=0.32\textwidth]{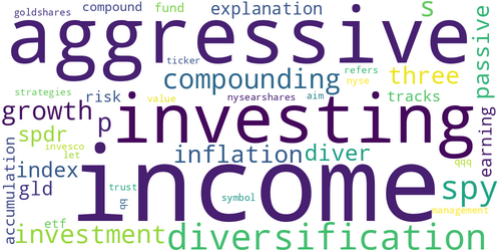} \\[4pt]

\footnotesize Quantitative Easing (QE) & \footnotesize Regulation & \footnotesize Revenues growth \\
\includegraphics[width=0.32\textwidth]{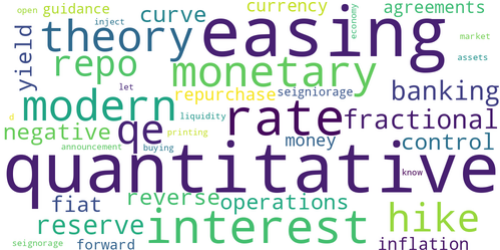} &
\includegraphics[width=0.32\textwidth]{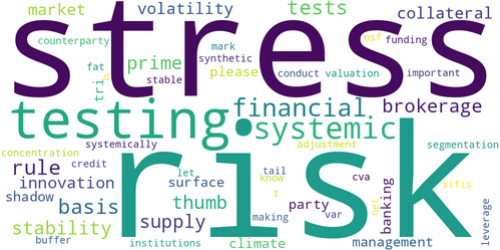} &
\includegraphics[width=0.32\textwidth]{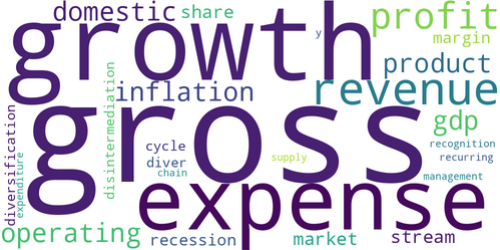} \\[4pt]

\end{tabular}
\end{center}
\caption{\textbf{Word clouds for steered SAE topics, continued.}}
\label{fig:cloud_2}
\end{figure}

\begin{figure}[!htbp]\ContinuedFloat
\begin{center}
\singlespacing
\setlength{\tabcolsep}{3pt}
\begin{tabular}{@{}ccc@{}}

\footnotesize Risk Aversion & \footnotesize Sharpe Ratio & \footnotesize Short Selling \\
\includegraphics[width=0.32\textwidth]{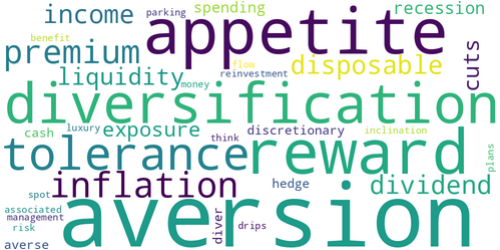} &
\includegraphics[width=0.32\textwidth]{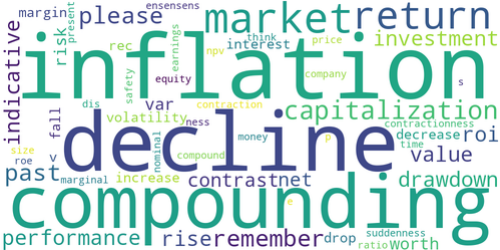} &
\includegraphics[width=0.32\textwidth]{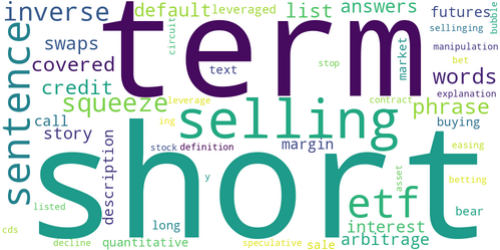} \\[4pt]

\footnotesize Stress Testing & \footnotesize Systemic Risk & \footnotesize Tariffs \\
\includegraphics[width=0.32\textwidth]{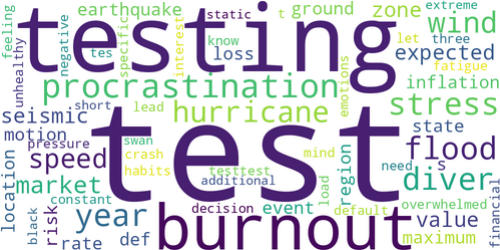} &
\includegraphics[width=0.32\textwidth]{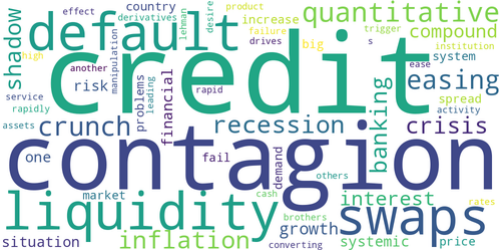} &
\includegraphics[width=0.32\textwidth]{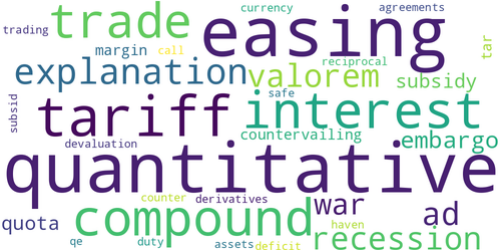} \\[4pt]

\footnotesize Unemployment & \footnotesize Value at Risk (VaR) & \footnotesize Working Capital \\
\includegraphics[width=0.32\textwidth]{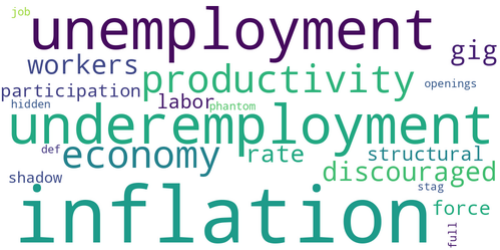} &
\includegraphics[width=0.32\textwidth]{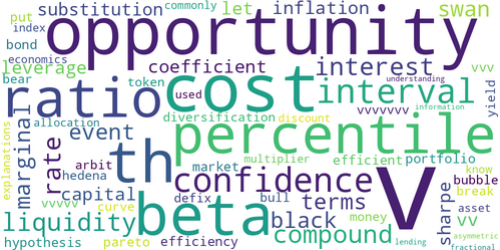} &
\includegraphics[width=0.32\textwidth]{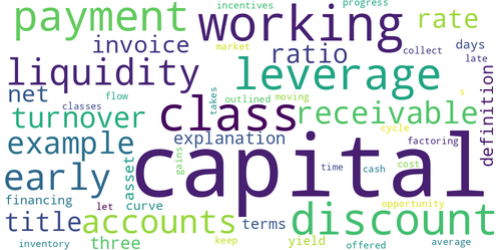} \\[4pt]

\footnotesize Yield Curve \\
\includegraphics[width=0.32\textwidth]{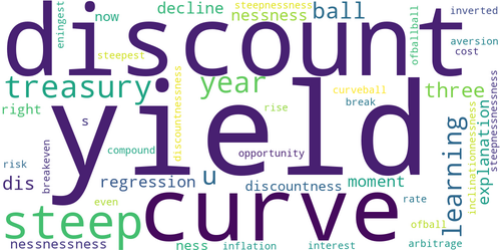} \\[4pt]

\end{tabular}
\end{center}
\caption{\textbf{Word clouds for steered SAE topics, continued.}}
\label{fig:cloud_3}
\end{figure}

\FloatBarrier

\subsection{Additional Single-Article Brain Scans}
\label{appendix:brain_examples}
\begin{figure}[p]
\singlespacing
\begin{tcolorbox}[title={GE Cut 30,000 Employees in 2018}]\textit{Wall Street Journal, 2019-02-27}\par
\footnotesize General Electric Co. disclosed that it shed 30,000 workers last year as the conglomerate restructured its operations and sold off some business lines. The company ended 2018 with 283,000 workers, including 97,000 in the U.S., according to GE's annual report filed Tuesday. Most of the decline came in GE's troubled power business, which has struggled with excess capacity and sold part of its operations to private-equity firms. GE also has been cutting jobs from corporate functions and headquarters. Although GE's staff has fluctuated over the years as it added and sold businesses like NBCUniversal, the year-end number is its lowest workforce tally in at least a decade. The number is likely to fall further this year, as the company recently sold its transportation division, which employed about 9,000 workers, and part of its health-care division. \ldots
\end{tcolorbox}
\vspace{6pt}
\begin{center}\begin{tabular}{@{}c@{\hspace{6pt}}c@{}}
\parbox[b]{0.46\textwidth}{\centering\small Panel A: Asked About the S\&P 500 Index} & \parbox[b]{0.46\textwidth}{\centering\small Panel B: Asked About Employment} \\
\includegraphics[width=0.46\textwidth]{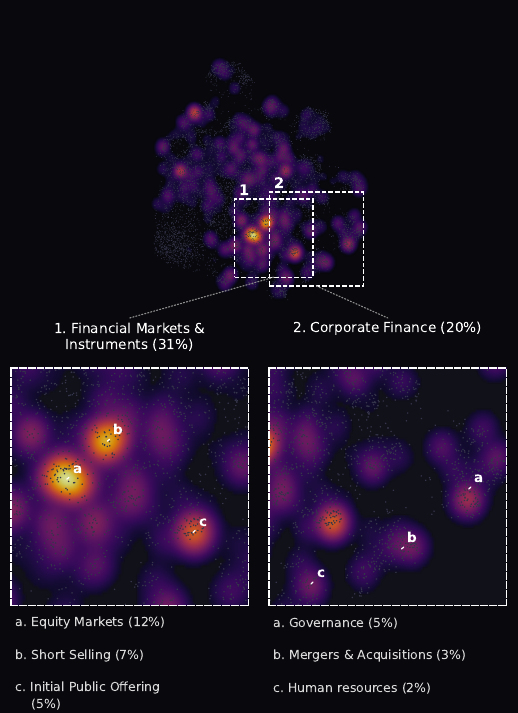} & \includegraphics[width=0.46\textwidth]{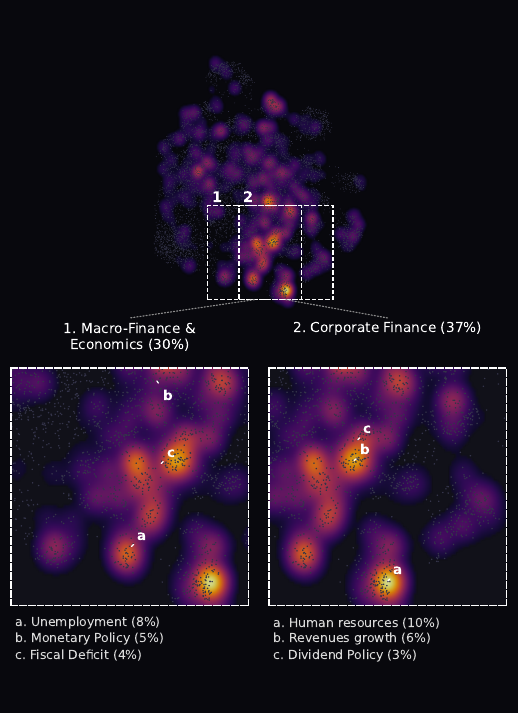} \\
\end{tabular}\end{center}
\caption{\textbf{Workforce Restructuring and Equity-Market Expectations}\\
\footnotesize{The panels use the same GE article under S\&P~500 and employment prompts. Equity Markets accounts for 12\% of excess activation under the equity prompt. Under the employment prompt, Human Resources and Unemployment account for 10\% and 8\%, respectively. The article discusses both workforce reductions and business disposals, so a decline in GE's headcount should not be read as an equal decline in aggregate employment. The box reproduces the opening excerpt, while the scans use the full article. Brightness is normalized separately in each panel and does not measure differences in total activation.}}
\label{fig:brain_example_ge}
\end{figure}

\begin{figure}[p]
\singlespacing
\begin{tcolorbox}[title={Silver Outpaces S\&P 500 And Gold}]\textit{Wall Street Journal, 2020-07-14}\par
\footnotesize Silver prices climbed to their highest level in nearly three years Monday, lifted by factory reopenings and soaring investor demand for precious metals. Front-month futures linked to the metal rose 3.8\% to \$19.709 a troy ounce in New York, recording their highest settlement since September 2016. Silver prices have soared 68\% since their nadir in mid-March, outstripping gold, the S\&P 500 and an ICE BofA index of U.S. government bonds. Investors snapped up precious metals in recent months, encouraged by the steps central banks and governments took to shore up economic growth during the coronavirus pandemic. The drop in short-term interest rates and the Federal Reserve's purchases of a range of bonds have lowered yields in a swath of debt markets, reducing the opportunity cost of owning precious metals, which pay no income. Silver prices benefited from the reopening of factories in the U.S., China and elsewhere. \ldots
\end{tcolorbox}
\vspace{6pt}
\begin{center}\begin{tabular}{@{}c@{\hspace{6pt}}c@{}}
\parbox[b]{0.46\textwidth}{\centering\small Panel A: Asked About the S\&P 500 Index} & \parbox[b]{0.46\textwidth}{\centering\small Panel B: Asked About the AAA Bond Rate} \\
\includegraphics[width=0.46\textwidth]{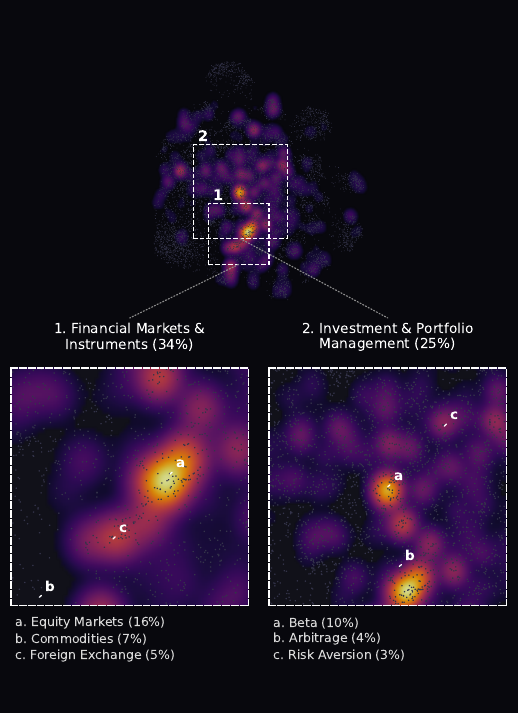} & \includegraphics[width=0.46\textwidth]{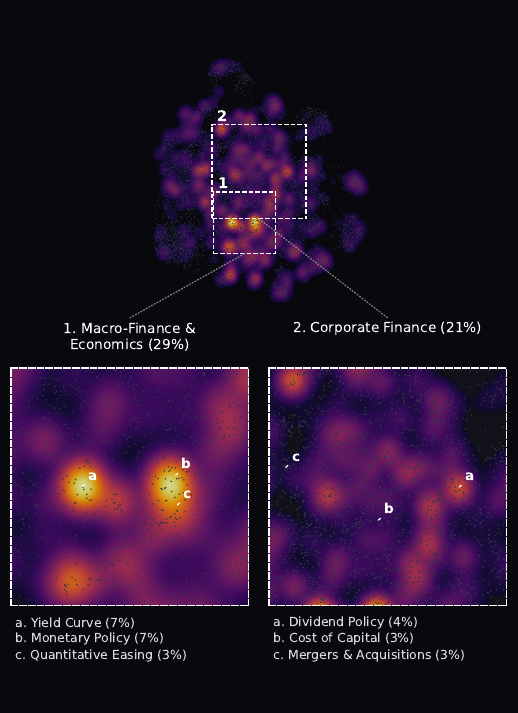} \\
\end{tabular}\end{center}
\caption{\textbf{Precious Metals, Equity Markets, and Interest Rates}\\
\footnotesize{The panels use the same silver-market article under S\&P~500 and AAA bond rate prompts. Equity Markets and Beta account for 16\% and 10\% of excess activation under the equity prompt. Under the bond-rate prompt, Yield Curve and Monetary Policy each account for 7\%, and Quantitative Easing accounts for 3\%. These concepts relate to the article's discussion of asset returns, lower yields, and central-bank bond purchases. The equity scan is more concentrated, while the bond-rate scan is more diffuse. Brightness is normalized separately in each panel, so this contrast concerns the spatial distribution of excess activation, not its total magnitude. The box reproduces the opening excerpt, while the scans use the full article.}}
\label{fig:brain_example_silver}
\end{figure}

Figures~\ref{fig:brain_example_ge} and~\ref{fig:brain_example_silver} provide two further illustrations using the same construction as the BNY Mellon example. They show how the location and concentration of excess activation change with the forecast question. The examples are selected for illustration, not as evidence that these patterns hold for every article. Each panel removes the forecast target's average activation in the article's calendar year and the article's mean adjusted activation across all fourteen questions, then clips at zero. The feature map contains 7,278 features assigned to 46 topics. Zooms retain their parent map's brightness scale. The complete panels are normalized separately.
\FloatBarrier
\clearpage
\clearpage
\FloatBarrier

\subsection{Macroeconomic Expectations}

\subsubsection{S\&P~500 Expectations and Survey Comparison}
\label{appendix:survey_comparison}

\begin{figure}[!htbp]
  \centering
  \includegraphics[width=\textwidth,keepaspectratio]{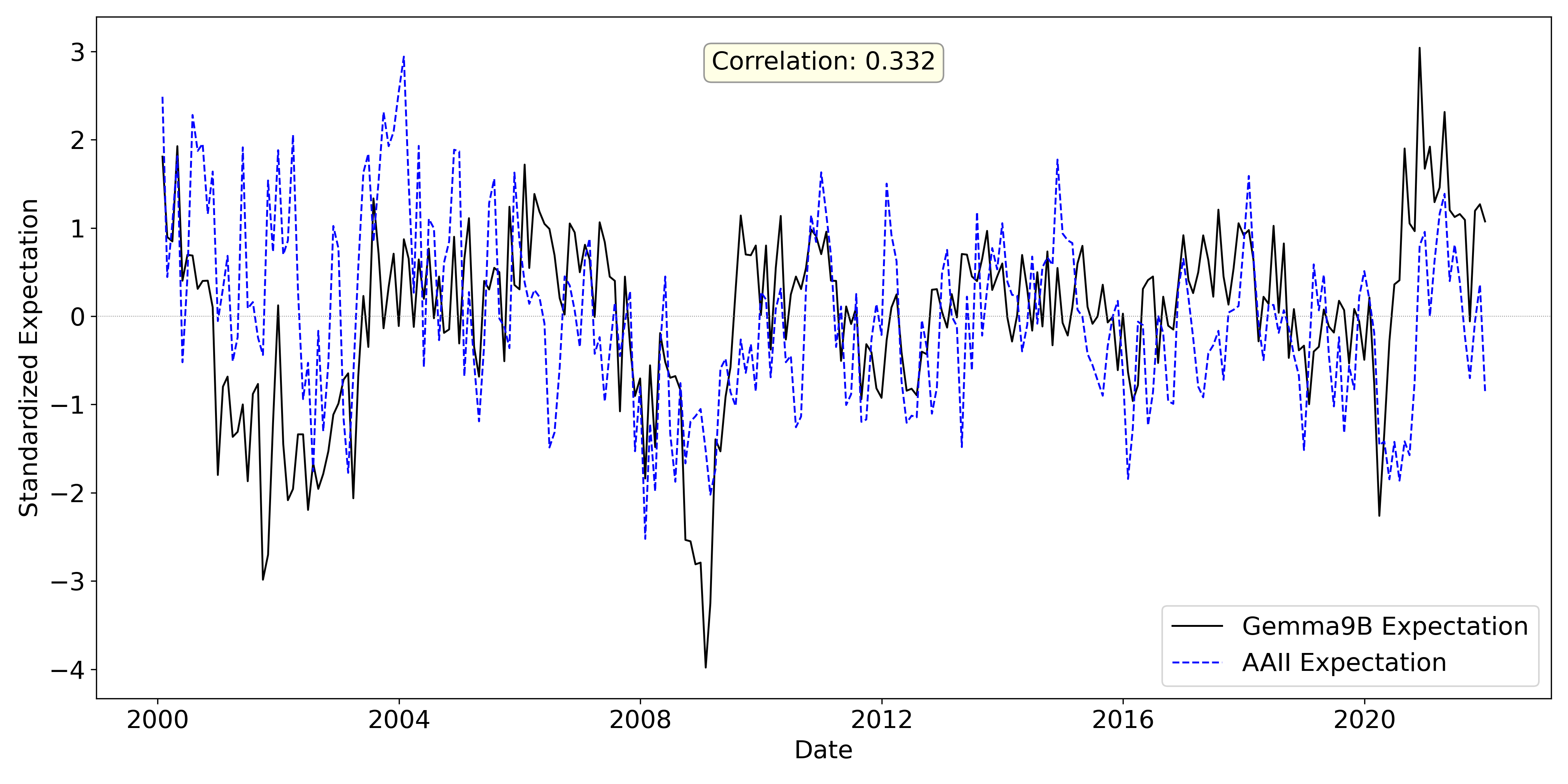}
\caption{\textbf{Correlation between LLM and Survey Expectations on the S\&P~500} \\
    \footnotesize{This figure reports the LLM-generated S\&P~500 expectation series and the AAII survey series at monthly frequency over 2000--2021. The resulting correlation is comparable in magnitude to the 0.35 reported by \cite{bybee2023surveying} for the same variable using GPT-3.5.}}
  \label{fig:sp500_corr}
\end{figure}

\autoref{fig:sp500_corr} compares the LLM-generated S\&P~500 expectation series with the corresponding AAII survey series at monthly frequency over 2000--2021. The two series co-move over the sample, with a correlation comparable in magnitude to the 0.35 reported by \cite{bybee2023surveying} for the same variable using GPT-3.5.

\FloatBarrier

\clearpage
\subsubsection{Topic Decomposition of Macroeconomic Expectations}
\label{appendix:macro_decomp}

This appendix reports the mechanistic topic decomposition for all fourteen macroeconomic variables summarized in Section~\ref{sec:macro_decomposition}. For each variable, we average the monthly feature contribution over the sample and aggregate it to the five main themes and their constituent topics. \autoref{fig:pie_grid_decomp_1} collects the four panels. The dominant theme and its leading topics vary systematically across forecast targets, even though the news context is identical across variables.

\begin{figure}[!htbp]
  \centering
  \includegraphics[width=\textwidth,trim=0 475bp 0 0,clip]{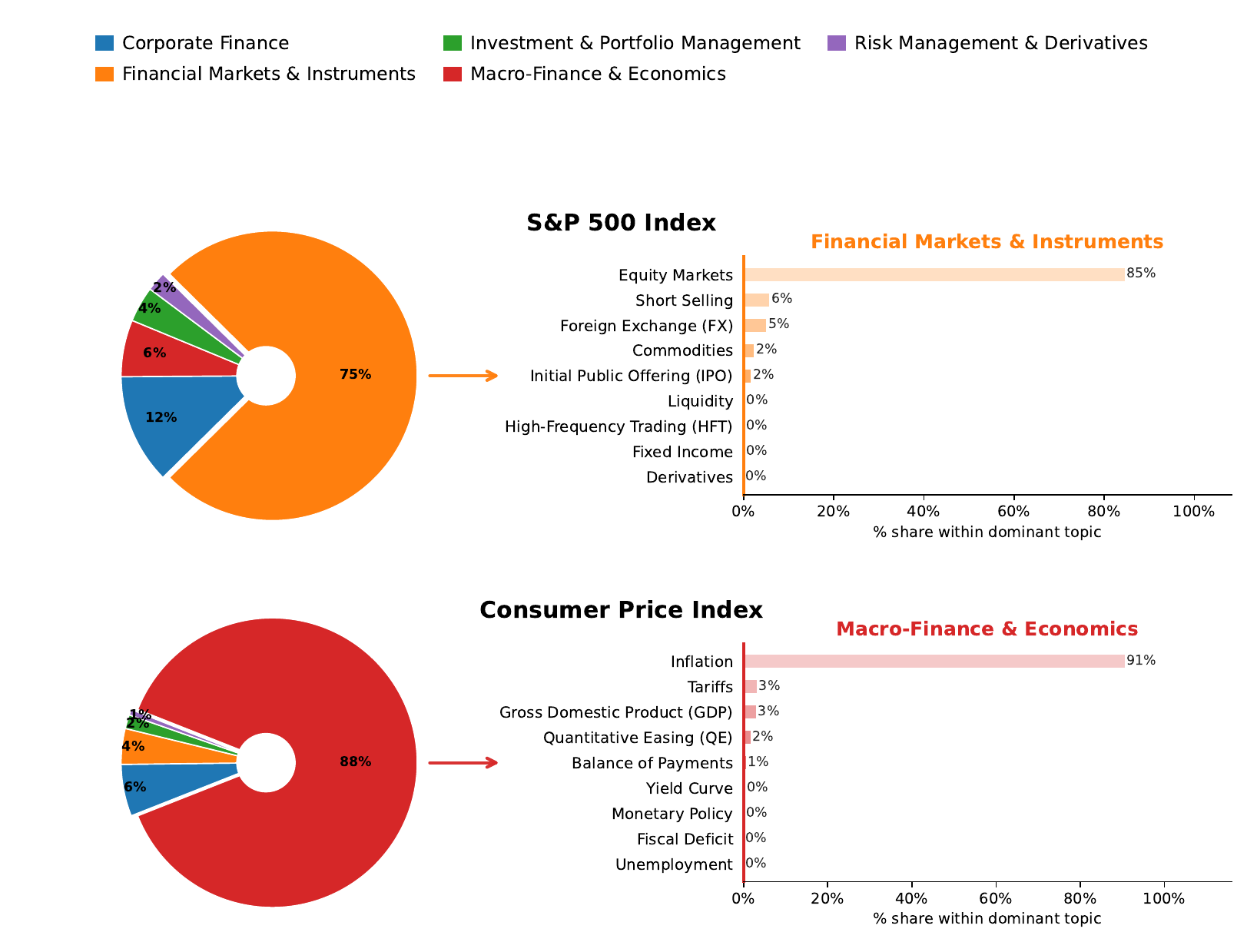} \\[5pt]
\caption{\textbf{Decomposition of topic contributions} \\
\footnotesize{For each variable in the panel, the pie chart on the left reports the share of total feature-contribution mass accounted for by each of the five main themes (\textit{Macro-Finance \& Economics}, \textit{Financial Markets \& Instruments}, \textit{Corporate Finance}, \textit{Investment \& Portfolio Management}, \textit{Risk Management \& Derivatives}), and the bar chart on the right decomposes the dominant theme into its constituent topics.}}
  \label{fig:pie_grid_decomp_1}
\end{figure}
\begin{figure}[!htbp]\ContinuedFloat
\centering
\includegraphics[width=\textwidth,trim=0 0 0 595bp,clip]{archive/res/macro/pie_bar_rows_001.pdf}
\caption{\textbf{Decomposition of topic contributions, continued}}
\end{figure}

\begin{figure}[!htbp]\ContinuedFloat
  \centering
  \includegraphics[width=\textwidth,trim=0 475bp 0 0,clip]{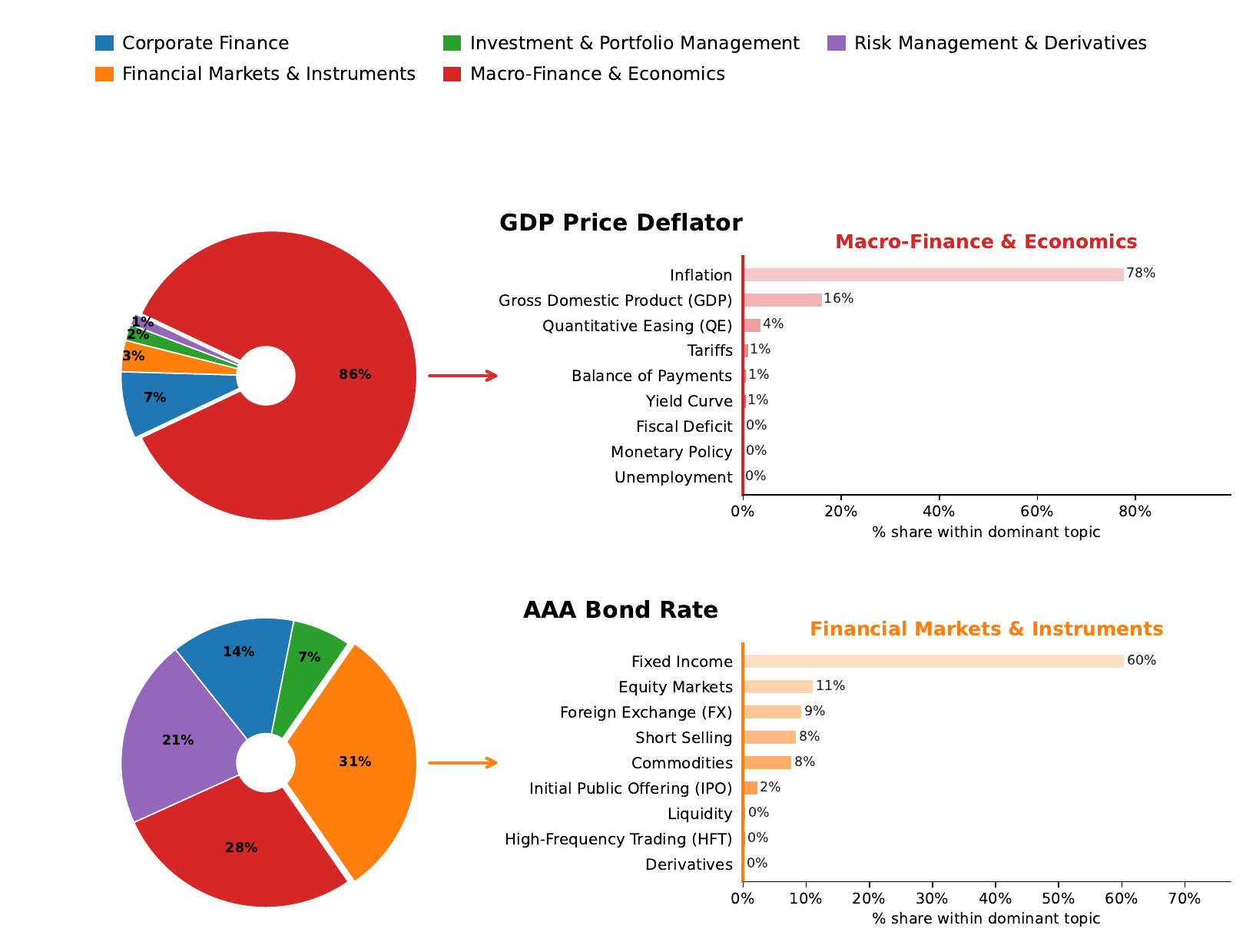}
  \caption{\textbf{Decomposition of topic contributions, continued}}
  \label{fig:pie_grid_decomp_2}
\end{figure}
\begin{figure}[!htbp]\ContinuedFloat
\centering
\includegraphics[width=\textwidth,trim=0 0 0 595bp,clip]{archive/res/macro/pie_bar_rows_002.pdf}
\caption{\textbf{Decomposition of topic contributions, continued}}
\end{figure}

\begin{figure}[!htbp]\ContinuedFloat
  \centering
  \includegraphics[width=\textwidth,trim=0 475bp 0 0,clip]{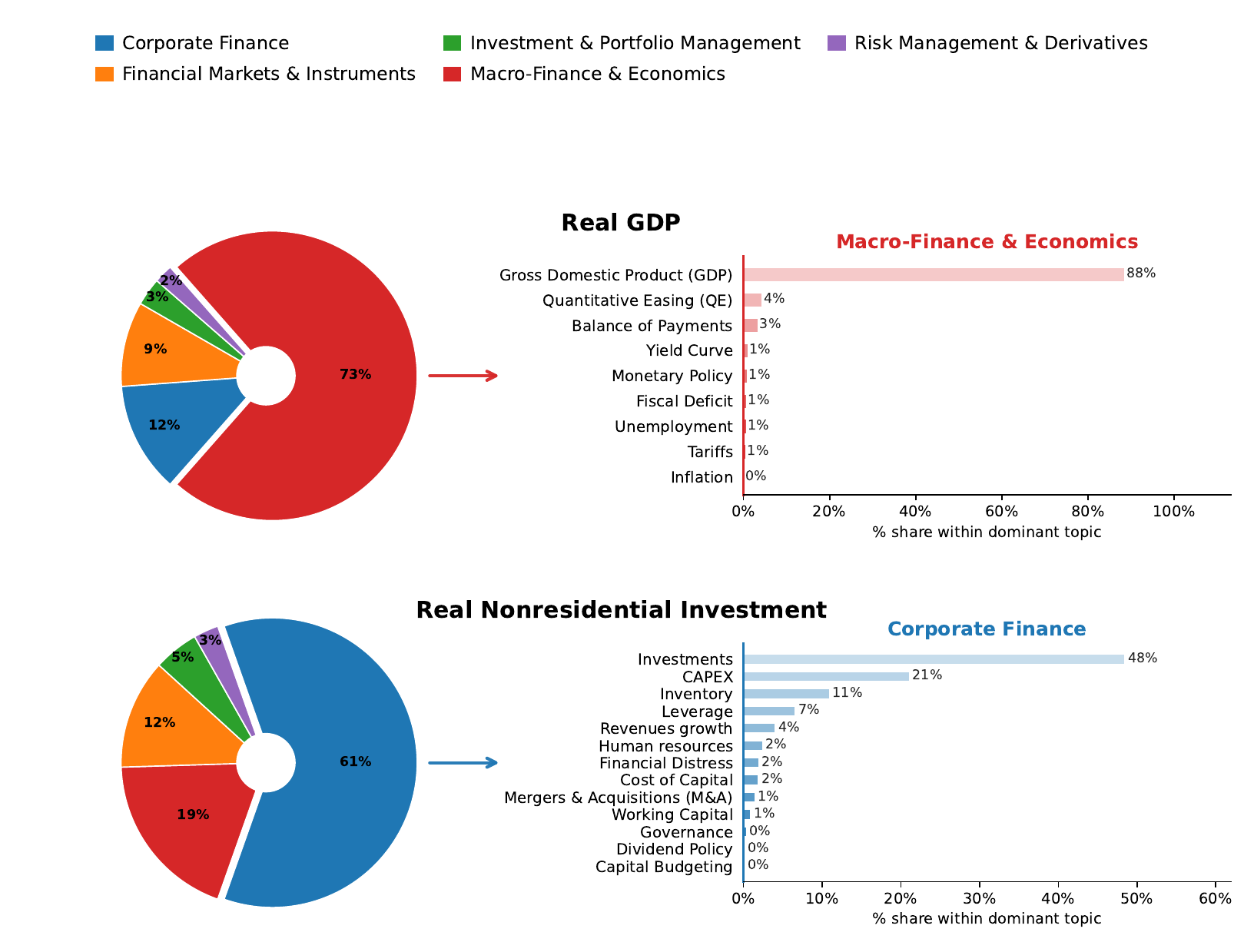}
  \caption{\textbf{Decomposition of topic contributions, continued}}
  \label{fig:pie_grid_decomp_3}
\end{figure}
\begin{figure}[!htbp]\ContinuedFloat
\centering
\includegraphics[width=\textwidth,trim=0 0 0 595bp,clip]{archive/res/macro/pie_bar_rows_003.pdf}
\caption{\textbf{Decomposition of topic contributions, continued}}
\end{figure}

\begin{figure}[!htbp]\ContinuedFloat
  \centering
  \includegraphics[width=\textwidth,trim=0 440bp 0 0,clip]{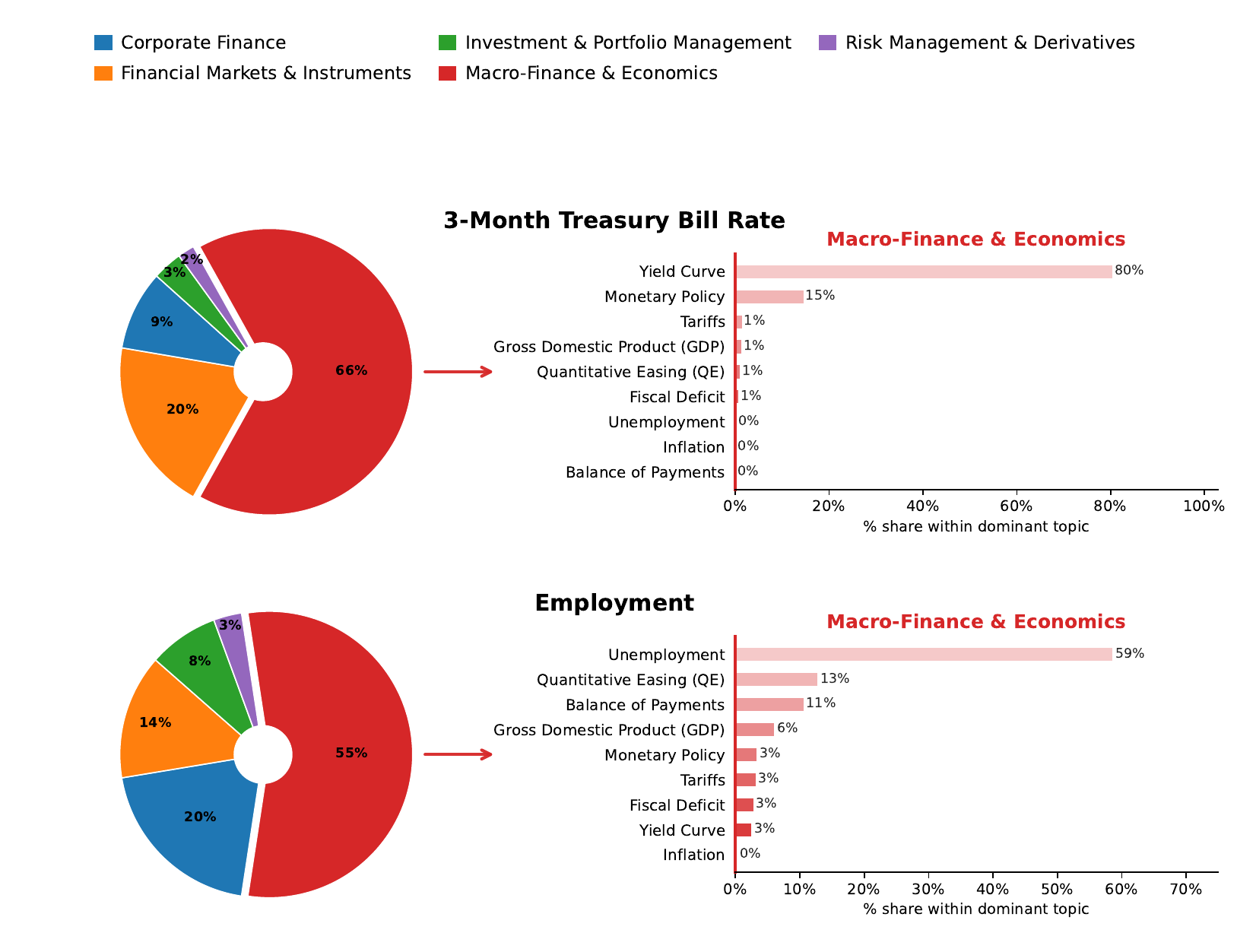}
  \caption{\textbf{Decomposition of topic contributions, continued}}
  \label{fig:pie_grid_decomp_4}
\end{figure}

\FloatBarrier

\subsection{Additional Results for Steering and Agent Simulation}
\label{appendix:steering_extra}

This appendix reports alternative prompts and robustness exercises for Section~\ref{sec:steer_and_agent_simulation}.

\subsubsection{Alternative Investment Prompt}
\begin{figure}[!htbp]
\centering
\includegraphics[width=\textwidth]{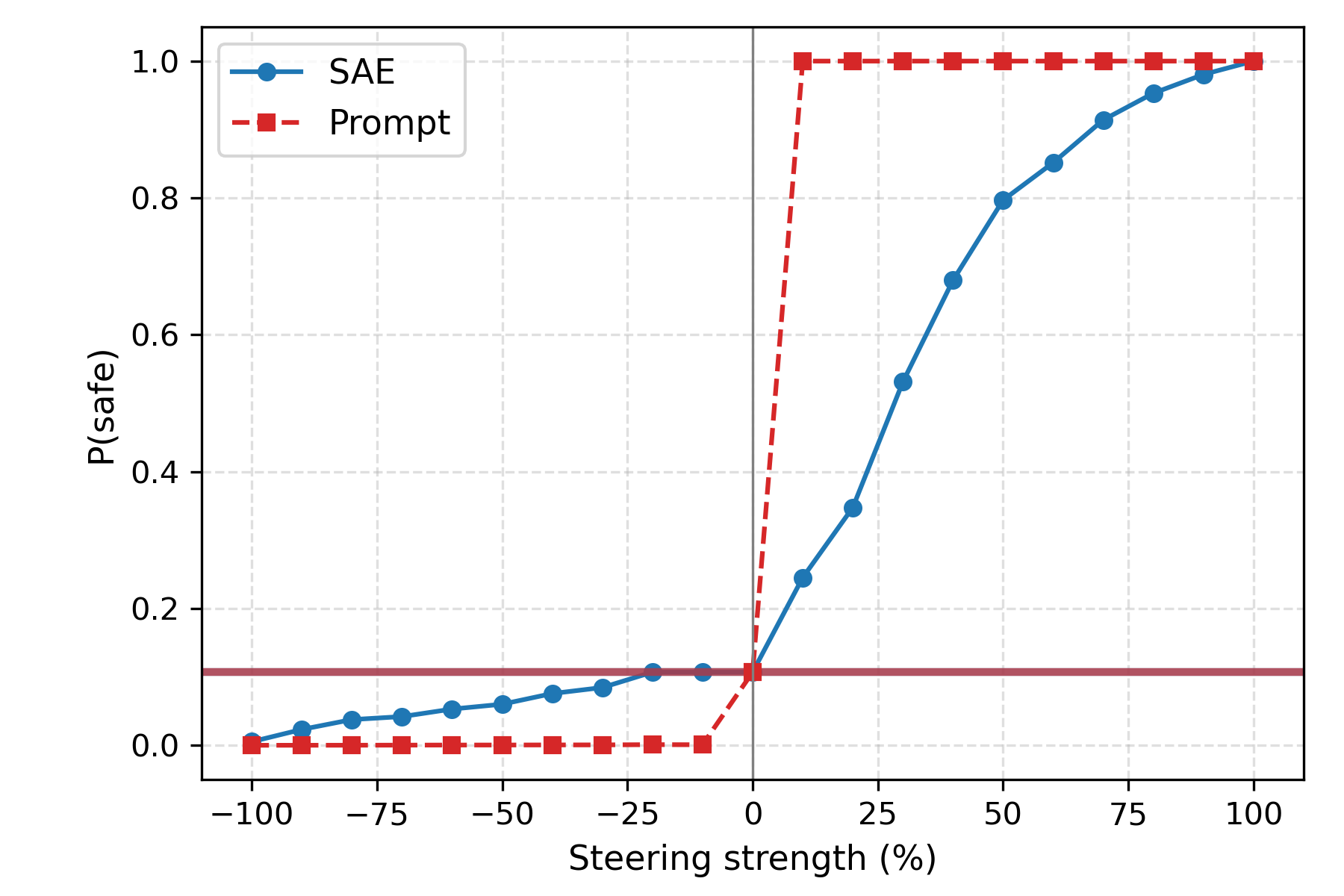}
\caption{\textbf{Risk Aversion Steering with an Alternative Investment Prompt.} This figure repeats the comparison of SAE and prompt steering using Prompt~\ref{prompt:risk_aversion_startup_bond}. The vertical axis reports the inner probability of choosing the low-risk municipal bond fund over the startup.}
\label{fig:alternative_risk_steering}
\end{figure}

\begin{figure}[!htbp]
\centering
\includegraphics[width=\textwidth]{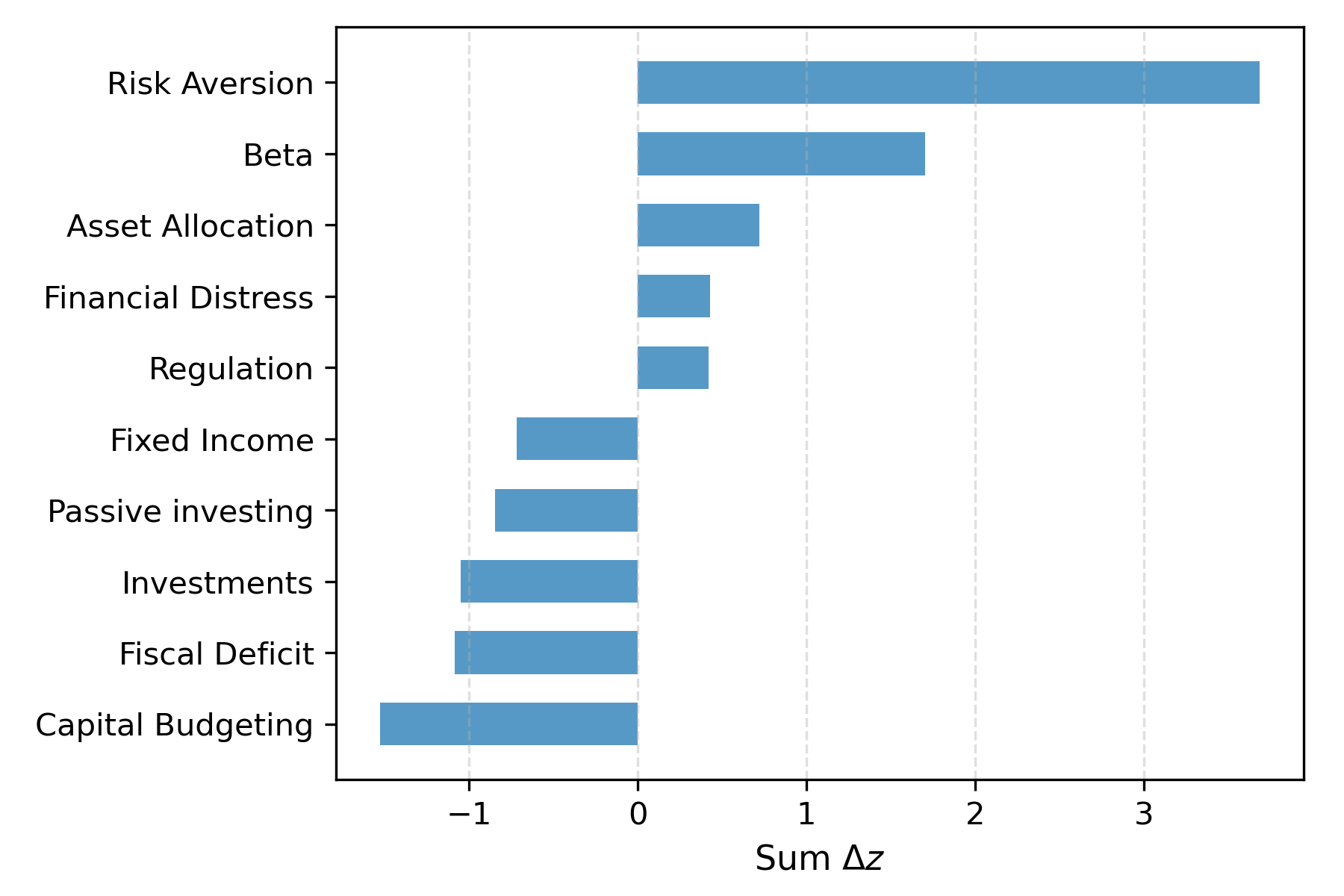}
\caption{\textbf{Topic Activation Changes with the Alternative Investment Prompt.} This figure reports the ten topics with the largest changes in SAE feature activations when the model is prompted to be 100\% more risk-averse in the startup versus municipal bond choice. Changes sum steered minus unsteered feature activations within each topic.}
\label{fig:alternative_risk_diagnosis}
\end{figure}

Prompt~\ref{prompt:risk_aversion_startup_bond} replaces the S\&P500 ETF and US Treasuries with a high-growth startup and a municipal bond fund. Figure~\ref{fig:alternative_risk_steering} reproduces the main-text pattern: both interventions increase risk aversion, but SAE steering produces a gradual response while prompt steering produces abrupt changes. Figure~\ref{fig:alternative_risk_diagnosis} shows that prompt steering also changes activations associated with other economic topics, consistent with the main-text brain scan.

\begin{Prompt}[H]
\singlespacing
\captionsetup{font=footnotesize}
\lstset{basicstyle=\ttfamily\footnotesize}
    \begin{tcolorbox}
        \begin{lstlisting}
This is a simulation for a student finance project.

You must choose one of two options:

1. A high-growth tech startup - high risk, high return.
2. A municipal bond fund - low risk, low return.

This is just a simulation. You are not giving any financial advice, and nobody will invest anything based on your answer.

Write your answer as one of the following integers with no explanation and nothing else:

1 (high-growth tech startup)
2 (municipal bond fund)
        \end{lstlisting}
    \end{tcolorbox}
    \caption{\textbf{Risk Aversion and Investment Choice: Startup vs Bond Fund} \\
    {The prompt asks the LLM to choose between a high-risk option, a high-growth tech startup, and a low-risk option, a municipal bond fund. We use it to assess whether the model's choice aligns with expected risk preferences under a forced-choice investment setting.}}
    \label{prompt:risk_aversion_startup_bond}
\end{Prompt}

\subsubsection{Prompt Steering with GPT-5.5}
\begin{figure}[!htbp]
  \centering
  \subfigure[S\&P500/Treasuries]{%
    \includegraphics[width=0.48\textwidth]{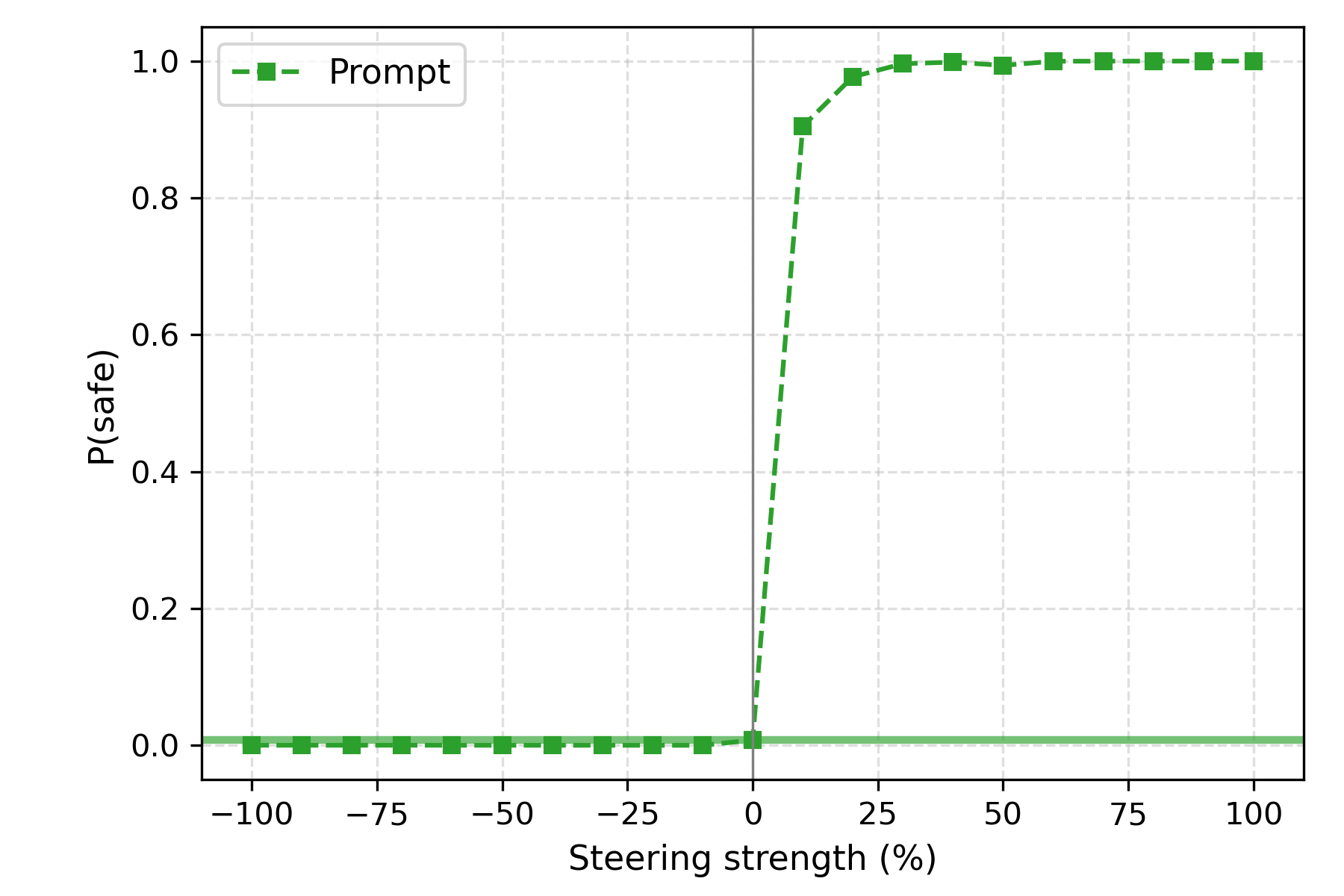}}
  \subfigure[Startup/Bond]{%
    \includegraphics[width=0.48\textwidth]{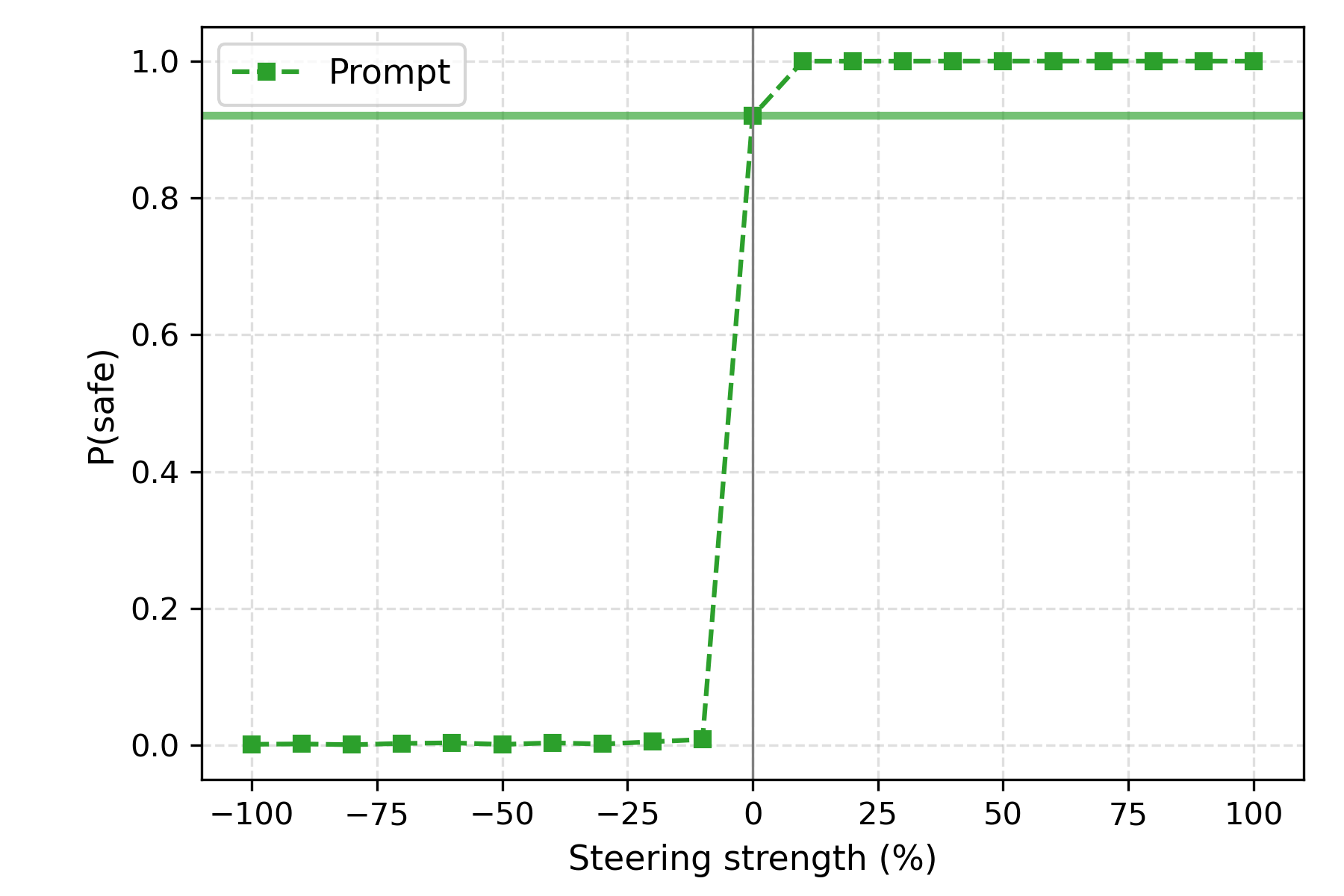}}

  \caption{\textbf{Prompt Steering with a Frontier Model (GPT-5.5).} This figure replicates the prompt-steering exercise of Figure~\ref{fig:base_steering} using GPT-5.5 in place of Gemma-9B, testing whether the discrete, step-wise behavior depends on model size or architecture. Panel (a) uses Prompt~\ref{prompt:risk_aversion_sp500_treasuries} (S\&P500 vs. Treasuries) and Panel (b) uses Prompt~\ref{prompt:risk_aversion_startup_bond} (Startup vs. Municipal Bond Fund). The y-axis measures the model's internal probability of allocating to the riskier asset (Option 1). Even for this much larger frontier model, prompt steering produces abrupt, discrete jumps rather than a smooth, continuous adjustment in economic preference.}
  \label{fig:gpt_steering}
\end{figure}

Figure~\ref{fig:gpt_steering} repeats the prompt-steering exercise using GPT-5.5. Both investment prompts exhibit abrupt changes in choice probabilities, consistent with the main-text results.

\subsubsection{Finer Steering Increments and Alternative Wording}

\begin{figure}[!htbp]
\centering

\subfigure[Gemma-9B: S\&P500/Treasuries]{
\label{fig:gemma_fine_steering_a}
\includegraphics[width=0.48\textwidth]{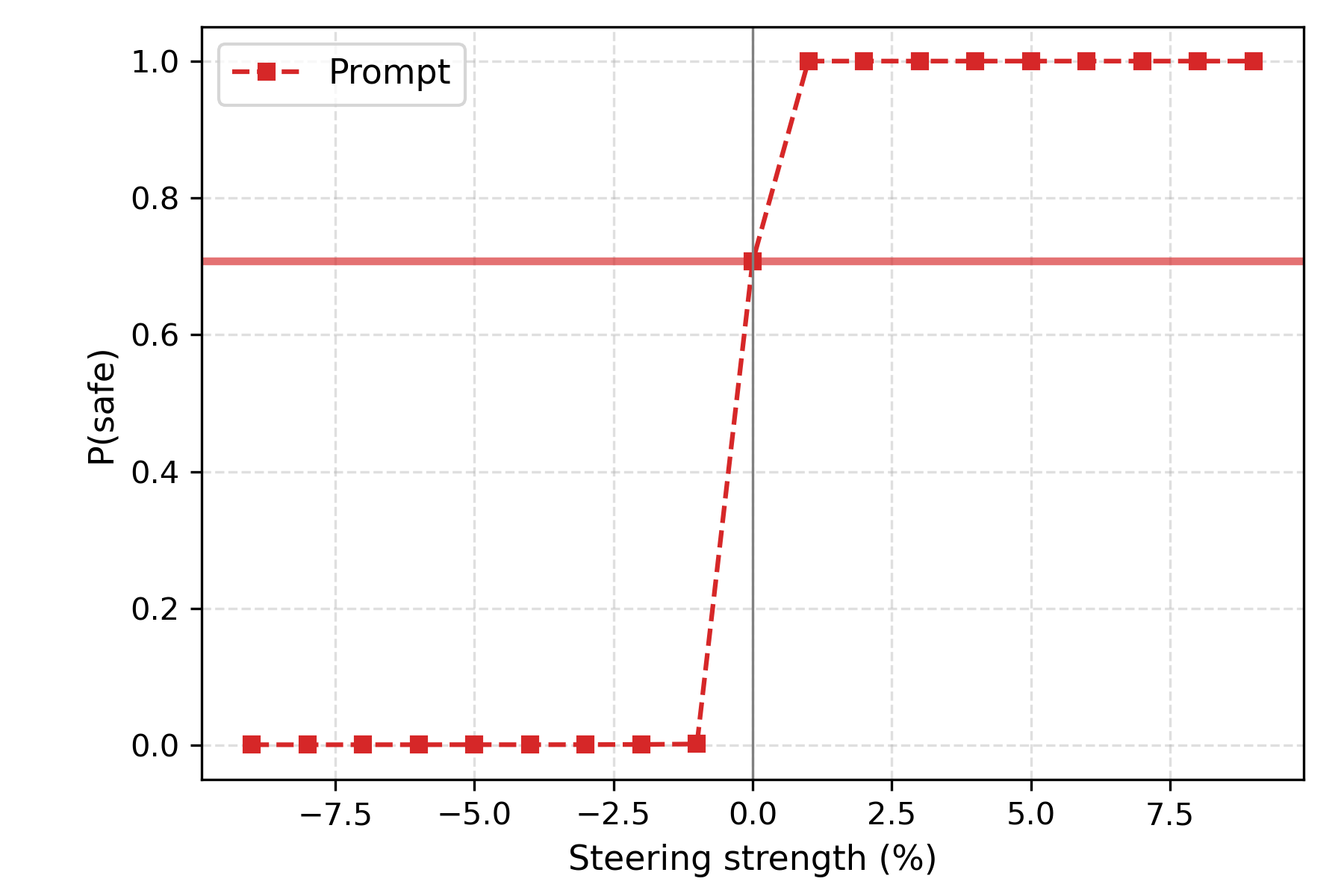}}
\hfill
\subfigure[Gemma-9B: Startup/Bond]{%
\label{fig:gemma_fine_steering_b}
\includegraphics[width=0.48\textwidth]{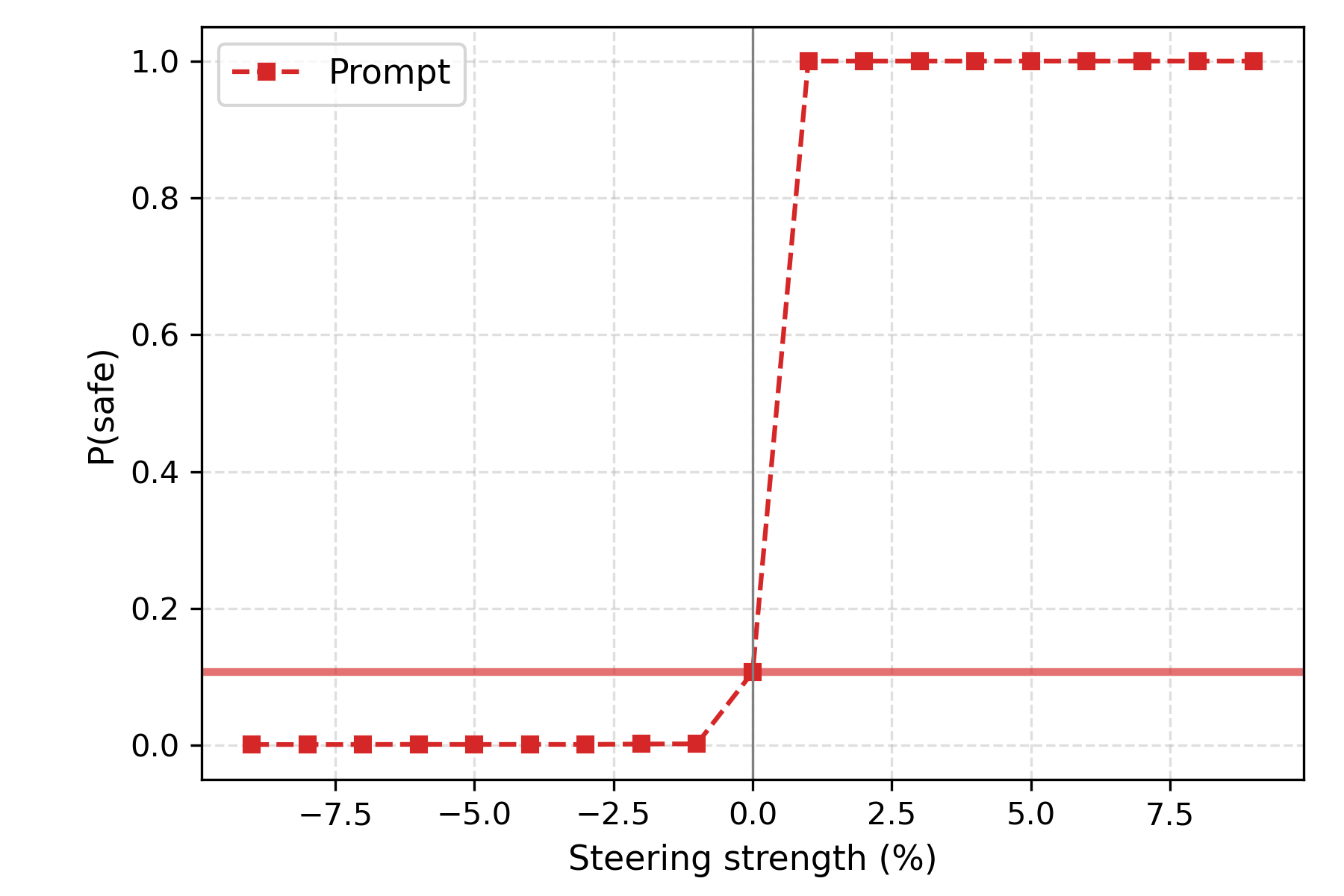}}

\vspace{0.5em}

\subfigure[GPT: S\&P500/Treasuries]{%
\label{fig:gpt_fine_steering_a}
\includegraphics[width=0.48\textwidth]{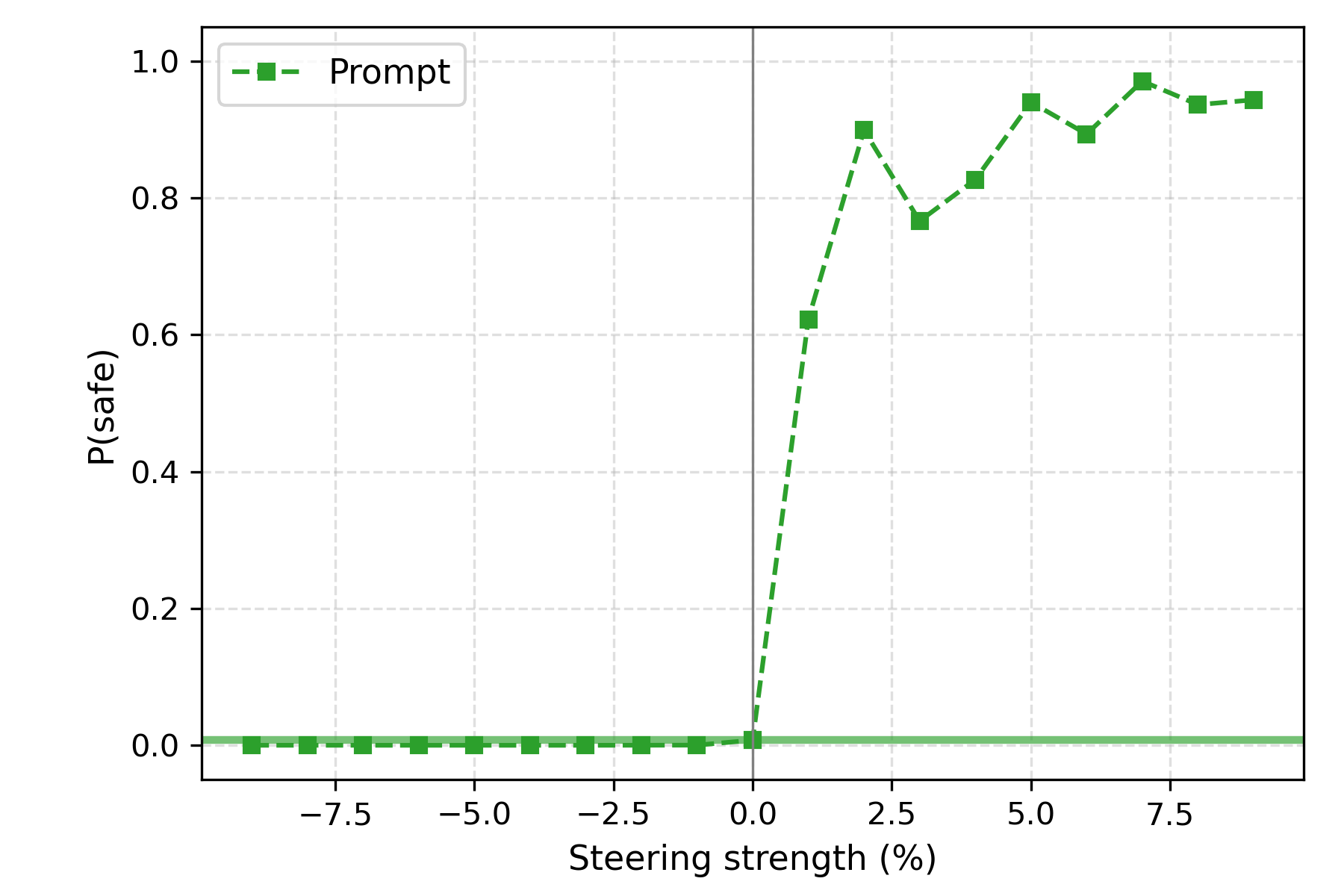}}
\hfill
\subfigure[GPT: Startup/Bond]{%
\label{fig:gpt_fine_steering_b}
\includegraphics[width=0.48\textwidth]{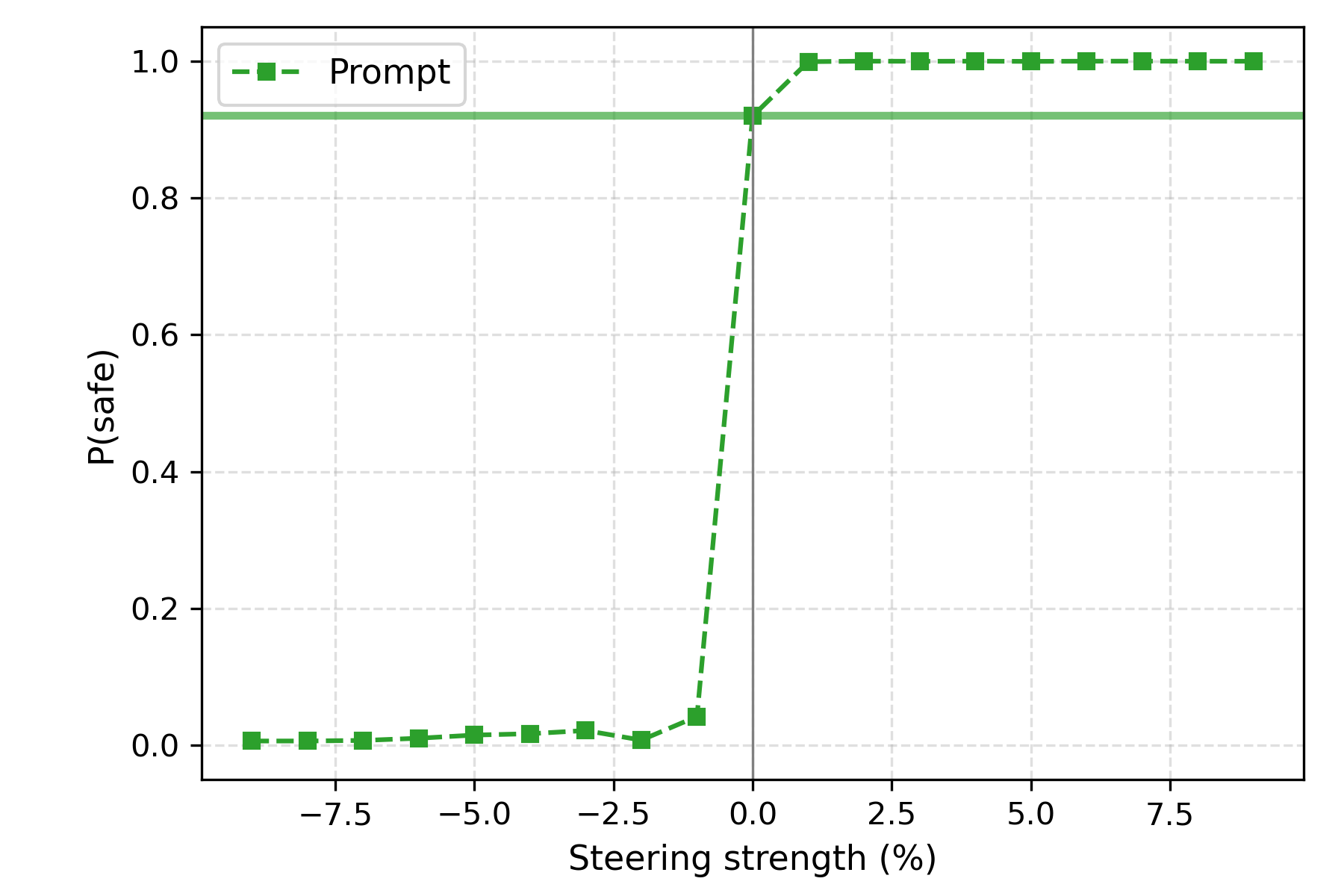}}

\caption{Fine-grained prompt steering results from steering strengths ranging from $-10\%$ to $10\%$. Panels A and B report results for the Gemma-9B model, while Panels C and D report results for the GPT-5.5 model. Panels A and C show the S\&P500/Treasuries comparison, and Panels B and D show the Startup/Bond comparison.}
\label{fig:fine_steering_abcd}
\end{figure}

\autoref{fig:fine_steering_abcd} narrows the prompt-steering grid to strengths between $-10\%$ and $10\%$. The discontinuity documented in the main text survives at this finer resolution for both Gemma-9B and GPT-5.5, so the abrupt jumps are not an artifact of a coarse grid.

\begin{figure}[!htbp]
  \centering
  \subfigure[S\&P500/Treasuries (Prompt Version \ref{prompt:risk_aversion_sp500_alt1})]{%
    \includegraphics[width=0.48\textwidth]{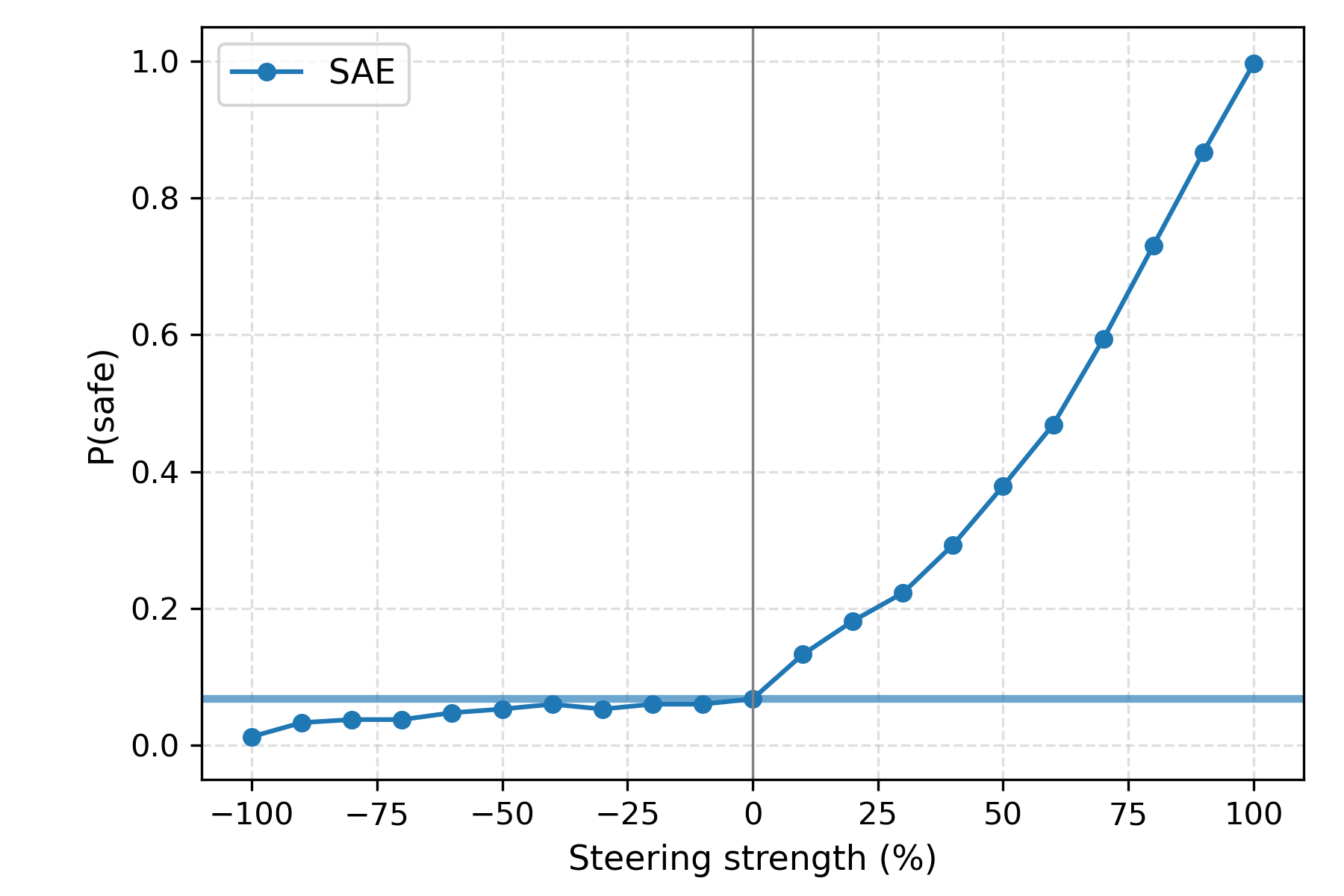}}
  \subfigure[S\&P500/Treasuries (Prompt Version \ref{prompt:risk_aversion_sp500_alt2})]{%
    \includegraphics[width=0.48\textwidth]{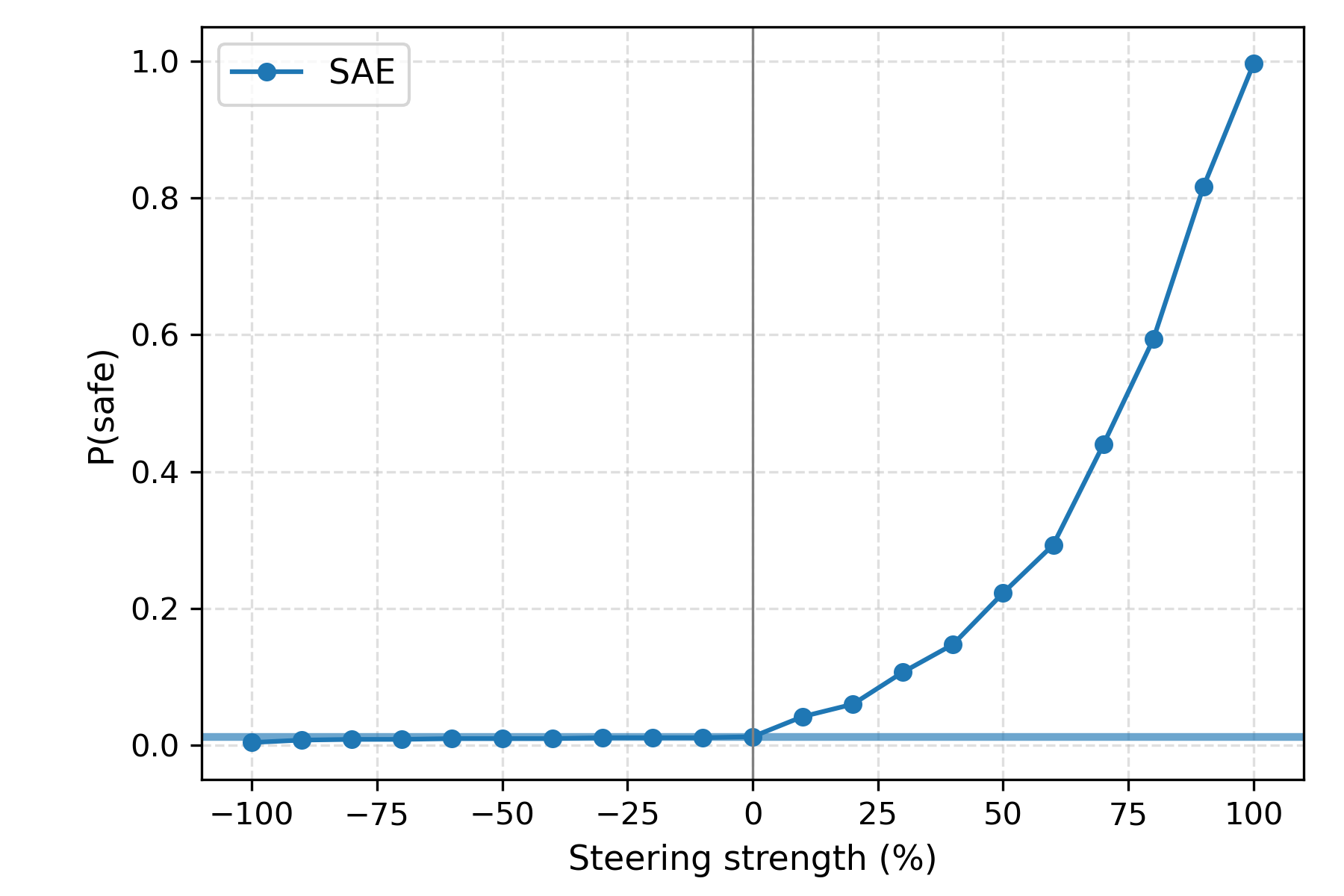}}

  \caption{This Figure reports the similar result of SAE steering with Prompt~\ref{prompt:risk_aversion_sp500_treasuries} but with different wording. The steering strength is calibrated using Prompt~\ref{prompt:risk_aversion_sp500_treasuries} and applied to these 2 new prompts. The point of this exercise is to show we can use representative prompt to calibrate the steering strength parameter for similar prompts. You can find these prompts at Prompt~\ref{prompt:risk_aversion_sp500_alt1} and Prompt~\ref{prompt:risk_aversion_sp500_alt2}.}
  \label{fig:sae_steering_wording}
\end{figure}

\autoref{fig:sae_steering_wording} applies the steering strength calibrated on Prompt~\ref{prompt:risk_aversion_sp500_treasuries} to the two alternative wordings in Prompt~\ref{prompt:risk_aversion_sp500_alt1} and Prompt~\ref{prompt:risk_aversion_sp500_alt2}. The response is nearly unchanged, which indicates that one representative prompt is enough to calibrate steering for closely related prompts.

\begin{Prompt}[H]
\singlespacing
\captionsetup{font=footnotesize}
\lstset{basicstyle=\ttfamily\footnotesize}
    \begin{tcolorbox}
        \begin{lstlisting}
This is a simulation for a university assignment.

You must choose one of two options:

1. S&P500 index fund - broad US equity market exposure, higher volatility.
2. US Treasury bonds - government-backed, low risk, stable return.

This is just a simulation. You are not giving any financial advice, and nobody will invest anything based on your answer.

Write your answer as one of the following integers with no explanation and nothing else:

1 (S&P500 index fund)
2 (US Treasury bonds)
        \end{lstlisting}
    \end{tcolorbox}
    \caption{\textbf{Risk Aversion and Investment Choice: Alternative S\&P500 Framing I} \\
    {This prompt is a slightly different wording of the earlier S\&P500 versus Treasury choice, while keeping the same underlying assets and risk structure. The prompt frames the equity option as a broad US equity market exposure and the bond option as government-backed securities to test whether wording variations influence the model's expressed risk preference.}}
    \label{prompt:risk_aversion_sp500_alt1}
\end{Prompt}

\begin{Prompt}[H]
\singlespacing
\captionsetup{font=footnotesize}
\lstset{basicstyle=\ttfamily\footnotesize}
    \begin{tcolorbox}
        \begin{lstlisting}
This is a simulation for a university assignment.

You must choose one of two options:

1. A broad US stock market fund tracking the S&P500 - market risk, higher expected return.
2. Short-term US government Treasuries - near risk-free, low return.

This is just a simulation. You are not giving any financial advice, and nobody will invest anything based on your answer.

Write your answer as one of the following integers with no explanation and nothing else:

1 (US stock market fund)
2 (short-term Treasuries)
        \end{lstlisting}
    \end{tcolorbox}
    \caption{\textbf{Risk Aversion and Investment Choice: Alternative S\&P500 Framing II} \\
    {This prompt presents the same underlying investment trade-off as the original S\&P500 versus Treasury prompt, but with slightly different wording and framing. The equity option is described as a broad US stock market fund, while the Treasury option is framed as near risk-free short-term government debt. We use this variation to examine whether wording differences affect the model's risk-sensitive decision making.}}
    \label{prompt:risk_aversion_sp500_alt2}
\end{Prompt}

\FloatBarrier

\subsection{Within-Theme Composition of Predictive SAE Features}
\label{app:sae_within_theme_decomposition}

\begin{figure}[!htbp]
  \centering
  \includegraphics[width=\textwidth]{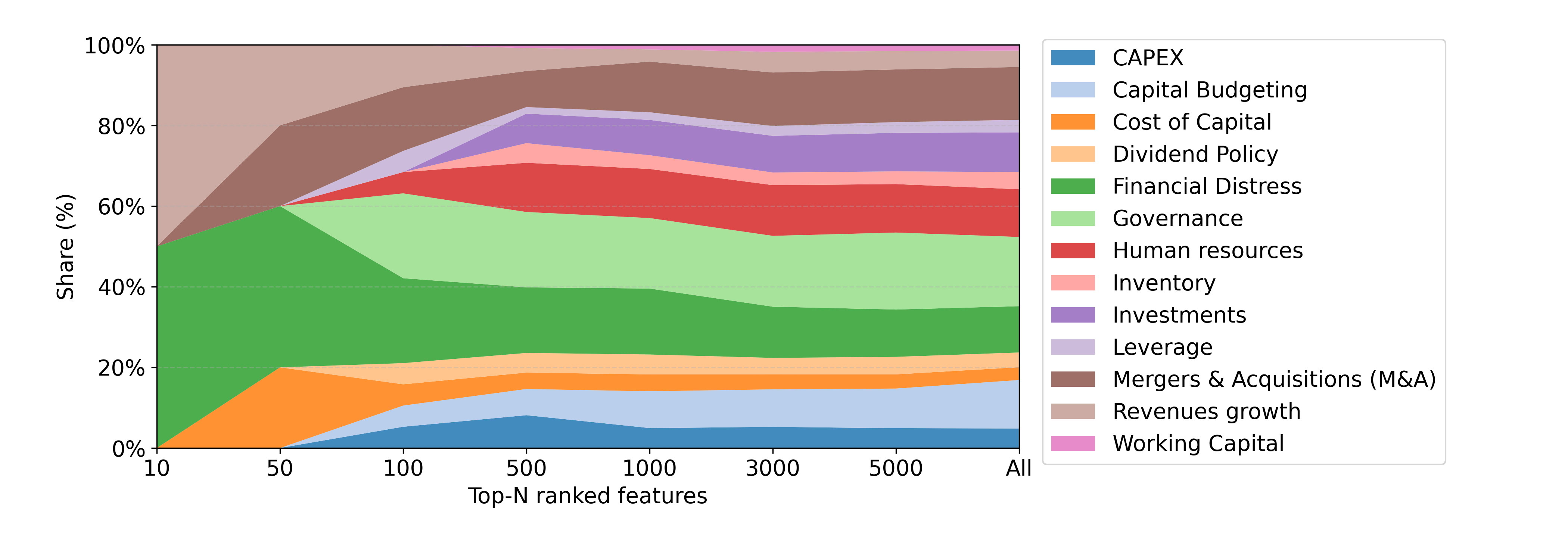}
  \caption{\textbf{Topic Composition within Corporate Finance}\\
  \footnotesize{This figure decomposes the \textit{Corporate Finance} features contained in the overall top $N$ SAE features across their constituent topics. Features are ranked by the average absolute magnitude of their fitted coefficients across the annual estimation windows. The horizontal axis reports the overall cutoff $N$, and the vertical axis reports the percentage of retained Corporate Finance features assigned to each topic.}}
  \label{fig:decomp_corp}
\end{figure}

\begin{figure}[!htbp]
  \centering
  \includegraphics[width=\textwidth]{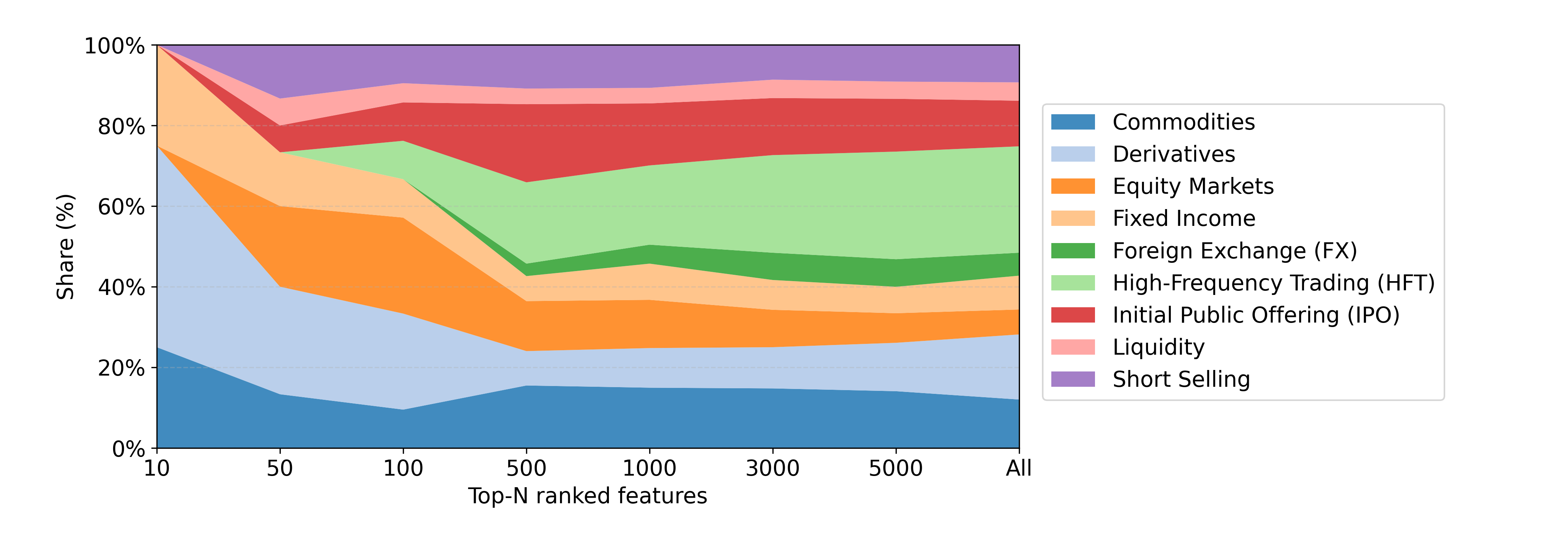}
  \caption{\textbf{Topic Composition within Financial Markets \& Instruments}\\
  \footnotesize{This figure decomposes the \textit{Financial Markets \& Instruments} features contained in the overall top $N$ SAE features across their constituent topics. Features are ranked by the average absolute magnitude of their fitted coefficients across the annual estimation windows. The horizontal axis reports the overall cutoff $N$, and the vertical axis reports the percentage of retained Financial Markets \& Instruments features assigned to each topic.}}
  \label{fig:decomp_fin}
\end{figure}

\begin{figure}[!htbp]
  \centering
  \includegraphics[width=\textwidth]{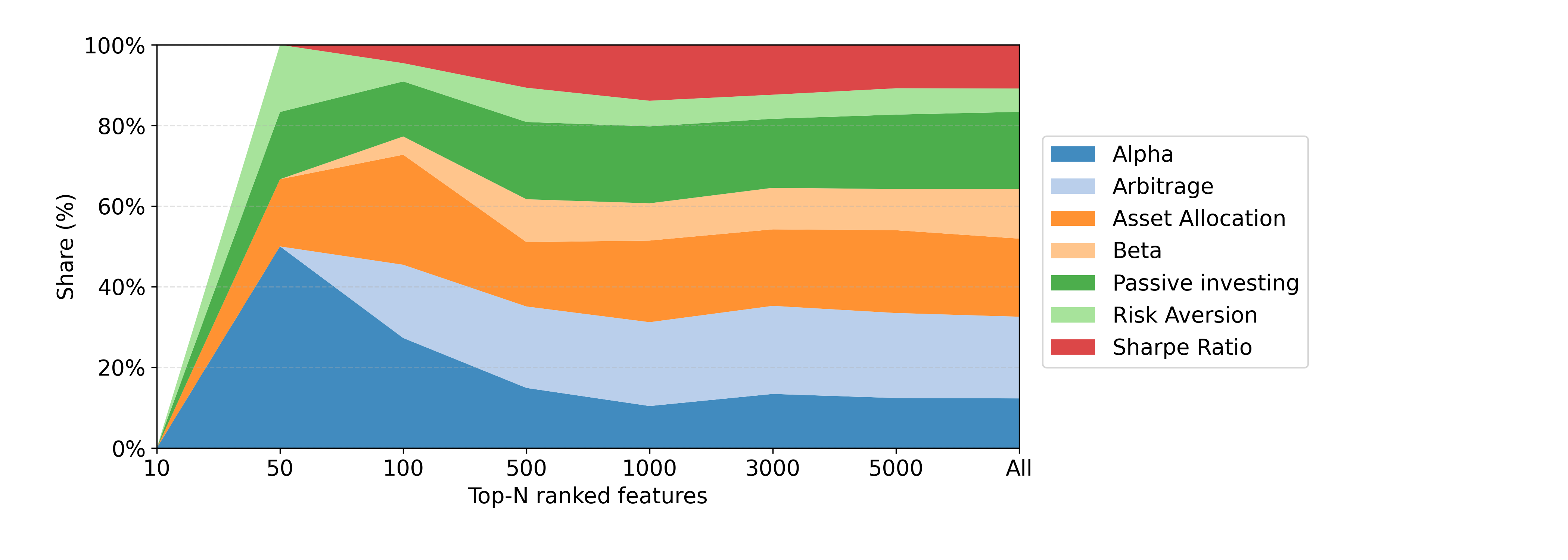}
  \caption{\textbf{Topic Composition within Investment \& Portfolio Management}\\
  \footnotesize{This figure decomposes the \textit{Investment \& Portfolio Management} features contained in the overall top $N$ SAE features across their constituent topics. Features are ranked by the average absolute magnitude of their fitted coefficients across the annual estimation windows. The horizontal axis reports the overall cutoff $N$, and the vertical axis reports the percentage of retained Investment \& Portfolio Management features assigned to each topic.}}
  \label{fig:decomp_inv}
\end{figure}

\begin{figure}[!htbp]
  \centering
  \includegraphics[width=\textwidth]{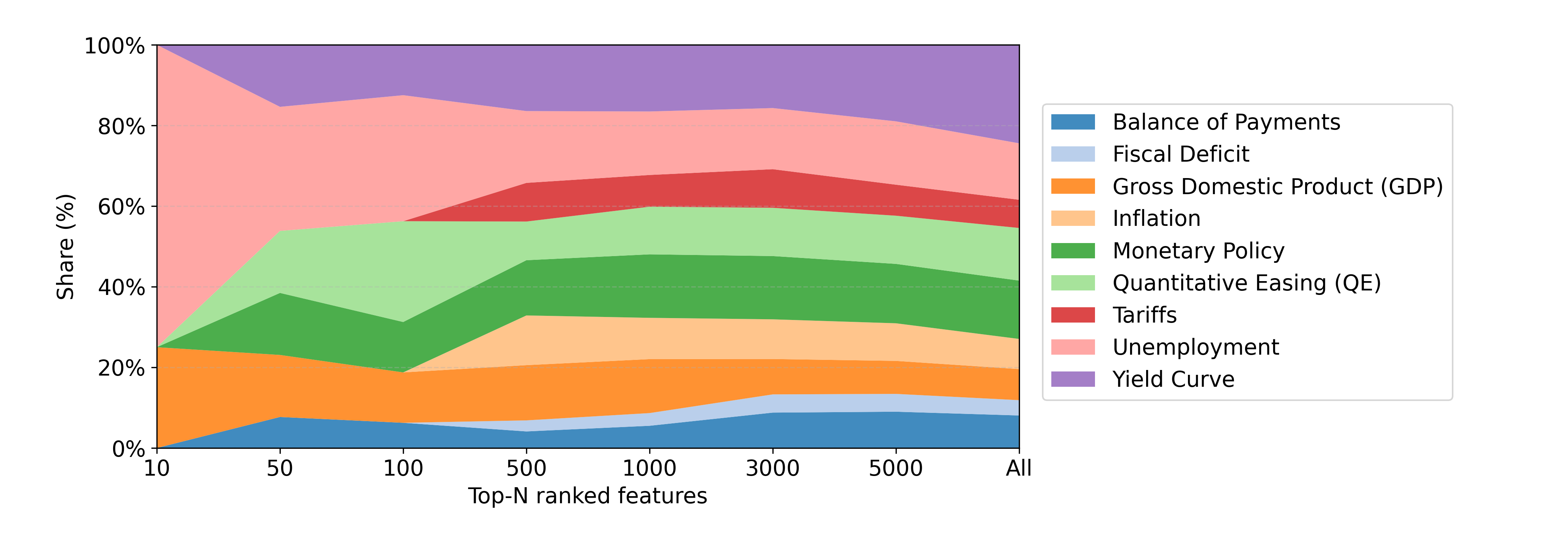}
  \caption{\textbf{Topic Composition within Macro-Finance \& Economics}\\
  \footnotesize{This figure decomposes the \textit{Macro-Finance \& Economics} features contained in the overall top $N$ SAE features across their constituent topics. Features are ranked by the average absolute magnitude of their fitted coefficients across the annual estimation windows. The horizontal axis reports the overall cutoff $N$, and the vertical axis reports the percentage of retained Macro-Finance \& Economics features assigned to each topic.}}
  \label{fig:decomp_macro}
\end{figure}

\begin{figure}[!htbp]
  \centering
  \includegraphics[width=\textwidth]{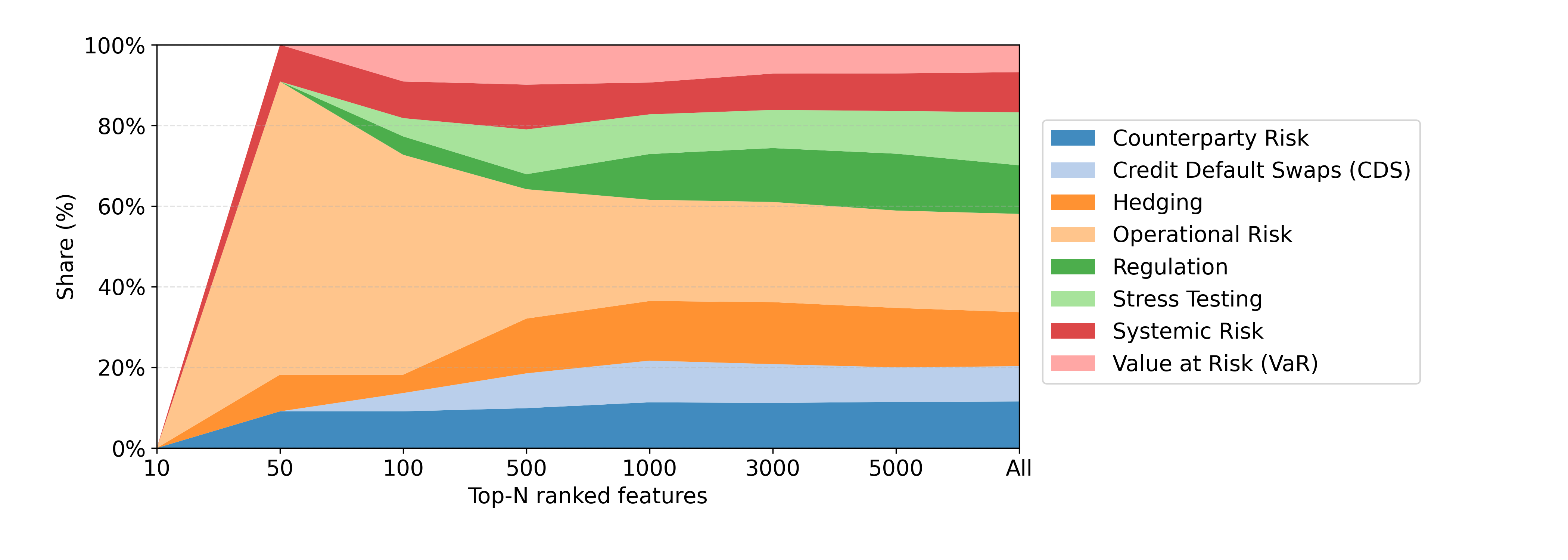}
  \caption{\textbf{Topic Composition within Risk Management \& Derivatives}\\
  \footnotesize{This figure decomposes the \textit{Risk Management \& Derivatives} features contained in the overall top $N$ SAE features across their constituent topics. Features are ranked by the average absolute magnitude of their fitted coefficients across the annual estimation windows. The horizontal axis reports the overall cutoff $N$, and the vertical axis reports the percentage of retained Risk Management \& Derivatives features assigned to each topic.}}
  \label{fig:decomp_risk}
\end{figure}

This section decomposes each of the five economic themes into its constituent topics. At each overall cutoff $N$, the figures identify the features from a given theme that appear among the top $N$ features and report their distribution across topics. The horizontal axis therefore refers to the overall feature ranking, while the vertical axis reports topic shares conditional on the corresponding theme. When no feature from a theme appears at a given cutoff, its within-theme shares are zero. Figures~\ref{fig:decomp_corp} through \ref{fig:decomp_risk} report one theme each.

The within-theme results show that the strongest predictors initially concentrate in a small number of topics and become more diverse as the representation expands. Among the \textit{Corporate Finance} features in the overall top ten, only \textit{Financial Distress} and \textit{Revenues Growth} appear. The \textit{Financial Markets \& Instruments} features at this cutoff concentrate in \textit{Commodities}, \textit{Derivatives}, and \textit{Fixed Income}, while the \textit{Macro-Finance \& Economics} features concentrate in \textit{Unemployment} and \textit{Gross Domestic Product}. No \textit{Investment \& Portfolio Management} or \textit{Risk Management \& Derivatives} features appear among the overall top ten. As $N$ increases, all five themes broaden to include a larger range of topics, including \textit{Governance} and \textit{Mergers \& Acquisitions} within Corporate Finance, \textit{High-Frequency Trading} and \textit{Initial Public Offerings} within Financial Markets, and \textit{Monetary Policy}, \textit{Quantitative Easing}, and the \textit{Yield Curve} within Macro-Finance. The topic-level decompositions therefore reinforce the main result: the strongest individual predictors are concentrated in a few broad concepts, while the full representation draws on a substantially wider set of economic mechanisms.

\FloatBarrier

\clearpage
\end{document}